\documentclass[twocolumn,rmp,aps,nofootinbib,amsmath,floatfix,preprintnumbers]{revtex4}

\usepackage{amssymb}  
\usepackage{graphicx}
\usepackage[dvips]{color}  

\newcommand{\al}{\alpha}
\newcommand{\be}{\beta}
\newcommand{\ga}{\gamma}

\newcommand{\de}{\delta}
\newcommand{\De}{\Delta}

\newcommand{\eps}{\epsilon}

\newcommand{\ka}{\kappa}

\newcommand{\Om}{\Omega}

\newcommand{\bm}[1]{\mbox{\boldmath${#1}$}}

\newcommand{\beq}{\begin{equation}}
\newcommand{\eeq}{\end{equation}}
\newcommand{\ba}{\begin{array}}
\newcommand{\ea}{\end{array}}
\newcommand{\bea}{\begin{eqnarray}}
\newcommand{\eea}{\end{eqnarray}}
\newcommand{\bi}{\begin{itemize}}  
\newcommand{\ei}{\end{itemize}}
\newcommand{\ben}{\begin{enumerate}} 
\newcommand{\een}{\end{enumerate}}
\newcommand{\bc}{\begin{center}}
\newcommand{\ec}{\end{center}}
\newcommand{\non}{\nonumber\\}

\newcommand{\p}{\partial}
\newcommand{\<}{\langle} 
\renewcommand{\>}{\rangle} 
\newcommand{\txt}{\textstyle}
\newcommand{\dsp}{\displaystyle}
\newcommand\eqn[1]{(\ref{#1})}      

\newcommand{\half} {{\txt \frac{1}{2}}}

\renewcommand{\O}{{\cal O}}

\newcommand{\MeV}{{\rm MeV}}

\newcommand{\Tr}{{\rm Tr}} 

\usepackage[normalem]{ulem}  

\begin{document}

\preprint{MIT-CTP-3861}

\title{Color superconductivity in dense quark matter}

\author{Mark G. Alford}
\email{alford@wuphys.wustl.edu}
\affiliation{Department of Physics, Washington University, 
  St Louis, MO 63130, USA}
\author{Krishna Rajagopal}
\email{krishna@lns.mit.edu}
\affiliation{Center for Theoretical Physics, Massachusetts Institute
of Technology, Cambridge, MA 02139, USA}
\author{Thomas Sch{\"a}fer}
\email{tmschaef@unity.ncsu.edu}
\affiliation{Department of Physics, North Carolina State University, 
  Raleigh, NC 27695, USA}
\author{Andreas Schmitt}
\email{aschmitt@wuphys.wustl.edu}
\affiliation{Department of Physics, Washington University, 
  St Louis, MO 63130, USA}

\begin{abstract}  
Matter at high density and low temperature is expected to be a color
superconductor, which is a degenerate Fermi gas of quarks with a
condensate of Cooper pairs near the Fermi surface that induces color
Meissner effects.
At the highest densities, where the QCD coupling is weak, rigorous
calculations are possible, and the ground state is a particularly
symmetric state, the color-flavor locked (CFL) phase. The CFL phase is a superfluid, an
electromagnetic insulator, and breaks chiral symmetry. The effective
theory of the low-energy excitations in the CFL phase is known and can
be used, even at more moderate densities, to describe its physical
properties.  At lower densities the CFL phase may be
disfavored by stresses that seek to separate the Fermi surfaces of the
different flavors, and comparison with the competing alternative
phases, which may break translation and/or rotation invariance, is
done using phenomenological models.  We review the calculations that
underlie these results, and then discuss transport properties of
several color-superconducting phases and their consequences for
signatures of color superconductivity in neutron stars.
\end{abstract}                                                                 

\date{21 Jan 2008}
\maketitle

\tableofcontents

\section{Introduction}
\label{sec:intro}
\subsection{General outline}
\label{sec:outline}

The study of matter at ultra-high density is the
``condensed matter physics of quantum chromodynamics''. It builds 
on our understanding of the strong interaction, derived from experimental
observation of few-body processes, to predict the behavior of macroscopic 
quantities in many-body systems where the fundamental particles of the 
standard model---quarks and leptons---become the relevant degrees of 
freedom. As in conventional condensed-matter physics, we seek to map the phase diagram
and calculate the properties of the phases. However, we are in the unusual
position of having a sector of the phase diagram where we can calculate many
properties of quark matter rigorously from first principles. This sector is
the region of  ``asymptotically high'' densities, where
quantum chromodynamics is weakly coupled.
We will review those rigorous results and 
describe the progress that has been made in building
on this solid foundation to extend our understanding to lower
and more phenomenologically relevant densities.
Quark matter occurs in various forms, depending on the temperature $T$
and quark chemical potential $\mu$ (see Fig.~\ref{fig:phase}).  At
high temperatures ($T\gg\mu$) entropy precludes any pattern of order
and there is only quark-gluon plasma (QGP), the phase of strongly
interacting matter that has no spontaneous symmetry breaking, and which
filled the universe for the first microseconds after the big
bang. Quark-gluon plasma is also being created in small, very
short-lived, droplets in ultrarelativistic heavy ion collisions at the
Relativistic Heavy Ion Collider.

In this review we concentrate on the regime of relatively low
temperatures, $T\ll\mu$, where we find a rich variety of spontaneous
symmetry breaking phases. To create such material in nature
requires a piston that can compress matter to super-nuclear
densities and hold it while it cools. The only known context
where this might happen is in the
interior of neutron stars, where gravity squeezes the star to an
ultra-high density state where it remains for millions of years.
This gives time for weak interactions to equilibrate, and for
the temperature of the star to drop far
below the quark chemical potential.
We do not currently know whether quark matter exists in the cores
of neutron stars. One of the reasons for studying
color superconductivity is to improve our understanding of how
a quark matter core would affect the observable behavior of
a neutron star, and thereby resolve this uncertainty.

When we speak of matter at the 
highest densities, we shall always take the high density limit 
with up, down and strange quarks only. We do so because neutron
star cores are not dense enough (by more than an order of magnitude) 
to contain any charm or heavier quarks, and
our ultimate goal is to gain insight into quark matter at 
densities that may be found in nature. 
For the same reason we focus on temperatures below about ten MeV,
which are appropriate for neutron stars that are more than a few seconds old.

As we will explain in some detail, at low temperatures and the highest 
densities we expect to find a degenerate liquid 
of quarks, with Cooper pairing near the Fermi surface that spontaneously 
breaks the color gauge symmetry (``color superconductivity''). Speculations 
about the existence of a quark matter phase at high density go back to the 
earliest days of the quark model of hadrons
\cite{Carruthers:1973,Ivanenko:1965,Pacini:1966,Boccaletti:1966,Itoh:1970uw}, 
and the possibility of quark Cooper pairing was noted even before
there was a comprehensive theory of the strong interaction
\cite{Ivanenko:1969gs,Ivanenko:1969bs}. After the development of 
quantum chromodynamics (QCD), with its property of asymptotic freedom 
\cite{Gross:1973id,Politzer:1973fx}, it became clear that a quark matter 
phase would exist at sufficiently high density 
\cite{Collins:1974ky,Kislinger:1975uy,Freedman:1976ub,Freedman:1977gz,Baym:1976yu,Chapline:1976gq,Chapline:1976gy} 
and the study of quark Cooper 
pairing was pioneered by Barrois and Frautschi 
\cite{Barrois:1977xd,Barrois:1979pv,Frautschi:1978rz}, who first used 
the term ``color superconductivity'', and by Bailin and Love 
\cite{Bailin:1979nh,Bailin:1983bm}, who classified many of the possible 
pairing patterns. Iwasaki and Iwado \cite{Iwasaki:1994ij,Iwasaki:1995uw} 
performed mean-field calculations of single-flavor pairing in a 
Nambu-Jona-Lasinio (NJL) model, but it was not until the prediction 
of large pairing gaps \cite{Alford:1997zt,Rapp:1997zu} and the
color-flavor locked (CFL) phase \cite{Alford:1998mk} that the phenomenology 
of color-superconducting quark matter became widely studied. At present 
there are many reviews of the subject from various stages in its development
\cite{Shovkovy:2004me,Nardulli:2002ma,Huang:2004ik,Buballa:2003qv,Ren:2004nn,Hsu:2000sy,Bailin:1983bm,Rajagopal:2000wf,Alford:2001dt,Hong:2000ck,Rischke:2003mt,Schafer:2003vz,Reddy:2002ri,Alford:2006fw}, 
and the reader may wish to consult them for alternative presentations 
with different emphases.
As these reviews make clear, the last decade has seen dramatic 
progress in our understanding of dense matter. We are now able to
obtain, directly from QCD,
rigorous and quantitative answers  to the basic
question: ``What happens to matter if you squeeze it to arbitrarily
high density?''. In Sec.\ \ref{sec:QCD}  we will show how QCD becomes
analytically tractable at arbitrarily high density: the coupling
is weak and the physics of confinement never arises, since
long-wavelength magnetic interactions are cut off, both by Landau 
damping and by the Meissner effect. 
As a result,  matter at the highest densities is known to be in the 
CFL phase, whose properties (see Sec.~\ref{sec:cfl})
are understood rigorously from first principles. There is
a well-developed effective field theory describing the low energy 
excitations of CFL matter (see Sec.~\ref{sec_eft}), so
that at any density at which the
CFL phase occurs, even if this density is not high enough for a 
weak-coupling QCD calculation to be valid, many phenomena can 
nevertheless be described quantitatively in terms of a few parameters, 
via the effective field theory.    

It should be emphasized that QCD at arbitrarily high density is more
fully understood than in any other context.  High energy scattering,
for example, can be treated by perturbative QCD, but making contact
with observables brings in poorly understood nonperturbative physics
via structure functions and/or fragmentation functions.  Or, in
quark-gluon plasma in the high temperature limit much of the physics
is weakly-coupled but the lowest energy modes remain strongly coupled
with nonperturbative physics arising in the nonabelian color-magnetic
sector. We shall see that there are no analogous difficulties in the
analysis of CFL matter at asymptotic densities.

If the CFL phase persists all the way down to the transition to
nuclear matter then we have an exceptionally good theoretical
understanding of  the properties 
of quark matter in nature. However, less 
symmetrically paired phases of quark matter may well intervene 
in the intermediate density region between nuclear and CFL matter 
(Sec.~\ref{sec:BCSstress}). 
We enumerate some of the possibilities in Sec.~\ref{sec:non_cfl}.
In principle this region could also be understood 
from first principles, using brute-force numerical methods (lattice QCD) to
evaluate the QCD path integral, but unfortunately current lattice QCD
algorithms are defeated by the fermion sign problem in the
high-density low-temperature regime \cite{Schmidt:2006us}.\footnote{
Condensation of Cooper pairs of quarks has been studied
on the lattice in 2-color QCD 
\cite{Nishida:2003uj,Kogut:1999iv,Kogut:2000ek,Fukushima:2007bj,Hands:1999md,Kogut:2001na,Hands:2006ve,Kogut:2002cm,Alles:2006ea}, 
for high isospin
density rather than baryon density \cite{Kogut:2002zg,Splittorff:2000mm,Son:2000xc} and
in NJL-type models \cite{Hands:2004uv}.}
This means
we have to use models, or try to derive information from astrophysical
observations.
In Sec.~\ref{sec:NJL}
we sketch an example of a (Nambu--Jona-Lasinio) model analysis 
within which one can compare some of the possible intermediate-density
phases suggested in Sec.~\ref{sec:BCSstress}.
We finally discuss the observational approach, which involves
elucidating the properties of the suggested phases of quark matter
(Secs.~\ref{sec:rigid} and \ref{sec_trans}), and then finding 
astrophysical signatures by
which their presence inside neutron stars might be established or ruled
out using astronomical observations (Sec.~\ref{sec:astro}).

\subsection{Inevitability of color superconductivity}
\label{sec:inevitability}

At sufficiently high density and low temperature it is a good starting point 
to imagine that quarks form a degenerate Fermi liquid. Because QCD is
asymptotically free --- the interaction becomes weaker as the momentum
transferred grows --- the quarks near the Fermi surface are almost free,
with weak QCD interactions between them. (Small-angle 
quark-quark scattering via a low-momentum gluon is no problem
because it is cut off by Landau damping, which, together with
Debye screening, keeps perturbation theory at high density
much better controlled than at high temperature
\cite{Pisarski:1998nh,Son:1998uk}.)
The quark-quark interaction is certainly attractive 
in some channels, since we know that quarks bind together to form baryons.
As we will now argue, these conditions are sufficient to guarantee
color superconductivity at sufficiently high density.

At zero temperature, the thermodynamic potential (which we will loosely refer
to as the ``free energy'') is $\Om = E-\mu N$, where $E$ is
the total energy of the system, $\mu$ is the chemical potential, and
$N$ is the number of fermions.
If there were no interactions then the energy required to add a particle
to the system would be the Fermi energy $E_F=\mu$, so adding or subtracting 
particles or holes near the Fermi surface would cost zero free energy. 
With a weak attractive interaction in any channel, if we
add a pair of particles (or holes) 
with the quantum numbers of the attractive channel, the free energy
is lowered by the potential energy of their attraction.
Many such pairs will therefore be created in the modes near the Fermi 
surface, and these pairs, being bosonic, will form a condensate. The 
ground state will be a superposition of states with different numbers 
of pairs, breaking the fermion number symmetry. This argument,
originally developed by Bardeen, Cooper, 
and Schrieffer (BCS) \cite{BCS} is 
completely general, and can be applied to electrons in a metal, nucleons 
in nuclear matter, $^3$He atoms, cold fermionic atoms in a trap, or 
quarks in quark matter.

The application of the BCS mechanism to pairing in dense quark matter
is in a sense more direct than in its original setting.  The dominant 
interaction between electrons in a metal is the repulsive Coulomb 
interaction, and it is only because this interaction is screened that 
the attraction mediated by phonons comes into play.  This means that 
the effective interactions that govern superconductivity in a metal 
depend on band structure and other complications and are very difficult 
to determine accurately from first principles.  
In contrast, in QCD the ``color Coulomb" interaction is attractive
between quarks whose color wave function is antisymmetric, meaning that
superconductivity arises as a direct consequence of the primary
interaction in the theory.  This has two important consequences.
First, at asymptotic densities where the QCD  interaction is weak
we can derive the gap parameter and other properties of color
superconducting quark matter rigorously from the underlying microscopic
theory.  Second, at accessible densities where the QCD interaction
is stronger the ratio of the gap parameter to the Fermi energy will
be much larger than in conventional BCS superconducting metals.
Thus, superconductivity in QCD is more robust, both in the
theoretical sense and in the phenomenological sense, than
superconductivity in metals.

It has long been known 
that, in the absence of pairing,
an unscreened static magnetic interaction results in a 
``non-Fermi-liquid" \cite{Holstein:1973,Ipp:2006ij,Chakravarty,Baym:1990uj,Vanderheyden:1996bw,Manuel:2000mk,Manuel:2000nh,Brown:2000eh,Boyanovsky:2000bc,Boyanovsky:2000zj,Ipp:2003cj,Gerhold:2004tb,Polchinski:1993ii,Nayak:1994ng}. However, in QCD the 
magnetic interaction is screened at nonzero frequency (Landau damping) 
and this produces a particularly mild form of non-Fermi-liquid behavior, 
as we describe in Sec.~\ref{sec_nfl}. In the absence of pairing but in the 
presence of interactions, there are still quark quasiparticles and 
there is  still a ``Fermi surface", and the BCS argument goes through.    
This argument is rigorous at high densities, where the QCD coupling $g$ 
is small. The energy scale below which non-Fermi liquid effects would 
become strong enough to modify the quasiparticle picture qualitatively 
is parametrically of order $\exp(-{\rm const}/g^2)$ whereas the BCS gap 
that results from pairing is parametrically larger, of order 
$\exp(-{\rm const}/g)$ as we shall see in Sec.~\ref{sec:QCD}.
This means that pairing occurs
in a regime where the basic logic of the BCS argument remains 
valid.

Since pairs of quarks cannot be color singlets, the Cooper pair
condensate in quark matter will break the local color symmetry
$SU(3)_c$, hence the term ``color superconductivity''. The quark pairs 
play the same role here as the Higgs particle does in the standard model: 
the color-superconducting phases can be thought of as Higgs phases of 
QCD. Here, the gauge bosons that acquire a mass through the process of 
spontaneous symmetry breaking are the gluons, giving rise to color Meissner effects. 
It is important to note 
that quarks, unlike electrons, have color and flavor as well as 
spin degrees of freedom, so many different patterns of pairing are 
possible. This leads us to expect a panoply of different possible 
color superconducting phases. 

As we shall discuss in Sec.~\ref{sec:cfl},
at the highest densities we can achieve an {\it ab initio} understanding 
of the properties of dense matter, and we find that its preferred state is
the CFL phase of three-flavor quark 
matter, which is unique in that {\it all} the
quarks pair (all flavors, all colors, all spins, all momenta 
on the Fermi surfaces) and all the nonabelian gauge bosons are massive.
The suppression of all of the infrared degrees of freedom
of the types that typically indicate either
instability toward further condensation or 
strongly coupled phenomena 
ensures that, at sufficiently high density,
the CFL ground state, whose only infrared 
degrees of freedom are Goldstone bosons
and an abelian photon, is stable.
In this regime, quantitative calculations of observable properties of
CFL matter can be done from first principles; there are no remaining
nonperturbative gaps in our understanding. 

As the density decreases, the effect of the strange quark mass becomes
more noticeable, imposing stresses that may modify the Cooper pairing
and the CFL phase may be replaced by other forms of color
superconducting quark matter. 
Furthermore, as the attractive interaction
between quarks becomes stronger at lower densities, correlations
beyond the two-body correlation that yields Cooper pairing may become
important, and at some point the ground state will no longer be a
Cooper-paired state of quark matter, but something quite
different. Indeed, by the time we decrease the density down to that of
nuclear matter, the average separation between quarks has increased to
the point that the interactions are strong enough to bind quarks into
nucleons. It is worth noting that quark matter is in this regard 
different from Cooper-paired ultracold fermionic atoms (to be
discussed in Sec.~\ref{sec_atoms}). For fermionic atoms, as the
interaction strength increases there is a crossover from BCS-paired
fermions to a Bose-Einstein condensate (BEC) of tightly-bound,
well-separated, weakly-interacting di-atoms (molecules).  In QCD,
however, the color charge of a diquark is the same as that of an
antiquark, so diquarks will interact with each other as strongly as
quarks, and there will not be a literal analogue of the BCS/BEC
crossover seen in fermionic atoms.  In QCD, the neutral bound states
at low density that are (by QCD standards) weakly interacting are
nucleons, containing three quarks not two.

We shall work with $N_c=3$ colors throughout. In the limit $N_c\to
\infty$ with fixed $\Lambda_{\rm QCD}$ (i.e fixed $g^2N_c$), Cooper
pairing is not necessarily energetically preferred. A strong
competitor for the large-$N_c$ ground state is the chiral density wave
(CDW), a condensate of quark-hole pairs, each with total momentum
$2p_F$ \cite{Deryagin:1992rw}. 
Quark-hole scattering is enhanced by a factor of $N_c$ over quark-quark 
scattering, but, unlike Cooper pairing, it only uses a small
fraction of the Fermi surface, and in the case of short range forces
the CDW phase is energetically favored in one-dimensional
systems, but not in two or more spatial dimensions
\cite{Shankar:1993pf}. However, in QCD in the large $N_c$ limit
the equations governing the CDW state become effectively one-dimensional
because the gluon propagator is not modified by the medium, 
so the quark-hole 
interaction is dominated by almost collinear scattering.
Since pairing gaps are exponentially small in the coupling but medium
effects only vanish as a power of $N_c$, the CDW state requires an
exponentially large number of colors. It is estimated that
for $\mu\sim 1$~GeV, quark-hole pairing becomes favored over
Cooper pairing when $N_c\gtrsim 1000$ \cite{Shuster:1999tn}.
Recent work \cite{McLerran:2007qj} discusses aspects
of physics at large $N_c$ at lower densities that may 
also be quite different from
physics at $N_c=3$.

Before turning to a description of CFL pairing in Sec.~\ref{sec:cfl} 
and less-symmetrically paired forms of color
superconducting quark matter in Sec.~\ref{sec:non_cfl}, we discuss
some generic topics that arise in the analysis of color-superconducting
phases: the gap equations, neutrality constraints, the resultant
stresses on Cooper pairing, and the expected overall form of the
phase diagram.

\subsection{Quark Cooper pairing}

The quark pair condensate can be characterized in a gauge-variant way
by the expectation value of the one-particle-irreducible quark-quark
two-point function, also known as the ``anomalous self-energy'',
\beq
\<\psi^\al_{ia} \psi^\be_{jb}\>=P^{\al\be}_{ij\,ab}\De 
\label{cooper}
\eeq
Here $\psi$ is the quark field operator, color indices $\al,\be$ range
over red, green, and blue ($r,g,b$), flavor indices $i,j$ range over 
up, down and strange ($u,d,s$), and $a,b$ are the spinor Dirac indices. 
The angle brackets denote the one-particle-irreducible part of the
quantum-mechanical ground-state expectation 
value. In general, both sides of this equation are functions of momentum.
The color-flavor-spin matrix $P$ characterizes a particular pairing channel, 
and $\De$ is the gap parameter which gives the strength of the pairing 
in this channel.  A standard BCS condensate is position-independent (so that in momentum 
space the pairing is between quarks with equal and opposite momentum) and 
a spin singlet (so that the gap is isotropic in momentum space). However, 
as we will see later, there is good reason to expect non-BCS condensates
as well as BCS condensates in high-density quark matter.

Although \eqn{cooper} defines a gauge-variant quantity, 
it is still of physical relevance. Just as electroweak symmetry breaking 
is most straightforwardly understood in the unitary gauge where the Higgs 
vacuum expectation value is uniform in space, so color superconductivity 
is typically analyzed in the unitary gauge where the quark pair operator 
has a uniform color orientation in space. We then relate the gap parameter 
$\De$ to the spectrum of the quark-like excitations above the ground 
state (``quasiquarks''), which is gauge-invariant.

In principle, a full analysis of the phase structure of quark matter
in the $\mu$-$T$ plane would be performed by writing down the 
free energy $\Om$, which is a function of the temperature, the chemical
potentials for all conserved quantities, and the gap parameters for
all possible condensates, including the quark pair condensates but also
others such as chiral condensates of the form $\<\bar q q\>$. We impose 
neutrality with respect to gauge charges (see Sect.~\ref{sec:neutrality} 
below) and then within the neutral subspace we minimize the free energy 
with respect to the strength of the condensate:
\beq
\frac{\p \Om}{\p \De} = 0, \qquad \frac{\p^2 \Om}{\p \De^2}>0\ .
\label{gap_eqn}
\eeq
We have written this gap equation and stability condition somewhat 
schematically since for many patterns of pairing there will be gap 
parameters with different magnitudes in different channels. The free 
energy must then be minimized with respect to each of the gap
parameters, yielding a coupled set of gap equations. The solution to 
\eqn{gap_eqn} with the lowest free energy that respects the neutrality 
constraints discussed below yields the favored phase.

\subsection{Chemical potentials and neutrality constraints}
\label{sec:neutrality}

Why do we describe ``matter at high density" by introducing a large
chemical potential $\mu$ for quark number but no chemical potentials
for other quantities?  The answer is that this reflects the
physics of neutron stars, which are the main physical arena that we consider.
Firstly, on the long timescales relevant to neutron stars,
the only global charges that are conserved
in the standard model are quark number and lepton number, so only these
can be coupled to chemical potentials (we shall discuss gauged charges
below). Secondly, a neutron star is permeable to 
lepton number because neutrinos are so light and weakly-interacting 
that they can quickly escape from the star, so the chemical potential
for lepton number is zero. 
Electrons are present because they carry
electric charge, for which there is a nonzero potential.
In the first few seconds of 
the life of a neutron star the neutrino mean free path may be short 
enough to sustain a nonzero lepton number chemical potential, see for 
instance~\cite{Ruester:2005ib,Kaplan:2001qk,Laporta:2005be,Berdermann:2004da}, but we will
not discuss that scenario.

Stable bulk matter must be neutral under all gauged charges, whether
they are spontaneously broken or not. Otherwise, the net charge
density would create large electric fields, making the energy
non-extensive.  In the case of the electromagnetic gauge symmetry,
this simply requires zero charge density, $Q=0$. The correct formal 
requirement concerning the color charge of a large lump of matter is 
that it should be a color {\em singlet}, i.e., its wavefunction should 
be invariant under a general color gauge transformation. However,
it is sufficient for us to impose color {\em neutrality},
meaning equality in the numbers of red, green, and 
blue quarks. This is a less stringent constraint 
(singlet$\,\Rightarrow\,$neutral but neutral$\,\not\Rightarrow\,$singlet)
but the projection of a color neutral 
state onto a color singlet costs no extra free energy in the 
thermodynamic limit~\cite{Amore:2001uf}. 
(See also~\cite{Elze:1983du,Elze:1984un}.)
In general there are 8 possible 
color charges, but because the
Cartan subalgebra of $SU(3)_c$ is two-dimensional 
it is always possible to transform to a gauge 
where all are zero except $Q_3$ and $Q_8$, the charges associated 
with the diagonal generators $T_3=\frac{1}{2}\,{\rm diag}(1,-1,0)$ 
and $T_8=\frac{1}{2\sqrt{3}}\,{\rm diag}(1,1,-2)$
in $(r,g,b)$ space 
\cite{Buballa:2005bv,Rajagopal:2005dg}. In this review, we only 
discuss such gauges. So to impose color neutrality we just require 
$Q_3=Q_8=0$. 

In nature, electric and color neutrality are enforced by the dynamics
of the electromagnetic and QCD gauge fields, whose zeroth components
serve as chemical potentials coupled to the charges $Q,Q_3,Q_8$,
and which are naturally driven to values that set these charges to 
zero~\cite{Iida:2000ha,Alford:2002kj,Gerhold:2003js,Kryjevski:2003cu,Dietrich:2003nu}.  
In an NJL model with fermions but no gauge fields (see Sec.~\ref{sec:NJL}) 
one has to introduce the chemical potentials $\mu_e$, 
$\mu_3$ and $\mu_8$ by hand in order to enforce color and electric 
neutrality. The neutrality conditions are then
\beq
\label{neutrality}
\ba{rcl}
Q  = \phantom{-}\dsp\frac{\partial \Omega}{\partial\mu_e}&=& 0 \\[2ex]
Q_3= -\dsp\frac{\partial \Omega}{\partial\mu_3}&=& 0 \\[2ex]
Q_8= -\dsp\frac{\partial \Omega}{\partial\mu_8}&=& 0\ .
\ea
\eeq
(Note that we define an electrostatic potential $\mu_e$ that is
coupled to the {\em negative} electric charge $Q$, so that in typical
neutron star conditions, where there is a finite density of electrons
rather than positrons, $\mu_e$ is positive.)  

Finally we should note that enforcing local neutrality is appropriate 
for uniform phases, but there are also non-uniform charge-separated 
phases (``mixed phases''), consisting of positively and negatively 
charged domains which are neutral on average. These are discussed
further in Sec.~\ref{sec_mixed}.

\subsection{Stresses on BCS pairing}
\label{sec:BCSstress}

The free energy argument that we gave in Sec.~\ref{sec:inevitability} 
for the inevitability of BCS pairing in the presence of an attractive 
interaction relies on the assumption that the quarks that pair with 
equal and opposite momenta can each be arbitrarily close to their 
common Fermi surface. However, as we will see in Sec.~\ref{sec:cfl}, 
the neutrality constraint, combined with the mass of the strange quark 
and the requirement that matter be in beta equilibrium, tends to pull 
apart the Fermi momenta of the different flavors of quarks, imposing 
an extra energy cost (``stress'') on the formation of Cooper pairs 
involving quarks of different flavors.  This raises the possibility 
of non-BCS pairing in some regions of the phase diagram.

To set the stage here, let us discuss a simplified example: consider 
two massless species of fermions, labeled $1$ and $2$, with different 
chemical potentials $\mu_1$ and $\mu_2$, and an attractive interaction 
between them that favors cross-species BCS pairing with a gap parameter 
$\De$. It will turn out that to a good approximation
the color-flavor locked pairing pattern contains three such sectors, so
this example captures the essential physics we will encounter in later 
sections. We define the average chemical potential and the stress parameter
\beq \label{deltamu}
\ba{rcl}
\bar\mu &=& \half(\mu_1+\mu_2) \\
\de\mu &=& \half(\mu_1-\mu_2)\ .
\ea
\eeq
As long as the stress $\de\mu$ is small enough relative to $\De$,
BCS pairing between species 1 and 2 can occur,
locking their Fermi surfaces together and ensuring that
they occur in equal numbers. 
At the Chandrasekhar-Clogston point \cite{Clogston:1962,Chandrasekhar:1962},
where $\de\mu=\De/\sqrt{2}$,
the two-species model undergoes a first-order transition to
the unpaired phase. 
At this point BCS pairing still exists as a locally stable
state, with a completely gapped spectrum of quasiparticles.
When $\de\mu$ reaches $\De$ the spectrum becomes gapless at momentum 
$p=\bar\mu$, indicating that
cross-species BCS pairing is no longer favored at all momenta 
\cite{Alford:2003fq}. 
If the two species are part of a larger pairing pattern, the
Chandrasekhar-Clogston transition can be shifted, and we shall
see that in the two-species subsectors of the CFL pattern it
is shifted to $\de\mu>\Delta$.
The onset of gaplessness is therefore the relevant threshold
for our purposes, and it always occurs at $\de\mu=\De$,
independent of the larger context in which the two flavors pair.
This follows from the fact that BCS pairing 
only occurs if the energy gained from turning a $1$ quark into a 
$2$ quark with the same momentum (namely $\mu_1-\mu_2$) is less 
than the cost of breaking the Cooper pair formed by these quarks, 
which is $2\De$ \cite{Rajagopal:2000ff}. Thus the $1$-$2$ Cooper 
pairs are energetically stable (or metastable) as long as $\de\mu
<\De$. A more detailed treatment of this illustrative example can 
be found in \cite{Alford:2005qw}.

This example uses massless quarks, but it can easily be modified to
include the leading effect of a quark mass $M$. A difference in the 
masses of the pairing quarks also stresses the pairing, because it 
gives them different Fermi momenta at the same chemical potential, 
so the quarks in a $1$-$2$ Cooper pair, which have equal and opposite 
momenta, will not both be close to their Fermi energies. The 
leading-order effect is easily calculated, since for a quark near 
its Fermi surface it acts like a shift in the quark chemical 
potential by $-M^2/(2\bar\mu)$ (given that Fermi momentum $p_F\approx
\bar\mu$ to this order).

 Returning from our toy model to realistic quark matter, the quark 
flavors that are potentially relevant at neutron-star densities are 
the light up ($u$) and down ($d$) quarks, with current masses $m_u$ 
and $m_d$ that are $\lesssim 5~{\rm MeV}$, and a medium-weight flavor, 
the strange ($s$) quark, with current mass $m_s\sim 90~{\rm MeV}$.   
Their effective ``constituent'' masses in the vacuum are hundreds 
of MeV larger, but are expected to decrease with increasing quark 
density. We shall refer to the density-dependent constituent masses 
as $M_{u,d,s}$ and shall typically neglect $M_u$ and $M_d$. As our 
toy model has illustrated, however, the strange quark mass $M_s$ 
will contribute to stresses on cross-flavor pairing, and those 
stresses will become more severe as the density (and hence $\bar\mu$) 
decreases. This will be a major theme of later sections.

\subsection{Overview of the quark matter phase diagram}
\label{subsec:overview}

\begin{figure}[t]
\begin{center} 
 \includegraphics[width=\hsize]{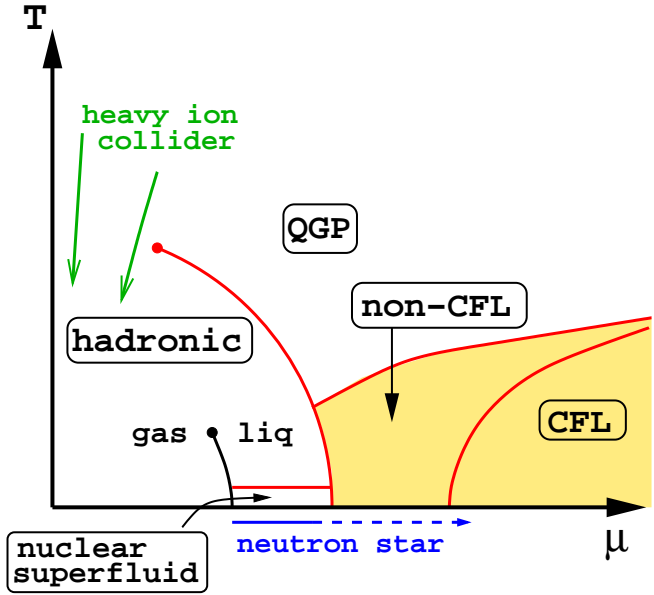}
\end{center}
\caption{(Color online) A schematic outline for the phase diagram of matter
at ultra-high density and temperature. The CFL phase is a superfluid
(like cold nuclear matter) and has broken chiral symmetry (like the
hadronic phase).}
\label{fig:phase}
\end{figure}

Fig.~\ref{fig:phase} shows a schematic phase diagram for QCD that is
consistent with what is currently known.
Along the horizontal axis the temperature is zero, and the density is zero
up to the onset transition where it jumps to nuclear density, and 
then rises with increasing $\mu$.
Neutron stars are in this region of the phase diagram, although it is 
not known whether their cores are dense enough to reach the quark matter 
phase. Along the vertical axis the temperature rises, taking us through 
the crossover from a hadronic gas to the quark-gluon plasma. This is the 
regime explored by high-energy heavy-ion colliders.

At the highest densities we find the color-flavor locked 
color-superconducting phase,\footnote{
As explained in Sec.~\ref{sec:outline}, we fix $N_f=3$
at all densities, to maintain relevance to neutron star interiors.
Pairing with arbitrary $N_f$ has been studied
\cite{Schafer:1999fe}. For $N_f$ a multiple 
of three one finds multiple copies of the CFL pattern; for $N_f=4,5$ 
the pattern is more complicated.}
in which the strange quark participates 
symmetrically with the up and down quarks in Cooper 
pairing. This is 
described in more detail in Secs.\ \ref{sec:cfl}, \ref{sec:QCD}, and \ref{sec_eft}. 
It is not yet clear what 
happens at intermediate density, and in Secs.\ \ref{sec:non_cfl} and \ref{sec:NJL} we will
discuss the factors that disfavor the CFL phase at intermediate 
densities, and survey the color superconducting phases that 
have been hypothesized to occur there.

Various aspects of color superconductivity at high temperatures 
have been studied, including the phase structure 
(see Sec.~\ref{sec:NJLModel}), spectral functions, pair-forming and -breaking fluctuations,
possible precursors to condensation such as pseudogaps,
and various collective phenomena
\cite{Kitazawa:2007zs,Kitazawa:2005vr,Kitazawa:2001ft,Kitazawa:2005pp,Fukushima:2005gt,Abuki:2001be,Kitazawa:2003cs,Voskresensky:2004jp,Yamamoto:2007ah,Hatsuda:2006ps}.
However, this review centers on quark matter at neutron star
temperatures, and
throughout Secs.~\ref{sec:cfl} and \ref{sec:non_cfl} we restrict
ourselves to the
phases of quark matter at zero temperature. This is because most of
the phases that we discuss are expected to persist up to critical
temperatures that are well above the core temperature of a typical
neutron star, which drops below 1 MeV within seconds of its birth
before cooling down through the keV range over millions of years.

\section{Matter at the highest densities}
\label{sec:cfl}
\subsection{Color-flavor locked (CFL) quark matter}
\label{subsec:cfl}

Given that quarks form Cooper pairs, the next question is who pairs with 
whom?  In quark matter at sufficiently high densities, where the up, down 
and strange quarks can be treated on an equal footing and the disruptive 
effects of the strange quark mass can be neglected, the most symmetric and
most attractive option is the color-flavor locked phase, where 
quarks of all three colors and all 
three flavors form conventional zero-momentum spinless Cooper pairs.
This pattern,
anticipated in early studies of
alternative condensates for zero-density chiral symmetry breaking
\cite{Srednicki:1981cu}, is encoded in
the quark-quark self-energy \cite{Alford:1998mk}
\begin{equation}
\label{CFLpattern}
\begin{array}{r} 
\langle \psi^\alpha_i C \gamma_5 \psi^\beta_j \rangle
\propto \Delta_{\rm CFL} (\kappa\!+\!1)\delta^\alpha_i\delta^\beta_j 
+ \Delta_{\rm CFL} (\kappa\!-\!1) \delta^\alpha_j\delta^\beta_i \\[1ex]
= \Delta_{\rm CFL} \eps^{\al\be A}\eps_{ij A} + \Delta_{\rm CFL} \ka
(\delta^\alpha_i\delta^\beta_j +\delta^\alpha_j\delta^\beta_i)
\end{array}
\end{equation}
The symmetry breaking pattern is
\begin{equation}
\label{CFLsymmetry}
\begin{array}{c}
 [SU(3)_c] \times U(1)_B \\[1ex]
 \times\,\,\underbrace{SU(3)_L \times SU(3)_R}_{\displaystyle\supset [U(1)_Q]}
\end{array}
\to \begin{array}{c}
  \phantom{SU(3)_c} \\[1ex]  
  \underbrace{SU(3)_{c+L+R}}_{\displaystyle\supset [U(1)_{{\tilde Q} }]}
  \times \,\mathbb{Z}_2
\end{array}
\end{equation} 
Color indices $\alpha,\beta$ and flavor indices $i,j$ run from 1 to 3,
Dirac indices are suppressed, and $C$ is the Dirac charge-conjugation
matrix.  Gauge symmetries are in square brackets. $\Delta_{\rm CFL}$
is the CFL gap parameter. The Dirac structure $C\ga_5$ is a Lorentz
singlet, and corresponds to parity-even spin-singlet pairing, so it is
antisymmetric in the Dirac indices. The two quarks in the Cooper pair
are identical fermions, so the remaining color+flavor structure must
be symmetric. The dominant color-flavor component in \eqn{CFLpattern}
transforms as $(\bar{\bf 3}_A,\bar{\bf 3}_A)$, antisymmetric in both. The
subdominant term, multiplied by $\kappa$, transforms as $({\bf 6}_S,{\bf 6}_S)$.
It is almost certainly not energetically favored on its own 
(all the arguments in Sec.~\ref{sec:CFLfavored} for the color triplet imply repulsion
for the sextet), 
but in the presence of the dominant pairing it
breaks no additional symmetries, so $\kappa$ is in general small but
not zero
\cite{Alford:1998mk,Schafer:1999fe,Shovkovy:1999mr,Pisarski:1999cn}.

\subsubsection{Color-flavor locking and chiral symmetry breaking}
\label{sec:CFLchiral}

A particularly striking feature of the CFL pairing pattern is that it breaks
chiral symmetry. Because of color-flavor locking, chiral symmetry
remains broken up to arbitrarily high densities in three-flavor quark
matter. The mechanism is quite different from the formation of the
$\<\bar\psi\psi\>$ condensate that breaks chiral symmetry in the
vacuum by pairing left-handed (L) quarks with right-handed (R) antiquarks.
The CFL condensate pairs L quarks with each other and R quarks with each 
other (quarks in a Cooper pair have opposite momentum, and zero net spin,
hence the same chirality) and so it might naively appear chirally symmetric.
However, the Kronecker deltas in \eqn{CFLpattern} connect color indices
with flavor indices, so that the condensate is not invariant under
color rotations, nor under flavor rotations, but only under
simultaneous, equal and opposite, color and flavor
rotations. Color is a vector symmetry, so the compensating flavor rotation
must be the same for L and R quarks, so
the axial part of the flavor group, which is
the chiral symmetry, is broken by the
locking of color and flavor
rotations to each other \cite{Alford:1998mk}.  Such locking is familiar from other
contexts, including the QCD vacuum, where a condensate of
quark-antiquark pairs locks $SU(3)_L$ to $SU(3)_R$ breaking chiral
symmetry ``directly'', and the B phase of superfluid ${}^3$He, where
the condensate transforms nontrivially under rotations of spin and
orbital angular momentum, but is invariant under simultaneous
rotations of both.

The breaking of the chiral symmetry is associated with an expectation
value for a gauge-invariant order parameter with the structure
$\bar\psi\bar\psi\psi\psi$ (see Sec.~\ref{sec_eft}).
There is also a subdominant ``conventional'' chiral
condensate $\langle{\bar\psi}\psi\rangle\ll \langle \psi C\gamma_5
\psi\rangle$ \cite{Schafer:1999fe}. 
These gauge-invariant observables distinguish the CFL phase from the QGP,
and if a lattice QCD algorithm applicable at high density ever becomes 
available, 
they could be used to map the presence of color-flavor locking in the
phase diagram.

We also expect massless Goldstone modes associated with chiral symmetry 
breaking (see Secs.~\ref{sec:CFLeffth} and \ref{sec_eft}).
In the real world there is small explicit breaking of chiral symmetry
from the current quark masses, so the order parameters will not go to zero
in the QGP, and the Goldstone bosons will be light but not massless.

\subsubsection{Superfluidity}
\label{sec:CFLsuperfluid}

The CFL pairing pattern spontaneously breaks
the exact global baryon number symmetry $U(1)_B$, 
leaving only a discrete $\mathbb{Z}_2$ symmetry under 
which all quark fields are multiplied by $-1$. There is
an associated gauge-invariant 6-quark order parameter with the
flavor and color structure of two Lambda baryons,
$\langle \Lambda \Lambda\rangle$
where $\Lambda=\epsilon^{abc}\epsilon_{ijk}\psi^a_i
\psi^b_j\psi^c_k$. This order parameter distinguishes the CFL phase 
from the QGP, and there is an associated massless Goldstone boson that makes 
the CFL phase a superfluid, see Sec.\ \ref{sec_eft_u1}. 
The vortices that result 
when CFL quark matter is rotated have been studied in \cite{Balachandran:2005ev,Nakano:2007dr,Iida:2002ev,Forbes:2001gj}.

\subsubsection{Gauge symmetry breaking and electromagnetism}
\label{sec:CFLgauge}

As explained above, the CFL condensate breaks the 
$SU(3)_c \times SU(3)_L \times SU(3)_R$ symmetry down to
the diagonal group $SU(3)_{c+L+R}$ of simultaneous color
and flavor rotations. Color is a gauge symmetry, and one of the
generators of $SU(3)_{L+R}$ is the electric charge, which
generates the $U(1)_Q$ gauge symmetry. This means that
the unbroken $SU(3)_{c+L+R}$ contains one gauged generator,
corresponding to an unbroken $U(1)_{\rm \tilde Q}$ which consists of
a simultaneous electromagnetic and color rotation. 
The rest of the color group is broken, so by the Higgs mechanism
seven gluons and 
one gluon-photon linear combination become massive via the Meissner 
effect. The orthogonal gluon-photon generator $\tilde Q$ remains
unbroken, because every diquark in the condensate has $\tilde Q=0$.
The mixing angle is $\cos\theta\equiv g/\sqrt{g^2 + 4 e^2/3}$ where
$e$ and $g$ are the QED and QCD couplings. Because
$e\ll g$ the angle is close to
zero, meaning that the $\tilde Q$ photon is mostly the original photon
with a small admixture of gluon. 

The $\tilde Q$ photon is massless. Given small but nonzero quark masses, 
there are no gapless $\tilde Q$-charged
excitations; the lightest ones are the 
pseudoscalar pseudo-Goldstone bosons $\pi^\pm$ and $K^\pm$ (see Secs.~\ref{sec:CFLeffth}
and \ref{sec_eft}), so for temperatures well below their masses
(and well below the electron mass~\cite{Shovkovy:2002sg})
the CFL phase is a transparent insulator,
in which $\tilde Q$-electric and magnetic  fields satisfy Maxwell's 
equations with a dielectric constant and index of refraction
that can be calculated directly from QCD \cite{Litim:2001mv},
\beq
n=1+\frac{e^2 \cos^2\theta}{9\pi^2} \frac{\mu^2}{\Delta_{\rm CFL}^2} \ .
\label{indexofrefraction}
\eeq
(This result is valid as long as $n-1 \ll 1$.)
Apart from the fact that $n\neq 1$, the emergence of the $\tilde Q$ photon
is an exact QCD-scale analogue of the TeV-scale 
spontaneous symmetry breaking that gave rise to the photon
as a linear combination of the $W_3$ and hypercharge 
gauge bosons, with the diquark condensate at the QCD scale
playing the role of the Higgs condensate at the TeV scale.

If one could shine a 
beam of ordinary light on a lump of CFL matter in vacuum, some would be 
reflected and some would enter, refracted, as a beam of $\tilde Q$-light.  
The reflection and refraction coefficients are known 
\cite{Manuel:2001mx} (see also \cite{Alford:2004ak}).
The static limit of this academic result is relevant: if a volume of CFL 
matter finds itself in a static magnetic field as within a neutron star, 
surface currents  are induced such that a fraction of this field is expelled 
via the Meissner effect for the non-$\tilde Q$ component of $Q$, while a 
fraction is admitted as $\tilde Q$-magnetic field \cite{Alford:1999pb}. The 
magnetic field within the CFL volume is not confined to flux tubes, and is 
not frozen as in a conducting plasma: CFL quark matter is a color 
superconductor but it is an electromagnetic insulator. 

All Cooper pairs have zero net  $\tilde Q$-charge, but some
have neutral constituents (both quarks $\tilde Q$-neutral)
and some have charged constituents (the two quarks have opposite
$\tilde Q$-charge). The  $\tilde Q$-component of an external magnetic field
will not affect the first type, but it will affect the pairing
of the second type, so external magnetic fields
can modify the CFL phase to the so-called magnetic CFL (``MCFL'')
phase. The MCFL phase has a different gap structure \cite{Ferrer:2005vd,Ferrer:2006vw}
and a different effective theory \cite{Ferrer:2007iw}. The original analyses of the MCFL phase 
were done for rotated magnetic fields $\tilde B$ large enough that all quarks are in the lowest Landau
level; solving the gap equations at lower $\tilde B$ shows that
the gap parameters in the MCFL phase exhibit de Haas-van Alphen
oscillations, periodic in $1/{\tilde B}$ \cite{Noronha:2007wg,Fukushima:2007fc}.

\subsubsection{Low-energy excitations}
\label{sec:CFLeffth}

The low-energy excitations in the CFL phase are: the 8 light 
pseudoscalars arising from broken chiral symmetry, the massless 
Goldstone boson associated with superfluidity, and the 
$\tilde Q$-photon. The 
pseudoscalars form an octet under the unbroken $SU(3)$ 
color+flavor symmetry, and can naturally be labeled according to their
$\tilde Q$-charges as pions, kaons, and an $\eta$.
The effective Lagrangian that describes their interactions, and the
QCD calculation of their masses 
and decay constants will be discussed in Sec.~\ref{sec_eft}.
We shall find, in particular, that even though
the quark-antiquark condensate is small, the pion decay constant 
is large, $f_\pi \sim \mu$. 

The symmetry breaking pattern \eqn{CFLsymmetry} does not include the 
spontaneous breaking of the
$U(1)_A$ ``symmetry'' because it is explicitly broken by instanton
effects. However, at large densities these effects become 
arbitrarily small, and the spontaneous  breaking of $U(1)_A$
will have an associated order parameter and a
ninth pseudo-Goldstone boson with the quantum numbers of the $\eta'$.
This introduces the possibility of a second type
of vortices \cite{Forbes:2001gj,Son:2000fh}.

Among the gapped excitations, we find the quark-quasiparticles which
fall into an ${\bf 8} \oplus {\bf 1}$ of the unbroken 
global $SU(3)_{c+L+R}$, so there are two gap parameters $\De_1$ and $\De_8$.
The singlet has the larger gap $\De_1 = (2+\O(\ka))\De_8$. 
We also find an octet of massive vector mesons, which are the
gluons that have acquired mass via the Higgs mechanism.
The symmetries of the 3-flavor CFL phase are the 
same as those one would expect for 3-flavor hypernuclear matter, 
and even the pattern of gapped excitations is remarkably similar, 
differing only in
the absence of a ninth massive vector meson. It is therefore 
possible that there is no phase transition between hypernuclear 
matter and CFL quark matter \cite{Schafer:1999pb}. This hadron-quark 
continuity can arise in nature only if the strange quark is so light 
that there is a hypernuclear phase, and this phase is 
characterized by proton-$\Xi^-$, neutron-$\Xi^0$ and $\Sigma^+$-$\Sigma^-$ 
pairing, which can then continuously evolve into CFL 
quark matter upon further increasing the density \cite{Alford:1999pa}.

\subsubsection{Why CFL is favored}
\label{sec:CFLfavored}

The dominant component of the CFL pairing pattern is 
the color $\bar{\bf 3}_A$, flavor $\bar{\bf 3}_A$, and Dirac $C \gamma_5$
(Lorentz scalar).
There are many reasons to expect the color $\bar{\bf 3}_A$ to be
favored. First, this is the most attractive channel
for quarks interacting via 
single-gluon exchange which is the dominant interaction at high densities 
where the QCD coupling is weak; second, it is also the most attractive channel for 
quarks interacting via the instanton-induced 't~Hooft interaction, which  
is important at lower densities; third, qualitatively, combining two quarks 
that are each separately in the color-${\bf 3}$ representation to obtain a 
diquark that is a color-$\bar{\bf 3}_A$ lowers the color-flux at large 
distances; and, fourth, phenomenologically, the idea that baryons can 
be modeled as bound states of a quark and a color-antisymmetric
diquark, taking advantage of the attraction in this diquark channel, 
has a long history and has had a recent renaissance 
\cite{Jaffe:2003sg,Selem:2006nd,Close:2002zu,Jaffe:1976ig,Anselmino:1992vg}.   

It is also easy to understand why pairing in the Lorentz-scalar channel
is favorable:  
it leaves rotational invariance unbroken, allowing for quarks at all angles
on the entire Fermi-sphere to participate coherently in the pairing.  
Many calculations have shown that pairing is weaker in channels 
that break rotational symmetry 
\cite{Iwasaki:1994ij,Alford:1997zt,Schafer:2000tw,Buballa:2002wy,Alford:2002rz,Schmitt:2002sc}. There is also a rotationally invariant 
pairing channel with negative parity described by the order parameter
$\langle \psi C \psi\rangle$. Perturbative gluon exchange interactions
do not distinguish between positive and negative parity diquarks, but 
non-perturbative instanton induced interactions do, favoring the 
positive parity channel \cite{Alford:1997zt,Rapp:1997zu,Rapp:1999qa}. 

Once we have antisymmetry in color and in 
Dirac indices, we are forced to antisymmetrize in flavor indices, and the
most general color-flavor structure that the arguments above imply
should be energetically favored is
\beq
\label{antisym_pairing}
\langle \psi^\alpha_i C 
\gamma_5 \psi^\beta_j\rangle \propto  \eps^{\al\be A}\eps_{ij B} \phi^A_B \ .
\eeq
CFL pairing corresponds to $\phi^A_B = \de^A_B$, and this is the only
pattern that pairs all the quarks
and leaves an entire $SU(3)$ global symmetry unbroken. 
The  2SC pattern is $\phi^A_B = \de^A_3 \de^3_B$,
in which only $u$ and $d$ quarks of two 
colors pair \cite{Barrois:1979pv,Bailin:1983bm,Alford:1997zt,Rapp:1997zu}, 
see Sec.\ \ref{subsec:2SC}.
As long as the
strange quark mass can be neglected (the parametric criterion turns out to 
be $ \Delta_{\rm CFL}\gg M_s^2/\mu$, see Sec.~\ref{sec:gCFL})
calculations comparing patterns of the structure 
\eqn{antisym_pairing} always find the CFL phase to have the highest
condensation energy, making it the favored pattern.
This has been confirmed in weak-coupling QCD calculations valid at 
high density \cite{Schafer:1999fe,Evans:1999at,Shovkovy:1999mr}, 
in the Ginzburg-Landau
approximation \cite{Iida:2000ha}, and in many calculations using
Nambu--Jona-Lasinio models
\cite{Alford:1998mk,Rapp:1999qa,Schafer:1999pb,Alford:1999pa,Malekzadeh:2006ik}.
In the high-density limit where $\Delta\gg M_s^2/\mu$ and
$\De\ll \mu$ we can expand in powers of $\De/\mu$ and 
explicitly compare CFL to 2SC pairing.
The CFL condensation energy is $(8\De_8^2+\De_1^2)\mu^2/(4\pi^2)$
which is $12\De_{\rm CFL}^2\mu^2/(4\pi^2)$ when $\ka\ll 1$
(see Sec.~\ref{sec:CFLeffth})
whereas the condensation energy in 
the 2SC phase is only $4\Delta_{\rm 2SC}^2\mu^2/(4\pi^2)$.
We shall see later that 
the 2SC gap parameter turns out to be larger than the CFL gap parameter
by a factor of $2^{1/3}$, so up to corrections of order $\ka$
the CFL condensation 
energy is larger than that in the 2SC phase by a factor of $3 \times 
2^{-2/3}$. At lower densities the condensation energies become
smaller, and we cannot neglect negative $M_s^4$ terms which
are energy penalties induced by the neutrality requirement.
Their coefficient is larger for CFL than for 2SC, partly (but
usually not completely) cancelling the
extra condensation energy---see Fig.~\ref{fig:energy}
and Sec.~\ref{subsec:2SC}.

\begin{figure}
\parbox[t]{0.33\hsize}{
\bc Unpaired\ec
\includegraphics[width=\hsize]{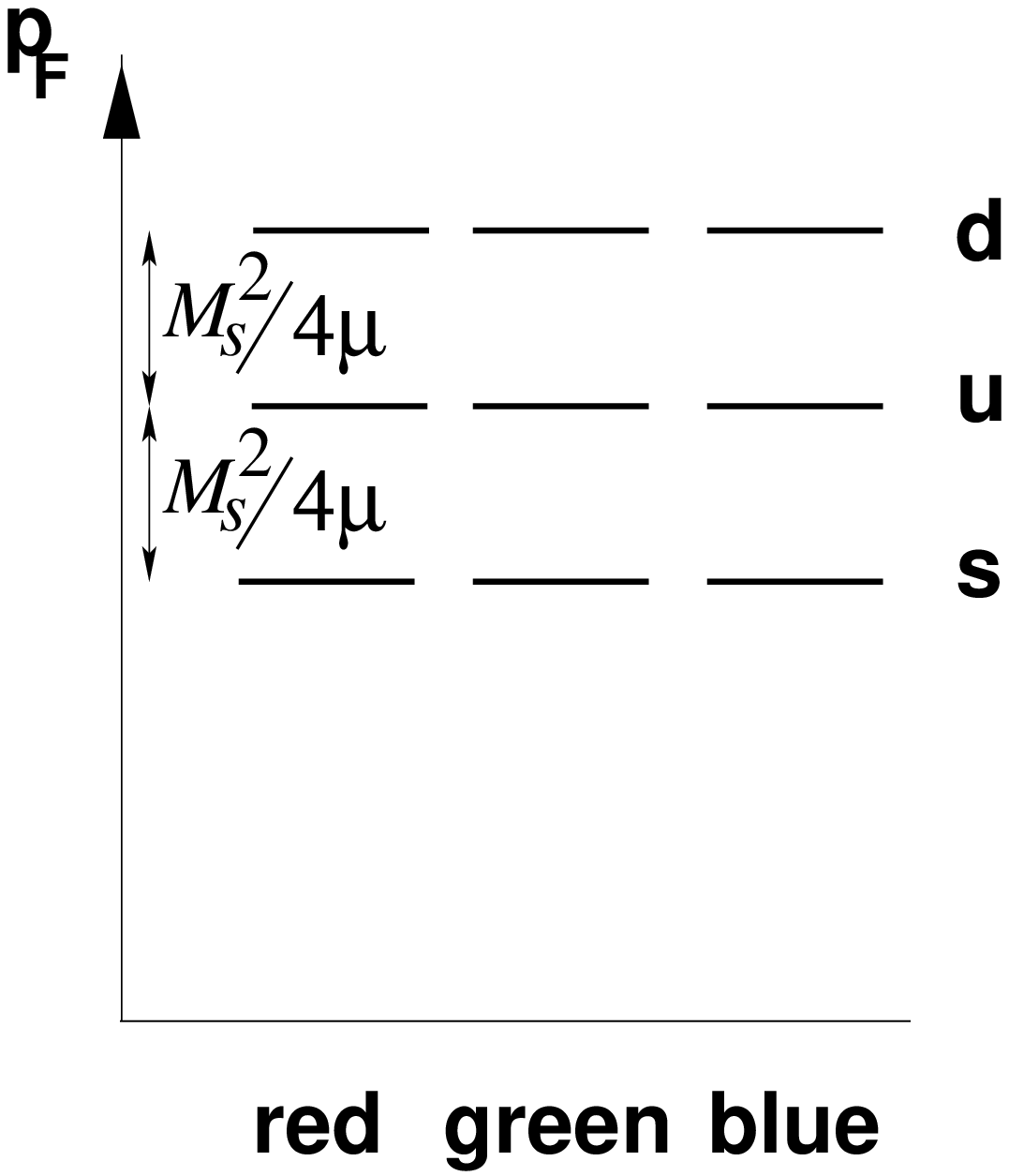}
}\parbox[t]{0.33\hsize}{
\bc 2SC pairing\ec
\includegraphics[width=\hsize]{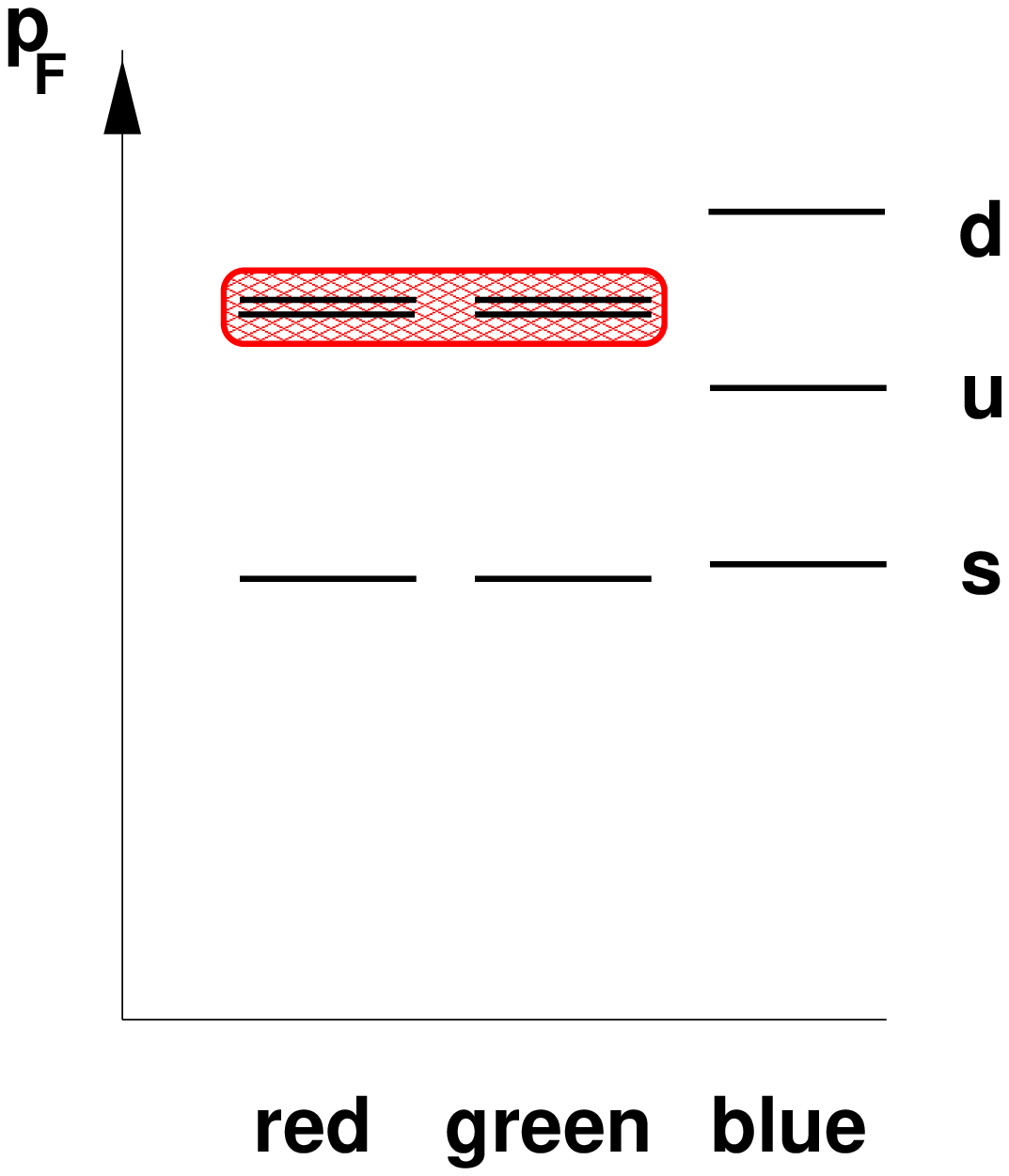}
}\parbox[t]{0.33\hsize}{
\bc CFL pairing\ec
\includegraphics[width=\hsize]{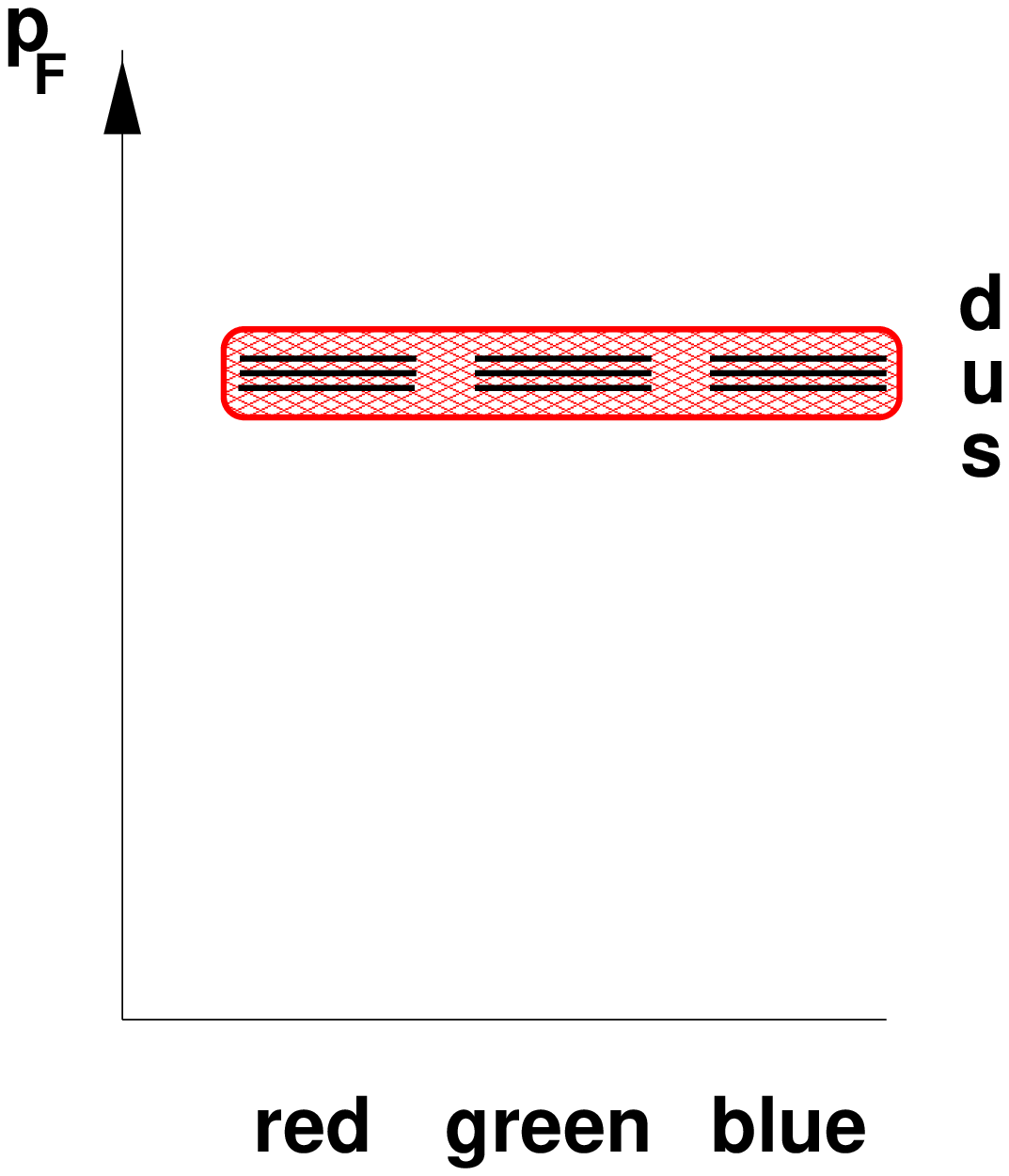}
}
\caption{(Color online)
Illustration of the splitting apart of the Fermi momenta of the
various colors and flavors of quarks (exaggerated for easy
visibility). In the unpaired phase,
requirements of neutrality and weak interaction equilibration
cause separation of the Fermi momenta of the various flavors.
The splittings increase with decreasing density, as $\mu$
decreases and $M_s(\mu)$ increases.
In the 2SC phase, up and down quarks of two colors pair,
locking their Fermi momenta together. In the CFL phase,
all colors and flavors pair and have a common Fermi momentum.
}
\label{fig:splitting}
\end{figure}

\subsection{Intermediate density: stresses on the CFL phase}
\label{sec:CFLstress}

As we noted in section \ref{sec:BCSstress}, BCS pairing between two
species is suppressed if their 
chemical potentials are sufficiently different. 
In real-world quark matter such stresses arise from the strange quark mass,
which gives the strange quark a lower Fermi momentum
than the down quark
at the same chemical potentials $\mu$ and $\mu_e$, and from the neutrality 
requirement, which gives the up quark a different chemical potential from 
the down and strange quarks at the same $\mu$ and $\mu_e$. Once flavor 
equilibrium under the weak interactions is reached, we find that all 
three flavors prefer to have different Fermi momenta at the same chemical
potentials. This is illustrated in Fig.~\ref{fig:splitting}, which shows 
the Fermi momenta of the different species of quarks.

In the unpaired phase (Fig.~\ref{fig:splitting}, left panel), the 
strange quarks have a lower Fermi momentum because they are heavier, 
and to maintain electrical neutrality the number of down quarks is 
correspondingly increased. To lowest order in the strange quark mass, 
the separation between the Fermi momenta is $\de p_F = M_s^2/(4\mu)$, 
so the 
splitting becomes larger as the density is reduced, and smaller as 
the density is increased. The phase space at the Fermi surface
is proportional to $\mu^2$, so the resultant difference in quark number
densities is $n_d\!-\!n_u = n_u\!-\!n_s \propto\mu^2\de p_F
\sim \mu M_s^2$.
Electrons are also present in weak equilibrium, with 
$\mu_e=M_s^2/(4\mu)$, so their charge density is parametrically of 
order $\mu_e^3\sim M_s^6/\mu^3\ll\mu M_s^2$, meaning that they are 
unimportant in maintaining neutrality.

In the CFL phase all the colors and flavors pair with each other, 
locking all their Fermi momenta together at a common value 
(Fig.~\ref{fig:splitting}, right panel). This is possible as long 
as the energy cost of forcing all species to have the same Fermi 
momentum is compensated by the pairing energy that is released by 
the formation of the Cooper pairs. Still working to lowest
order in $M_s^2$, we can say that parametrically the cost is
$\mu^2\de p_F^2 \sim M_s^4$, and the pairing energy is 
$\mu^2\De_{\rm CFL}^2$, so we expect CFL pairing to become
disfavored when $\De_{\rm CFL} \lesssim M_s^2/\mu$.
In fact, the CFL phase remains favored over the unpaired phase as long as 
$\De_{\rm CFL}> M_s^2/4\mu$ \cite{Alford:2002kj}, but already becomes
unstable against unpairing when $\De_{\rm CFL}\gtrsim M_s^2/2\mu$
(see Sec.~\ref{sec:gCFL}).
NJL model calculations
\cite{Alford:2002kj,Alford:2004hz,Fukushima:2004zq,Abuki:2004zk,Blaschke:2005uj,Ruster:2005jc}
find that if the attractive interaction were strong enough to induce
a 100~MeV CFL gap when $M_s=0$
then the CFL phase would survive all the way down to the transition
to nuclear matter. Otherwise, there must be a transition to some other
quark matter phase: this is the ``non-CFL'' region shown schematically 
in Fig.~\ref{fig:phase}. When the stress is small, the CFL pairing can 
bend rather than break, developing a condensate of $K^0$ mesons, 
described in Sec.~\ref{subsec:kaon} below. When the stress is 
larger, however, CFL pairing becomes disfavored. A comprehensive 
survey of possible BCS pairing patterns shows that all of them suffer 
from the stress of Fermi surface splitting \cite{Rajagopal:2005dg}, 
so in the intermediate-density ``non-CFL'' region we expect more 
exotic non-BCS pairing patterns. In Sec.~\ref{sec:non_cfl} we 
give a survey of possibilities that have been explored.

\subsection{Kaon condensation: the CFL-$K^0$ phase}
\label{subsec:kaon}

 Bedaque and Sch\"afer \cite{Bedaque:2001je} showed that when the 
stress is not too large (high density), it may  simply modify the 
CFL pairing pattern by inducing a flavor rotation of the condensate.
This modification can be interpreted as a condensate of ``$K^0$'' 
mesons. The $K^0$ meson carries negative strangeness (it has the 
same strangeness as a $\bar{s}$ quark), so forming a $K^0$ condensate  
relieves the stress on the CFL phase by reducing its strangeness
content. At large density kaon condensation occurs for $M_s \gtrsim 
m^{1/3}\De^{2/3}$, where $m$ is mass of the light ($u$ and $d$) 
quarks. At moderate density the critical strange quark mass is 
increased by instanton contribution to the kaon mass \cite{Schafer:2002ty}.
Kaon condensation was initially demonstrated using an effective theory 
of the Goldstone bosons, but with some effort can also be seen in an 
NJL calculation \cite{Buballa:2004sx,Forbes:2004ww}. The CFL-$K^0$ 
phase is a superfluid; it is a neutral insulator; all its quark 
modes are gapped (as long as $M_s^2/(2\mu)<\De$); it breaks chiral 
symmetry.  In all these respects it is similar to the CFL phase. 
Once we turn on small quark masses, different for all flavors, the 
$SU(3)_{c+L+R}$ symmetry of the CFL phase is reduced by explicit 
symmetry breaking to just $U(1)_{\tilde Q}\times U(1)_{\tilde Y}$, 
with $\tilde Y$ a linear combination of a diagonal color generator 
and hypercharge. In the CFL-$K^0$ phase, the kaon condensate breaks 
$U(1)_{\tilde Y}$ spontaneously.  This modifies the spectrum
of both quarks and Goldstone modes, and thus can affect transport 
properties.

\section{Below CFL densities}
\label{sec:non_cfl}

As we discussed in the introduction (end of Sec.~\ref{sec:outline})
and above (Sec.~\ref{sec:CFLstress}), at intermediate densities the
CFL phase suffers from stresses induced by the strange quark mass,
combined with beta-equilibration and neutrality requirements.  It can
only survive down to the transition to nuclear matter (occurring at
quark chemical potential $\mu=\mu_{\rm nuc}$) if the pairing is 
strong enough: roughly $\De_{\rm CFL}>M_s(\mu_{\rm nuc})^2/2\mu_{\rm nuc}$,
ignoring strong interaction corrections, which are presumably
important in this regime. It is therefore quite possible that other
pairing patterns occur at intermediate densities, and in this section
we survey some of the possibilities that have been suggested.

\begin{figure}[t]
 \bc
 \includegraphics[width=\hsize]{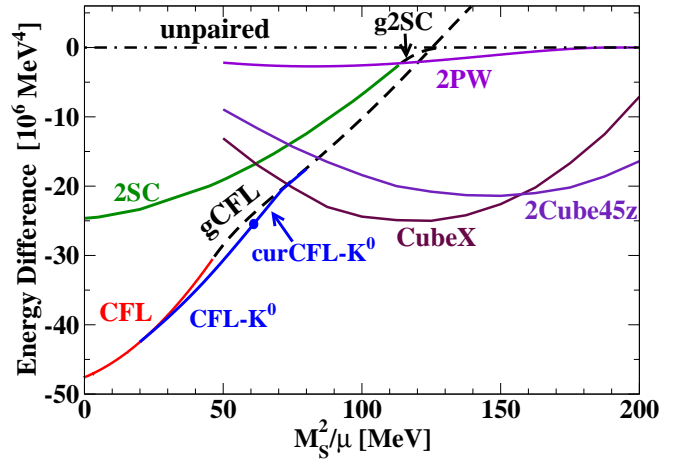}
 \ec
\caption{(Color online)
Free energy of various phases of dense 3-flavor quark matter, assuming
$\De_{\rm CFL}=25~\MeV$. The homogeneous phases are CFL and 2SC, their gapless analogs gCFL and
g2SC, and the kaon-condensed phase CFL-$K^0$.  The true ground state
must have a free energy below that of the gCFL phase, which is known
to be unstable. The inhomogeneous phases are curCFL-$K^0$, which is CFL-$K^0$
with meson supercurrents, and 2PW, CubeX, and 2Cube45z, which are
crystalline color superconducting phases.  The transition from
CFL-$K^0$ to curCFL-$K^0$ is marked with a dot.  In 2PW the condensate is a
sum of only two plane waves.  CubeX and 2Cube45z involve more plane
waves, their condensation energies are larger but less reliably
determined, so their curves should be assumed to have error bands
comparable in size to the difference between them.
}
\label{fig:energy}
\end{figure}

Fig.~\ref{fig:energy} shows a comparison of the free energies of some
of these phases. We have chosen $\De_{\rm CFL}=25~\MeV$, so there is
a window of non-CFL pairing
between nuclear density and the region where the CFL
phase becomes stable. (For stronger pairing,
$\De_{\rm CFL} \sim 100~\MeV$, there would be no such window.)
The curves for the CFL, 2SC,
gCFL, g2SC, and crystalline phases (2PW, CubeX and 2Cube45z) are
obtained from an NJL model as described in Sec.~\ref{sec:NJL}. 
The curves for the CFL-$K^0$ and meson supercurrent 
(curCFL-$K^0$) phases
are calculated using the CFL effective theory with parameters chosen
by matching to weak-coupling QCD, as described in Sec.~\ref{sec_eft}, 
except that the gap was chosen to match $\De_{\rm
CFL}=25~\MeV$.
The phases displayed in Fig.~\ref{fig:energy} are 
discussed in the following sections.

\subsection{Two-flavor pairing: the 2SC phase}
\label{subsec:2SC}

After CFL, 2SC is the most straightforward less-symmetrically paired
form of quark matter, and was one of the first patterns to be analyzed
\cite{Barrois:1979pv,Bailin:1979nh,Bailin:1983bm,Alford:1997zt,Rapp:1997zu}.
In the 2SC phase, quarks with two out of three colors (red and
green, say) and two out of three flavors, pair in the standard BCS
fashion.  The flavors with the most phase space near their Fermi
surfaces, namely $u$ and $d$, are the ones that pair, leaving the
strange and blue quarks unpaired (middle panel of Fig.~\ref{fig:splitting}).
According to NJL models, if the coupling is weak then there is no 2SC region 
in the phase diagram \cite{Steiner:2002gx}. 
This can be understood by an 
expansion in powers of $M_s$, which finds that the CFL$\to$2SC transition
occurs at the same point as the 2SC$\to$unpaired transition, leaving no 
2SC window \cite{Alford:2002kj} (this is the
situation in Fig.~\ref{fig:energy}).
However, NJL models with stronger coupling 
leave open the possibility of a 2SC window in the ``non-CFL'' region of the 
phase diagram \cite{Ruster:2005jc,Abuki:2005ms}. (These calculations have 
to date not included the possibility of meson current or crystalline color 
superconducting phases, discussed below, that may prove more favorable.)

The 2SC pairing pattern, corresponding to
$\phi^A_B = \de^A_3 \de^3_B$ in \eqn{antisym_pairing}, is
$\< \psi^\al_i C \ga_5 \psi^\be_j \>^{\phantom\dagger} 
\propto  \Delta_{\rm 2SC} \epsilon_{ij3}\epsilon^{\al\be 3}$,
where the symmetry breaking pattern, assuming massless up and down quarks, is
\beq
\label{2SCsymmetry}
\ba{r}
 [SU(3)_c] \times  \underbrace{SU(2)_L \times SU(2)_R 
   \times U(1)_B\times U(1)_{S}}_{\displaystyle\supset [U(1)_Q]} \\[7ex]
 \to [SU(2)_{rg}] 
   \times  \underbrace{SU(2)_L \times SU(2)_R \times U(1)_{\tilde B}
     \times U(1)_{S} }_{\displaystyle\supset [U(1)_{\tilde Q}]}
\ea
\eeq
using the same notation as in Eq.~\eqn{CFLsymmetry}. The unpaired massive
strange quarks introduce a $U(1)_S$ symmetry.
The color $SU(3)_c$ gauge symmetry is broken down 
to an $SU(2)_{rg}$ red-green gauge symmetry, whose confinement distance 
rises exponentially with density,
as $\Delta^{-1} \exp({\rm const}\,\mu/(g \Delta) )$
\cite{Rischke:2000cn} (see also \cite{Casalbuoni:2001ha,Ouyed:2001fv}). 
An interesting 
feature of 2SC pairing is that no global symmetries are broken. The 
condensate is a singlet of the $SU(2)_L\times SU(2)_R$ flavor symmetry, 
and baryon number survives as $\tilde B$, a linear combination of the 
original baryon number and the broken diagonal $T_8$ color generator. 
Electromagnetism, originally a linear combination of $B$, $S$,
and $I_3$ (isospin), survives as an unbroken linear combination $\tilde Q$
of $\tilde B$, $S$, and $I_3$.
2SC quark matter is therefore a color superconductor but 
is neither a superfluid nor an electromagnetic superconductor, and there
is no order parameter that distinguishes it from the unpaired phase or 
the QGP \cite{Alford:1997zt}. With respect to the unbroken $U(1)_{\tilde Q}$
gauge symmetry, the 2SC phase is a conductor not an insulator because
some of the ungapped blue and strange quarks are $\tilde Q$-charged.

\subsection{The unstable gapless phases}
\label{sec:gCFL}

As was noted in Sec.~\ref{sec:CFLstress}, and can be seen in
Fig.~\ref{fig:energy},  the CFL phase becomes unstable
when $\mu \approx \half M_s^2/\De_{\rm CFL}$. At this point the pairing
in the $gs$-$bd$ sector suffers the instability 
discussed in Sec.~\ref{sec:BCSstress}, and it becomes
energetically favorable to convert $gs$ quarks into $bd$ quarks
(both near their common Fermi momentum).\footnote{The
onset of gaplessness occurs  at the $\mu$ at which
$\frac{1}{2}(\mu_{bd}-\mu_{gs})=\De_{\rm CFL}$, as explained
in Sec.~\ref{sec:BCSstress}.  Note that in the CFL phase
$(\mu_{bd}-\mu_{gs})=M_s^2/\mu$, twice its value in unpaired
quark matter because of the nonzero color chemical potential 
$\mu_8 \propto M_s^2/\mu$ required by color
neutrality in the presence of CFL pairing \cite{Steiner:2002gx,Alford:2002kj}.  
}
If we restrict ourselves to diquark condensates that are 
spatially homogeneous, the result is a modification of the pairing
in which there is still pairing in all the color-flavor channels
that characterize CFL, but $gs$-$bd$ Cooper pairing
ceases to occur in a range of momenta near the Fermi surface
\cite{Alford:2003fq,Alford:2004hz,Fukushima:2004zq}.
In this range of momenta there are $bd$ quarks but no $gs$ quarks,
and quark modes at the edges of this range are ungapped,
hence this is called a gapless phase (``gCFL'').
Such a phenomenon was first proposed for
two flavor quark matter (``g2SC'') \cite{Shovkovy:2003uu}, see also 
\cite{Gubankova:2003uj}. It has been confirmed in
NJL analyses such as those in 
\cite{Alford:2003fq,Alford:2004hz,Alford:2004nf,Ruster:2004eg,Fukushima:2004zq,Alford:2004zr,Ruster:2005jc,Abuki:2005ms,Abuki:2004zk},
which predict that at densities too low for CFL pairing there will be gapless 
phases. 

In Fig.~\ref{fig:energy}, where $\De_{\rm CFL}=25~\MeV$, we see the transition
from CFL to gCFL at $M_s^2/\mu \approx 2\De_{\rm CFL}=50~\MeV$.
(It is interesting to note that, whereas the CFL phase is a $\tilde
Q$-insulator, the gCFL phase is a $\tilde Q$-conductor, because it has
a small electron density, balanced by unpaired $bu$ quarks from a
very thin momentum shell of broken $bu$-$rs$ pairing; the
CFL$\to$gCFL transition is the analogue of an insulator to metal
transition at which a ``band'' that was unfilled in the insulating
phase drops below the Fermi energy, making the material a metal.)  
The gCFL phase then remains favored beyond the value $M_s^2/\mu \approx
4 \De_{\rm CFL} = 100~\MeV$ at which the CFL phase would become
unfavored relative to completely unpaired quark
matter \cite{Alford:2002kj}.

However, it turns out that in QCD the gapless phases,
both g2SC \cite{Huang:2004am,Giannakis:2004pf} and gCFL 
\cite{Casalbuoni:2004tb,Fukushima:2005cm}, are unstable
at zero temperature. (Increasing the temperature above a critical
value removes the instability; the critical value varies dramatically
between phases, from a fraction of an MeV to of order 10 MeV
\cite{Fukushima:2005cm}.)  The instability manifests itself in an
imaginary Meissner mass $m^{\phantom{2}}_M$ for some of the
gluons. $m_M^2$ is the low-momentum current-current two-point
function, and $m_M^2/ (g^2\De^2)$ (where the strong interaction
coupling is $g$) is the coefficient of the gradient term in the
effective theory of small fluctuations around the ground-state
condensate, so a negative value indicates an instability towards
spontaneous breaking of translational invariance
\cite{Reddy:2004my,Huang:2005pv,Iida:2006df,Hashimoto:2006mn,Fukushima:2006su}.
Calculations in a simple two-species model \cite{Alford:2005qw} show
that gapless charged fermionic modes generically lead to imaginary
$m^{\phantom{2}}_M$.

The instability of the gapless phases indicates that there must be 
other phases of even lower free energy, that occur in their place
in the phase diagram. The nature of those phases is not reliably determined
at present; likely candidates are discussed below.

\subsection{Crystalline color superconductivity}
\label{subsec:crystal}

The Meissner instability of the gCFL phase points to a breaking
of translational invariance, and crystalline color superconductivity
represents a possible resolution of that instability. The basic idea,
first proposed in condensed matter physics
\cite{LarkinOvchinnikov,FuldeFerrell} and more recently analyzed
in the context of color superconductivity
\cite{Alford:2000ze,Bowers:2002xr,Casalbuoni:2003wh}, is to allow
the different quark flavors to have different Fermi momenta, thus accommodating
the stress of the strange quark mass, and to form
Cooper pairs with nonzero momentum, each quark lying close to its
respective Fermi surface. The price one must pay for this arrangement
is that only fermions in certain regions on the Fermi surface
can pair. Pairs with nonzero momenta chosen from some set of wave vectors 
${\bf q}_a$ yield condensates that vary in position space like $\sum_a 
\exp(i {\bf q}_a \cdot {\bf x})$, forming a crystalline pattern whose 
Bravais lattice is the set of ${\bf q}_a$.

Analyses to date have focused on $u$-$d$ and $u$-$s$ pairing,
neglecting pairing of $d$ and $s$ because the separation of their Fermi momenta
is twice as large (Fig.~\ref{fig:splitting}).
If the 
$\langle ud \rangle$ condensate includes only pairs with a single 
nonzero momentum ${\bf q}$, this means that  in position space the
condensate is a single plane-wave and means that in momentum space
pairing is allowed on a single ring on the $u$ Fermi surface and a 
single ring on the opposite side of the $d$ Fermi surface. The simplest 
``crystalline" phase of three-flavor quark matter that has been analyzed 
\cite{Casalbuoni:2005zp,Mannarelli:2006fy} includes two such single-plane
wave condensates (``2PW"), one  $\langle ud \rangle$ and one $\langle us 
\rangle$. The favored orientation of the two ${\bf q}$'s is parallel, 
keeping the  two ``pairing rings" on the $u$ Fermi surface (from the 
$\langle us \rangle$ and $\langle ud \rangle$  condensates) as far 
apart as possible \cite{Mannarelli:2006fy}. This simple pattern of 
pairing leaves much of the Fermi surfaces unpaired, and it is much 
more favorable to choose a pattern in which the $\langle us \rangle$
and $\langle ud \rangle$ condensates each include pairs with more 
than one ${\bf q}$-vector, thus more than one ring and more than one 
plane wave. Among such more realistic pairing patterns, the two that 
appear most favorable have either 
four ${\bf q}$'s per condensate
that together point at the eight corners of a cube in momentum space 
(``CubeX") or 
eight ${\bf q}$'s per condensate
that each point at the 
corners of separate cubes, rotated relative to each other by 45 
degrees (``2Cube45z") \cite{Rajagopal:2006ig}. 
It has been shown that the chromomagnetic instability is no longer
present in these phases \cite{Ciminale:2006sm}.
The free energies
of the 2PW, CubeX and 2Cube45z phases as calculated within an NJL model
(see Sec.~\ref{sec:NJL}) are shown in Fig.~\ref{fig:energy}.  
The calculation is an expansion in powers of $(\De/\de p_F)^2$
which in the CubeX and 2Cube45z phases turns out to be of order
a tenth to a quarter. According to results obtained in a calculation
done to third order in this expansion parameter,
the CubeX and 2Cube45z condensation energies are large enough that
one or other of them is favored over a wide range of $M_s^2/\mu$ 
as illustrated in Fig.~\ref{fig:energy}. The uncertainty in each is of
the same order as the difference between them, so one cannot yet
say which is favored, but the overall scale is plausible (one would
expect condensation energies an order of magnitude
bigger than that of the 2PW state).
We discuss crystalline color superconductivity in greater detail
in Sec.~\ref{sec:NJL}.

\subsection{Meson supercurrent (``curCFL-$K^0$'')}
\label{sec_kcur}

 Kaon condensation alone does not remove the gapless modes that
occur in the CFL phase when $M_s$ becomes large enough, but it
does affect the number of gapless modes and the onset value of
$M_s$. In the CFL-$K^0$ phase, the electrically charged ($bs$) 
mode becomes gapless at $M_s^2/\mu\approx 8\Delta/3$ (compared 
to $2\Delta$ in the CFL phase), and the electrically neutral 
($bd$) mode becomes gapless for $M_s^2/\mu\approx 4\Delta$
\cite{Kryjevski:2004jw,Kryjevski:2004kt}. (In an NJL model 
analysis \cite{Forbes:2004ww}, the charged mode
in the CFL-$K^0$ phase becomes gapless at $M_s^2/\mu \approx 
2.44 \Delta$ for $\Delta=25$ MeV as in Fig.~\ref{fig:energy}).
The gapless CFL-$K^0$ phase has an instability which is similar 
to the instability of the gCFL phase. This instability can be 
viewed as a tendency towards spontaneous generation of Goldstone 
boson (kaon) currents \cite{Schafer:2005ym,Kryjevski:2005qq}. The 
currents correspond to a spatial modulation of the kaon condensate.
There is no net transfer of any charge because the Goldstone
boson current is counterbalanced by a backflow of 
ungapped fermions. The meson supercurrent ground state is lower in
energy than the CFL-$K^0$ state and the magnetic screening
masses are real \cite{Gerhold:2006np}. Because the ungapped 
fermion mode is electrically charged, both the magnitude of the
Goldstone boson current needed to stabilize the phase 
and the magnitude of the resulting energy gain relative to
the phase without a current are very small. 
Goldstone boson currents can also be generated in the gCFL phase
without $K^0$ condensation. In this case gauge invariance implies that
the supercurrent state is equivalent to a single plane-wave LOFF
state, but the analyses can be carried out in the limit that the gap
is large compared to the magnitude of the
current \cite{Gerhold:2006dt}. This analysis is valid near the onset
of the gCFL phase, but not for larger mismatches, where states with
multiple currents are favored.

\subsection{Single-flavor pairing}
\label{subsec:single}

If the stress due to the strange quark mass is large enough then there
may be a range of quark matter densities where no pairing between
different flavors is possible, whether spatially uniform or
inhomogeneous.  From Fig.~\ref{fig:energy} we can estimate that this
will occur when $M_s^2/(\mu\De_{\rm CFL}) \gtrsim 10$, so it requires
a large effective strange quark mass and/or small CFL pairing gap.
The best available option in this case is Cooper pairing of each
flavor with itself.  Single-flavor pairing may also arise 
among the strange quarks in a 2SC
phase, since they are not involved in two-flavor
pairing. We will discuss these cases separately below.

To maintain fermionic antisymmetry of the Cooper pair wavefunction,
single-flavor pairing phases have to either be symmetric in color,
which greatly weakens or eliminates the attractive interaction,
or symmetric in Dirac indices, which compromises the uniform
participation of the whole Fermi sphere.
As a result, they
have much lower critical temperatures than multi-flavor phases 
such as the CFL or 2SC phases, perhaps as large as a few MeV, more 
typically in the eV to many keV range
\cite{Alford:1997zt,Schafer:2000tw,Buballa:2002wy,Alford:2002rz,Schmitt:2002sc,Schmitt:2004et}.

Matter in which each flavor only pairs with itself
has been studied using NJL models and weakly-coupled QCD.
These calculations agree that the energetically favored state
is color-spin-locked (CSL) pairing for each flavor
\cite{Bailin:1979nh,Schafer:2000tw,Schmitt:2004et}. CSL pairing
involves all 3 colors, with
the color direction of each Cooper pair correlated with its spin
direction, breaking $SU(3)_c\times SO(3)_{\rm rot}\to SO(3)_{c+\rm rot}$.
The phase is isotropic, with
rotational symmetry surviving as a group of simultaneous spatial and color 
rotations. Other possible phases exhibiting spin-one, single-flavor,
pairing include the polar, planar, and A phases described in 
\cite{Schmitt:2004et,Schafer:2000tw} (for an NJL model treatment see
\cite{Alford:2002rz}).  Some of these phases exhibit point 
or line nodes in the energy gap at the Fermi surface, and 
hence do break rotational symmetry.

If 2SC pairing occurs with strange quarks present, one might expect 
the strange quarks of all three colors to undergo CSL self-pairing,
yielding an isotropic ``2SC+CSL'' pattern.
However, the 2SC pattern breaks the color symmetry, and 
in order to maintain color neutrality, a color chemical potential
is generated, which splits
the Fermi momentum of the blue strange quarks away from that of the red 
and green strange quarks.
This is a small effect, but so is the CSL pairing gap, and NJL model 
calculations indicate that the color chemical potential typically 
destroys CSL pairing of the strange quarks \cite{Alford:2005yy}.
The system falls back on the next best alternative, which is spin-one 
pairing of the red and green strange quarks.

Because their gaps and critical temperatures can range as low as
the eV scale, single-flavor pairing phases in compact stars would
appear relatively late in the life of the star, and might cause
dramatic changes in its behavior. For example, unlike the CFL
and 2SC phases, many single-flavor-paired phases are electrical
superconductors \cite{Schmitt:2003xq}, so their appearance
could significantly affect the magnetic field dynamics of the star.

\subsection{Gluon condensation}
\label{sec_glue_cond}

 In the 2SC phase (unlike in the CFL phase) the magnetic instability
arises at a lower value of the stress on the BCS pairing than that
at which the onset of gapless pairing occurs. In this 2SC regime, 
analyses done using a Ginzburg-Landau approach indicate that the 
instability can be cured by the appearance of a chromoelectric 
condensate \cite{Gorbar:2005rx,Gorbar:2006up,Hashimoto:2007ut,Gorbar:2007vx}. The 2SC condensate 
breaks the color group down to the $SU(2)_{rg}$ red-green
subgroup, and five of the gluons become massive vector bosons via the
Higgs mechanism. The new condensate involves some of these massive
vector bosons, and because they transform non-trivially under
$SU(2)_{rg}$ it now breaks that gauge symmetry. Because they are
electrically charged vector particles, rotational symmetry is also broken, 
and the phase is an electrical superconductor. 
Alternatively, it has been suggested \cite{Ferrer:2007uw}
that the gluon condensate may be inhomogeneous with a large
spontaneously-induced $\tilde Q$ magnetic field.

\subsection{Secondary pairing}
\label{sec_sec_pair}

 Since the Meissner instability is generically associated with the 
presence of gapless fermionic modes, and the BCS mechanism implies 
that any gapless fermionic mode is unstable to Cooper pairing in the 
most attractive channel, one may ask whether the instability could
be resolved without introducing spatial inhomogeneity simply by 
``secondary pairing'' of the gapless quasiparticles, which would 
then acquire their own gap $\De_s$ \cite{Hong:2005jv,Huang:2003xd}. 
Furthermore, there is a mode in the gCFL phase whose dispersion relation 
is well approximated as quadratic, $\epsilon \propto (k - {\rm const})^2$, 
yielding a greatly increased density of states at low energy (diverging 
as $\epsilon^{-1/2}$), so its secondary pairing is much stronger than 
would be predicted by BCS theory: $\Delta_s \propto G^2$ for an effective 
four-fermion coupling strength $G$, as compared with the standard BCS 
result $\Delta \propto \exp(-{\rm const}/G)$ \cite{Hong:2005jv}. This 
result is confirmed by an NJL study in a two-species model 
\cite{Alford:2005kj}, but the secondary gap $\De_s$ was found to be 
still much smaller than the primary gap $\De_p$, so it does not 
generically resolve the magnetic instability (in the temperature 
range $\De_s \ll T \ll \De_p$, for example).

\subsection{Mixed phases}
\label{sec_mixed}

Another way for a system to deal with a stress on its pairing pattern
is to form a mixed phase, which is a charge-separated state
consisting of positively and negatively 
charged domains which are neutral on average. The coexisting phases
have a common
pressure and a common value of the charge chemical potential
which is not equal to the neutrality value for either phase
\cite{Ravenhall:1983uh,Glendenning:1992vb}.
The size of the domains is determined by a balance between
surface tension (which favors large domains) and electric field
energy (which favors small domains).
Separation of color charge is expected 
to be suppressed by the very high energy cost of color electric fields, 
but electric charge separation is quite possible, and may occur at the 
interface between color-superconducting quark matter and nuclear matter
\cite{Alford:2001zr} and an interface between quark matter and the 
vacuum \cite{Jaikumar:2005ne,Alford:2006bx}, just as it occurs at 
interfaces between nuclear matter and a nucleon gas
\cite{Ravenhall:1983uh}. Mixed phases are a 
generic phenomenon, since, in the approximation where 
Coulomb energy costs are neglected,
any phase can always lower its free 
energy density by becoming charged (this follows from the fact 
that free energies are concave functions of chemical potentials).
In this approximation, if two phases A and B can coexist at the same 
pressure with opposite charge densities then
such a mixture will always be favored over a 
uniform neutral phase of either A or B. For a pedagogical discussion, 
see \cite{Alford:2004hz}. Surface and Coulomb energy costs can cancel this 
energy advantage, however, and have to be calculated on a case-by-case 
basis.

In quark matter it has been found that as long
as we require local color neutrality such mixed phases are not the
favored response to the stress imposed by the strange quark mass
\cite{Alford:2003fq,Alford:2004nf}. Phases involving color charge
separation have been studied \cite{Neumann:2002jm} but it seems likely
that the energy cost of the color-electric fields will disfavor them.


\subsection{Relation to cold atomic gases}
\label{sec_atoms}

An interesting class of systems in which stressed superconductivity
can be studied experimentally is trapped atomic gases in which two
different hyperfine states (``species'') of the atom
pair with each other \cite{Giorgini:2007}.  
This is a useful experimental model
because the stress and interaction strength are both under
experimental control, unlike quark matter where one physical variable
($\mu$) controls both the coupling strength and the stress.  
The atomic pairing stress can be adjusted by changing the relative
number of atoms of the two species (``polarization'').  The scattering
length of the atoms can be controlled using Feshbach resonances,
making it possible to vary the strength of the inter-atomic attraction
from weak (where BCS pairing occurs) through the unitarity limit
(where a bound state forms) to strong (Bose-Einstein condensation of
diatomic molecules).

The theoretical expectation is that, in the weak coupling limit, there
will be BCS pairing as long as $\delta\mu$, the chemical potential
difference between the species, is small enough.  The BCS phase is
unpolarized because the Fermi surfaces are locked together.  A
first-order transition from BCS to crystalline (LOFF) pairing is
expected at $\delta\mu=\Delta_0/\sqrt{2}$, where $\De_0$ is the BCS
gap at $\delta\mu=0$; then at $\delta\mu_c$ a continuous
transition to the unpaired
phase \cite{Clogston:1962,Chandrasekhar:1962,LarkinOvchinnikov,FuldeFerrell}.
For the single plane wave LOFF state $\delta\mu_c\simeq 0.754\Delta_0$,
but for multiple plane wave states $\delta\mu_c$ may be larger.

Experiments with cold trapped atoms near the unitary limit
(strong coupling) have seen phase separation between an unpolarized
superfluid and a polarized normal state \cite{Zwierlein:2006,Zwierlein:2005,Partridge:2005}.
If one ignores the crystalline phase (perhaps only favored
at weak coupling \cite{Sheehy:2007,Sheehy:2006,Mannarelli:2006hr}) 
this is consistent with the theoretical expectation for the
BCS regime: the net polarization forces the system to
phase separate, yielding a mixture of BCS and unpaired phases
with $\delta\mu$ fixed at the first order transition
between them \cite{Bedaque:2003hi,Carlson:2005kg}.
It remains an exciting possibility that crystalline superconducting
(LOFF) phases of cold atoms may be observed: this may require experiments closer
to the BCS regime.


In the strong coupling limit the superfluid consists of tightly
bound molecules. Adding an extra atom requires energy $\Delta$.
For $|\delta\mu|> \Delta$ the atomic gas is a homogeneous mixture of
an unpolarized superfluid and a fully polarized Fermi gas, so the
system is a stable gapless superfluid. This means that in strong
coupling polarization can be carried by a gapless superfluid, whereas
in weak coupling even a small amount of polarization leads to the
appearance of a mixed BCS/LOFF phase. It is not known what happens
at intermediate coupling, but one possibility is that the gapless
superfluid and the LOFF phase are connected by a phase transition
\cite{Son:2005qx}. This transition would correspond to a magnetic instability of the
gapless superfluid.

\section{Weak-coupling QCD calculations}
\label{sec:QCD}
% ...

We have asserted in Secs.\ \ref{sec:intro} and \ref{sec:cfl} that at sufficiently high 
densities it is possible to do controlled calculations of properties 
of CFL quark matter directly from the QCD Lagrangian. We describe how 
to do such calculations in this section. We shall focus on the 
calculation of the gap parameter, but shall also discuss the critical 
temperature $T_c$ for the transition from  the CFL phase to the 
quark-gluon plasma and the Meissner and Debye masses that control 
color-magnetic and color-electric effects in the CFL phase. Phenomena 
that are governed by the massless Goldstone bosons and/or the light 
pseudo-Goldstone bosons are most naturally described by first constructing 
the appropriate effective theory and then, if at sufficiently high 
densities, calculating its parameters directly from the QCD Lagrangian.  
We defer these analyses to Sec.\ \ref{sec_eft}.  

Although the weak-coupling calculations that we describe in this 
section are only directly applicable in the CFL phase, we shall present 
them in a sufficiently general formalism that they can be applied to 
other spatially homogeneous phases also, including for example the 2SC 
and CSL phases. These phases can be analyzed at weak-coupling either 
just by ansatz, or by introducing such a large strange quark mass that 
CFL pairing is disfavored even at enormous densities. Such calculations 
provide insights into the properties of these  phases, even though they 
do not occur  in the QCD phase diagram at high enough densities for
a weak-coupling approach to be applicable. To keep our notation general, 
we shall refer to the gap parameter as $\Delta$; in the CFL phase, 
$\Delta_{\rm CFL}\equiv \Delta$.

We shall see that at weak coupling the expansion parameter that 
controls the calculation of $\log(\Delta/\mu)$  is at best $g$, 
certainly not $g^2$. (The leading term is of order $1/g$; the
$\log g$ and $g^0$ terms have also been calculated. The ${\cal O}(g\log g)$ 
and ${\cal O}(g)$ terms are nonzero, and have not yet 
been calculated.  Beyond ${\cal O}(g)$, it is possible that fractional 
powers of $g$ may arise in the series.) We therefore expect the 
weak-coupling calculations to be quantitatively reliable only at 
densities for which $g(\mu)< 1$, which corresponds to densities
many orders of magnitude greater than that at the centers of neutron 
stars.  Indeed, it has been shown \cite{Rajagopal:2000rs} that 
some of the ${\cal O}(g)$ contributions start to decline in 
magnitude relative to the $g^0$ term only for $g(\mu) \lesssim 0.8$ 
which corresponds, via the two-loop QCD beta function, 
to $\mu\gtrsim 10^8$~MeV meaning densities 15-16 orders of 
magnitude greater than those at the centers of compact stars.  
The reader may therefore be tempted to see this section as academic.  
{}From a theoretical point of view, it is exceptional to have an instance 
where the properties of a superconducting phase can be calculated 
rigorously from a fundamental short-distance theory, making this 
exploration a worthy pursuit even if academic. From a practical 
point of view, the quantitative understanding that we derive from 
calculations reviewed in this section provides a completely solid
foundation from which we can extrapolate downwards in $\mu$.  The
effective field theory described in Sec.\ \ref{sec_eft} gives us a well-defined way of
doing so as long as we stay within the CFL phase, meaning that we 
can come down from $\mu > 10^8~{\rm MeV}$ all the way down to
$\mu\sim M_s^2/(2\Delta_{\rm CFL})$.   Finally, we shall gain qualitative
insights into the CFL phase and other color superconducting phases,
insights that guide our thinking at lower densities.  

The QCD Lagrangian is given by
\beq \label{QCDL}
{\cal L} = \overline{\psi}(i\ga^\mu D_\mu+\hat{\mu}\ga_0-\hat{m})\psi
   -\frac{1}{4}G_a^{\mu\nu}G^a_{\mu\nu} \, .
\eeq
Here, $\psi$ is the quark spinor in Dirac, color, and flavor space, i.e., a 
$4N_cN_f$-component spinor, and $\overline{\psi}\equiv \psi^\dag\ga_0$. The 
covariant derivative acting on the fermion field is $D_\mu=\partial_\mu 
+igT_a A_\mu^a$, where $g$ is the strong coupling constant, $A_\mu^a$ are 
the gauge fields, $T^a=\lambda^a/2$ ($a=1,\ldots,8$) are the generators of 
the gauge group $SU(3)_c$, and $\lambda^a$ are the Gell-Mann matrices. The 
field strength tensor is $G_a^{\mu\nu}=\partial_\mu A_\nu^a -\partial_\nu 
A_\mu^a+g f^{abc}A_\mu^bA_\nu^c$ with  the $SU(3)_c$ structure constants 
$f^{abc}$. The chemical potential $\hat{\mu}$ and the quark mass 
$\hat{m}={\rm diag}(m_u,m_d,m_s)$ are diagonal matrices in flavor space. 
If weak interactions are taken into account flavor is no longer conserved 
and there are only two chemical potentials, one for quark (baryon) number, 
$\mu$,  and one for electric charge, $\mu_e$. At the very high densities 
of interest in this section, the constituent quark masses are essentially 
the same as the current quark masses $m_u$, $m_d$ and $m_s$ meaning
that we need not distinguish between them.  Furthermore, at asymptotic 
densities we can neglect even the strange quark mass, so throughout most of
this section we shall set $m_u=m_d=m_s=0$.


If the coupling is small then the natural starting point is a free Fermi gas of 
quarks. In a degenerate quark gas all states with momenta $p<p_F=(\mu^2-m_q^2)^{1/2}$
are occupied, and all states with $p>p_F$ are empty. Because of Pauli-blocking, 
interactions mainly modify states in the vicinity of the Fermi surface. Since the 
Fermi momentum is large, typical interactions between quarks near the Fermi surface
involve large momentum transfer  and are governed by the weak coupling
$g(\mu)$. Interactions in which quarks scatter by only a small angle involve only a
small momentum transfer and are therefore potentially dangerous. However, small 
momenta correspond to large distances, and medium modifications of the exchanged 
gluons are therefore important. In a dense medium, electric gluons are Debye screened 
at momenta $q\sim g\mu$. The dominant interaction for momenta below the screening 
scale is due to unscreened, almost static, magnetic gluons. In a hot quark-gluon
gas, interactions between magnetic gluons become nonperturbative for momenta
less than $g^2T$. This phenomenon does not take place in a very dense quark 
liquid, and gluon exchanges with arbitrarily small momenta remain perturbative.
On a qualitative level this can be attributed to the absence of Bose enhancement
factors in soft gluon propagators. A more detailed explanation will be given in 
Sec.~\ref{sec_nfl}.  The unscreened magnetic interactions nevertheless make the 
fluid a ``non-Fermi liquid'' at temperatures above the critical temperature for 
color superconductivity. We shall discuss this also in Sec.~\ref{sec_nfl}, where 
we shall see that these non-Fermi liquid effects do not spoil the basic logic
of the BCS argument that diquark condensation must occur in the presence of an 
attractive interaction, but are crucial in the calculation of the gap that results.

\subsection{The gap equation}
\label{sec_qcd_gap}

As discussed in Sec.~\ref{sec:inevitability}, any attractive interaction in a 
many-fermion system leads to Cooper pairing. QCD at high density provides an 
attractive interaction via one-gluon exchange. In terms of quark 
representations of $SU(3)_c$, the attractive channel is the antisymmetric 
anti-triplet ${\bf \bar{3}}_A$,  appearing by ``pairing'' two color triplets: 
${\bf 3} \otimes {\bf 3} = {\bf\bar{3}}_A \oplus {\bf 6}_S$. Consequently, 
only quarks of different colors form Cooper pairs. There is an induced 
pairing in the symmetric sextet channel ${\bf 6}_S$. However, this pairing 
is much weaker \cite{Alford:1998mk,Schafer:1999fe,Shovkovy:1999mr,Pisarski:1999cn,Alford:1999pa}, 
and we shall largely neglect it in the following. As in an electronic 
superconductor, Cooper pairing results in an energy gap in the 
quasiparticle excitation 
spectrum. Its magnitude at zero temperature $\Delta$ is crucial for the 
phenomenology of a superconductor. In addition, it also sets the scale for 
the critical temperature $T_c$ of the phase transition which can be expected 
to be of the same order as $\Delta$ (in BCS theory, $T_c=0.57\Delta$). Over 
the course of the next five subsections, we shall discuss the QCD gap equation, 
which is used to determine both $\Delta$ and $T_c$.

Our starting point is the partition function
\beq
{\cal Z} = \int {\cal D}A\,{\cal D}\overline{\psi}\,{\cal D}\psi\; e^{i\cal S} \, , 
\eeq
with the action ${\cal S}=\int d^4x\, {\cal L}$ and the Lagrangian (\ref{QCDL}). 
In the following we shall only sketch the derivation of the gap equation. Details 
following the same lines can be found in 
\cite{Manuel:2000nh,Pisarski:1999av,Pisarski:1999tv,Schmitt:2002sc,Rischke:2003mt,Schmitt:2004hg,Schmitt:2004et}.

We begin by introducing Nambu-Gorkov spinors. This additional two-dimensional 
structure proves convenient in the theoretical description of a superconductor 
or a superfluid, see for instance \cite{abrikosov,fetter}. It allows for the 
introduction of a source that couples to quark bilinears (as opposed to 
quark-anti-quark bilinears). Spontaneous symmetry breaking is realized by taking 
the limit of a vanishing source. The Nambu-Gorkov basis is given by 
\beq
\Psi=\left(\begin{array}{c}\psi \\ \psi_C\end{array}\right) \, , \qquad 
\overline{\Psi}=(\overline{\psi},\overline{\psi}_C) \, , 
\label{NambuGorkov}
\eeq
where $\psi_C =C\overline{\psi}^T$ is the charge-conjugate spinor, obtained by 
multiplication with the charge conjugation matrix $C\equiv i\ga^2\ga^0$. In a 
free fermion system, the new basis is a pure doubling of degrees of freedom with 
the inverse fermion propagator consisting of the original free propagators,
\beq
S^{-1}_0 = \left(\begin{array}{cc}
  [G_0^+]^{-1} & 0 \\ 0 & [G_0^-]^{-1} 
  \end{array}\right)
\eeq    
where $[G_0^\pm]^{-1}(X,Y)\equiv -i\,(i\ga^\mu\partial_\mu \pm \mu\ga_0 )
\,\delta^{(4)}(X-Y)$. Here and in the following capital letters denote four-vectors, 
e.g., $X\equiv(x_0,{\bf x})$. The effect of a nonzero diquark condensate can now be 
taken into account through adding a suitable source term to the action and computing 
the effective action $\Gamma$ as a functional of the gluon and fermion propagators 
$D$ and $S$
\cite{Abuki:2003ut,Ruster:2003zh,Rischke:2003mt,Schmitt:2004hg,Miransky:2001sw,Takagi:2002vj}:
\bea 
\label{effectiveaction}
\Gamma[D,S] &=& -\frac{1}{2}\Tr\log D^{-1}- \frac{1}{2} \Tr(D_0^{-1}D -1)\non
&&+\,\frac{1}{2}\Tr\log S^{-1} +\, \frac{1}{2} \Tr (S_0^{-1}S -1) \non
&&+\,\Gamma_2[D,S]\, .
\eea
This functional is called the  ``2PI effective action'' since the contribution 
$\Gamma_2[D,S]$ consists of all two-particle irreducible diagrams 
\cite{Luttinger:1960ua,Baym:1962sx,Cornwall:1974vz}. This formalism is particularly 
suitable for studying spontaneous symmetry breaking in a self-consistent way. The 
ground state of the system is obtained by finding the stationary point of the 
effective action. The stationarity conditions yield Dyson-Schwinger equations for 
the gauge boson and fermion propagators,
\begin{subequations} \label{dysonschwinger}
\bea  
D^{-1} &=& D_0^{-1} + \Pi \,\, , \label{dysonschwinger1} \\ 
S^{-1} &=& S_0^{-1} + \Sigma \,\, , \label{dysonschwinger2}
\eea
\end{subequations}
where the gluon and fermion self-energies are the functional derivatives of $\Gamma_2$ 
at the stationary point, 
\beq \label{selfenergydef} 
\Pi\equiv -2\frac{\delta\Gamma_2}{\delta D} 
\,\, , \qquad  
\Sigma \equiv 2\frac{\delta\Gamma_2}{\delta S} \,\, . 
\eeq 
Writing the second of these equations as $\Gamma_2[S]=(1/4)\,{\rm Tr}(\Sigma\,S)$, 
we can then use the Dyson-Schwinger equation (\ref{dysonschwinger2}) to evaluate
the fermionic part of the effective action at the stationary point,  obtaining the 
pressure
\beq \label{pressure}
P= \frac{1}{2}\Tr\log S^{-1} - 
\frac{1}{4}\Tr\left(1-S_0^{-1}S\right) \, .
\eeq
We shall return to this expression for the pressure in Sec.~\ref{subsubsec:pressure}. 

Here, we proceed to analyze 
the Dyson-Schwinger equation (\ref{dysonschwinger2}) for the fermion propagator. 
We denote the entries of the 2$\times$2 matrix $\Sigma$ in Nambu-Gorkov space as
\beq \label{sigmanambu} 
\Sigma\equiv\left(\begin{array}{cc}\Sigma^+ & \Phi^- \\ \Phi^+ & \Sigma^- 
\end{array}\right) \,\, ,
\eeq
where the off-diagonal elements are related via $\Phi^-=\ga_0(\Phi^+)^\dag\ga_0$.
One can invert the Dyson-Schwinger equation formally to obtain the full fermion 
propagator in the form
\beq \label{fullquark}
S = \left(\begin{array}{cc} G^+ & F^- \\ F^+ & G^- 
\end{array}\right) \, ,
\eeq   
where the fermion propagators for quasiparticles and charge-conjugate 
quasiparticles are
\beq \label{Gpm}
G^\pm = \left\{[G_0^\pm]^{-1} + \Sigma^\pm - 
\Phi^\mp([G_0^\mp]^{-1}+\Sigma^\mp)^{-1}
\Phi^\pm\right\}^{-1} \, ,
\eeq 
and the so-called anomalous propagators, typical for a superconducting
system, are given by
\beq \label{Xpm}
F^\pm = -([G_0^\mp]^{-1} + 
\Sigma^\mp)^{-1}\Phi^\pm G^\pm \, .
\eeq
They can be thought of as describing the propagation of a charge-conjugate particle 
(i.e., a hole) with propagator $([G_0^-]^{-1} + \Sigma^-)^{-1}$ that is converted 
into a particle  with propagator $G^+$, via the condensate $\Phi^+$. (Or, a particle 
that is converted into a hole via the condensate.)  The essence of superconductivity 
or superfluidity is the existence of a difermion condensate that makes the quasiparticle 
excitations superpositions of elementary states with fermion-number $\ldots,-5,-3,-1,1,
3,5,\ldots$; we see the formalism accommodating this phenomenon here.  

\begin{figure}[t]
\begin{center} 
\includegraphics[width=\hsize]{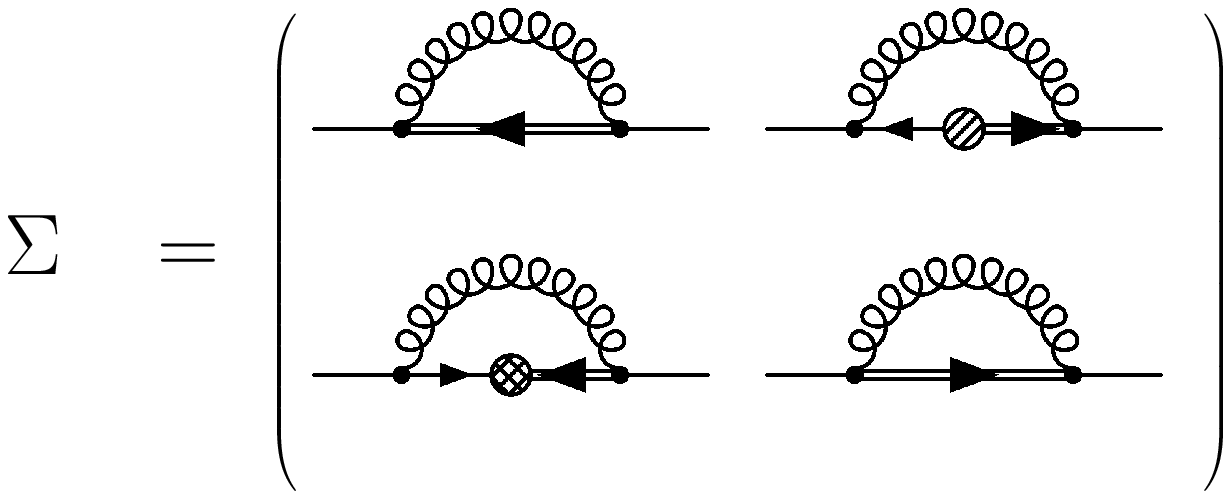}
\vskip 0.3in
\includegraphics[width=0.7\hsize]{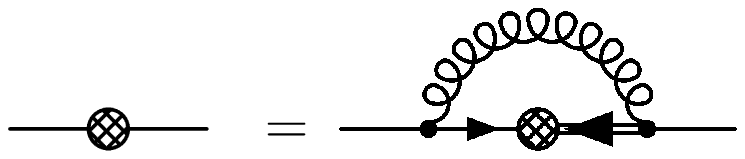}
\end{center}
\caption{Upper panel: Diagrammatic representation of the quark self-energy 
in Nambu-Gorkov space. Curly lines correspond to the gluon propagator $D$.  
The quasiparticle propagators $G^+$ and $G^-$ are denoted by double lines 
with an arrow pointing to the left and right, respectively. The anomalous 
propagators $F^\pm$ in the off-diagonal entries are drawn according to 
their structure given in Eq.~(\ref{Xpm}): thin lines correspond to the 
term $([G_0^\mp]^{-1} + \Sigma^\mp)^{-1}$, while the cross-hatched and hatched circles 
denote the gap matrices $\Phi^+$ and $\Phi^-$, respectively. Lower panel:
The QCD gap equation (\ref{gapeq}) is obtained by equating $\Phi^+$ with the 
lower left entry of the self-energy depicted in the upper panel (the other 
off-diagonal component yields an equivalent equation for $\Phi^-$).
}
\label{fig:sigma}
\end{figure}

We shall approximate $\Gamma_2$ by only taking into account two-loop diagrams. Upon 
taking the functional derivative with respect to $S$, this corresponds to a one-loop 
self-energy $\Sigma$. We show $\Sigma$ diagrammatically in Fig.~\ref{fig:sigma}. We 
shall argue later that this approximation is sufficient to calculate $\log(\Delta/\mu)$ 
up to terms of order $g^0$. Upon making this approximation, the gap equation takes the 
form shown in the lower panel of Fig.~\ref{fig:sigma}, namely
\beq
\Phi^+(K) = g^2\int_Q \ga^\mu \, T_a^T \, F^+(Q) \, \ga^\nu \, T_b \, 
D_{\mu\nu}^{ab}(K-Q)
\, , \label{gapeq}
\eeq
in momentum space, where $D_{\mu\nu}^{ab}(K-Q)$ is the gluon propagator. 

Note that in the derivation of the gap equation we have assumed the system 
to be translationally invariant. This assumption fails for crystalline color 
superconductors, see Sec.~\ref{sec:NJL}.  There has been some work on analyzing 
a particularly simple crystalline phase in QCD at 
asymptotically high densities and weak coupling \cite{Leibovich:2001xr}, but the formalism we are 
employing does not allow us to incorporate it into our presentation
and, anyway, this subject remains to date largely unexplored.

\subsection{Quasiparticle excitations}
\label{sec_qp}

Before we proceed with solving the gap equation, it is worthwhile to
derive the dispersion relations for the fermionic  quasiparticle excitations 
in a color superconductor.  That is, we suppose that the gap parameter(s)
$\Delta$ have been obtained in the manner that we shall describe below
and ask what are the consequences for the quasiparticle dispersion relations.
Based on experience with ordinary superconductors or superfluids, we expect 
(and shall find) gaps in the dispersion relations for the fermionic 
quasiparticles. We may also expect that in some color superconducting 
phases, quasiparticles with different colors and flavors, or different 
linear combinations of color and flavor, differ in their gaps and 
dispersion relations. Indeed, some gaps may vanish or may be nonzero 
only in certain directions in momentum space. 

The quasiparticle dispersion relations are encoded within the anomalous
self-energy $\Phi^+$, defined in (\ref{sigmanambu}), which satisfies
the gap equation (\ref{gapeq}). We shall assume that $\Phi^+$ can be 
written in the form 
\beq \label{gapmatrix}
\Phi^+(K)=\sum_{e=\pm}  \Delta^{(e)}(K)\, {\cal M} \, 
   \Lambda^{(e)}_{\bf k} \, ,
\eeq
where ${\cal M}$ is a matrix in  color, flavor and Dirac space, 
and $\Lambda_{\bf k}^{(e)}\equiv (1+e\gamma_0 \bm{\gamma} 
\cdot \hat{\mathbf{k}})/2$ are projectors onto states of positive 
($e=+$) or negative ($e=-$) energy. The corresponding gap functions are 
denoted as $\Delta^{(e)}(K)$ and will be determined by the gap 
equation. Here and in the following the energy superscript is denoted
in parentheses to distinguish it from the superscript that denotes
components in Nambu-Gorkov space. 
In our presentation we shall assume that ${\cal M}$ is 
momentum-independent, corresponding to a condensate of Cooper pairs 
with angular momentum $J=0$, but the formalism can easily be extended 
to allow a momentum-dependent ${\cal M}_{\bf k}$ as required for 
example in the analysis of the CSL phase and we shall quote results 
for this case also. Note that in Eq.~(\ref{gapmatrix}) we are assuming 
that every nonzero entry in ${\cal M}$ is associated with the same 
gap functions $\Delta^{(e)}$; the formalism would have to be generalized 
to analyze phases in which there is more than one independent gap 
function, as for example in the gCFL phase.


We shall analyze color superconducting phases whose color, flavor and Dirac
structure takes the form
\beq \label{calM}
{\cal M}_{ij}^{\alpha\beta} =   \phi_{A}^{B}  \epsilon^{\alpha\beta A} \, \epsilon_{ijB} \,\ga_5\, , 
\eeq
where the $\gamma_5$ Dirac structure selects a positive parity condensate,
where, as described in Secs.~\ref{sec:intro} and \ref{sec:cfl}, the 
antisymmetric color matrix is favored since QCD is attractive
in this channel and the antisymmetric flavor matrix is then required, 
and where $\phi$ is a $3\times 3$ matrix.  
We note that because the full flavor symmetry is the 
chiral $SU(3)_L\times SU(3)_R$ symmetry,  the matrix $\phi$ is actually a pair 
$(\phi^L,\phi^R)$. In this section we shall assume $\phi^R=\phi^L$. The case 
$\phi^R\neq\phi^L$, which corresponds to a meson condensate in the CFL phase, 
is discussed in Sec.~\ref{sec_gbs}. 

The excitation spectrum is given by the poles of the propagator $S$
in (\ref{fullquark}). (We shall see that the diagonal and the off-diagonal
entries in $S$ have the same poles.)
It will turn out that the Hermitian matrix ${\cal M}{\cal M}^\dag$
determines which quasiparticles are gapped and determines the ratios
among the magnitudes of (possibly) different gaps.  It is convenient
to write this matrix via its spectral representation
\beq\label{spectralrep}
{\cal M}{\cal M}^\dag=\sum_r\lambda_r\,{\cal P}_r \, , 
\eeq 
where $\lambda_r$ are the eigenvalues and ${\cal P}_r$ the projectors onto the 
corresponding eigenstates.  

The final preparation that we must discuss prior to computing the 
propagator is that we approximate the diagonal elements of the 
quark self-energy as \cite{Brown:1999yd,Manuel:2000nh,Gerhold:2005uu}
\beq 
\label{sigmaapp}
\Sigma^\pm \simeq \ga_0\Lambda^{(\pm)}_{\bf k} \frac{g^2}{18\pi^2}k_0
    \log\frac{48e^2m_g^2}{\pi^2k_0^2}  \,\, ,
\eeq
where $m_g^2=N_fg^2\mu^2/(6\pi^2)$ is the square of the effective gluon 
mass at finite density, and $e$ is the Euler constant. The expression 
(\ref{sigmaapp}) is the low energy approximation to the one-loop 
self-energy, valid for $k_0\sim \Delta \ll m_g$, for the positive 
energy ($e=+$) states. Taking the low energy approximation to 
$\Sigma^{\pm}$ and neglecting the self-energy correction
to the negative energy states will prove sufficient to determine 
$\log(\Delta/\mu)$ up to order $g^0$.

With all the groundwork in place, we now insert $G_0$, $\Sigma$ from
(\ref{sigmaapp}), and $\Phi$ from (\ref{gapmatrix}) into (\ref{Gpm}) and
(\ref{Xpm}), and hence (\ref{fullquark}), and use (\ref{spectralrep}) to 
simplify the result. We find that the diagonal entries in the fermion 
propagator $S$ are given by
\beq 
\label{prop}
G^\pm = \left(\left[G_0^\mp\right]^{-1} + \Sigma^\mp\right)
  \sum_{e,r} 
  \frac{ {\cal P}_r\, \Lambda_{\bf k}^{(\mp e)}}
       { \big[k_0/Z^{(e)}(k_0)\big]^2 - 
         \big[\epsilon_{k,r}^{(e)}\big]^2} \, ,
\eeq
while the anomalous propagators are 
\begin{subequations} 
\label{anomalous}
\bea \label{S212SC}
F^+(K)&=&-\sum_{e,r} 
  \frac{ \gamma_0 \, {\cal M}\, \gamma_0\, {\cal P}_r 
         \Lambda_{\bf k}^{(-e)} \, \Delta^{(e)}}
       {\big[k_0/Z^{(e)}(k_0)\big]^2- 
        \big[\epsilon_{k,r}^{(e)}\big]^2} 
    \,\, , \\
F^-(K)&=&-\sum_{e,r} 
  \frac{ {\cal M}^\dag\, {\cal P}_r \Lambda_{\bf k}^{(e)}
         \, (\Delta^{(e)})^*}
       {\big[k_0/Z^{(e)}(k_0)\big]^2-
        \big[\epsilon_{k,r}^{(e)}\big]^2} 
     \,\, .
\eea
\end{subequations}
In writing these expressions, we have defined the wave function 
renormalization factor 
\beq 
Z^{(+)}(k_0) \equiv \left(1 + \frac{g^2}{18\pi^2}
      \log\frac{48e^2m_g^2}{\pi^2 k_0^2}\right)^{-1} \, , 
\eeq
for the positive energy $e=+$ components, originating from the 
self-energy (\ref{sigmaapp}). (By neglecting the negative energy 
contribution to $\Sigma^{\pm}$ in (\ref{sigmaapp}), we are 
setting  the negative energy wave function renormalization 
$Z^{(-)}(k_0)= 1$.) We have furthermore defined
\beq \label{excite}
\epsilon_{k,r}^{(e)} \equiv 
   \sqrt{(ek-\mu)^2+\lambda_r \, |\Delta^{(e)}|^2} \, .
\eeq
The $r$'th quasiparticle and antiquasiparticle energies are then given
by solving $k_0 = Z^{(e)}(k_0) \epsilon_{k,r}^{(e)}$ for $k_0$. To 
leading order in $g$, wave function renormalization can be neglected 
and the quasiparticle and antiquasiparticle energies are given by
the $\epsilon_{k,r}^{(e)}$ themselves. We see from (\ref{excite}) that 
the antiparticles have $\epsilon > \mu$ --- in fact, for $k$ near $\mu$
they have $\epsilon \sim  2 \mu$.  They therefore never play an important
role at high density.  This justifies our neglect of the negative energy 
$\Sigma^{\pm }$ and hence of $Z^{(-)}$. And, it justifies the further 
simplification that we shall henceforth employ, setting the antiparticle 
gap to zero, $\Delta^{(-)}=0$, and denoting $\Delta\equiv\Delta^{(+)}$. 
We shall also use the notation $Z(k_0)\equiv Z^{(+)}(k_0)$ and 
$\epsilon_{k,r} \equiv\epsilon^{(+)}_{k,r}$. We then see that the 
minimum value of $\epsilon_{k,r}$ occurs at the Fermi surface, where 
$k=\mu$, and is given by $\sqrt{\lambda_r}\Delta$ which is conventionally 
referred to as the gap, again neglecting wave function renormalization.  
We see that although we must solve the gap equation in order to determine 
the magnitude of the gap parameter $\Delta$, as we will do in
Secs.~\ref{sec_eliash} and \ref{sec_qcdgaps}, the ratios among the 
actual gaps in the quasiparticle spectra that result are determined 
entirely by the $\lambda_r$'s, namely the eigenvalues of ${\cal M}
{\cal M}^\dag$.

We close this subsection by evaluating the pattern of quasiparticle
gaps explicitly for the CFL and 2SC phases. We list the order parameters 
$\phi^B_A$, eigenvalues $\lambda_r$, and corresponding projectors 
${\cal P}_r$ for these two phases in Table \ref{table2SCCFL}. In the 
CFL phase, one finds the eigenvalues $\lambda_1=4$ with degeneracy 
${\rm Tr}[{\cal P}_1]=1$ and $\lambda_2=1$ with degeneracy ${\rm Tr}
[{\cal P}_2]=8$. This means that all nine quasiparticles are gapped.
There is an octet with gap $\Delta$, and a singlet with gap $2\Delta$. 
The octet Cooper pairs are $gu$-$rd$, $bd$-$gs$, $bu$-$rs$, as well 
as two linear combinations of the three quarks $ru$-$gd$-$bs$. (Here, 
$gu$ refers to a green up quark, etc.) The singlet Cooper pair with 
twice the gap is the remaining orthogonal combination of $ru$-$gd$-$bs$.   
In the 2SC phase, on the other hand, we find four quasiparticles
with $\lambda_1=1$ and hence gap $\Delta$ and 5 quasiparticles with 
$\lambda_2=0$ that are unpaired. The gapped quasiparticles involve 
the first two colors, red and green, and the first two flavors, 
up and down. The Cooper pairs have color-flavor structure $ru$-$gd$ 
and $gu$-$rd$. (Note that all these color-flavor combinations depend on the 
chosen basis of the color and flavor (anti)triplets. This basis is fixed by 
Eq.\ (\ref{calM}); applying color (flavor) rotations to $\epsilon^{\alpha\beta A}$
($\epsilon_{ijB}$) would change the basis and yield different, physically equivalent, 
color-flavor combinations for the 2SC and CFL phases.)

\begin{table*}[th]
\begin{tabular}{|c||c||c|c||c|c|} 
\hline
\;\; phase \;\; & $\phi_{A}^{B}$ &  
\;\; $\lambda_1$  \;\; & 
\;\; $\lambda_2$  \;\; & $({\cal P}_1)_{\alpha\beta}^{ij}$ & 
$({\cal P}_2)_{\alpha\beta}^{ij}$ \\ 
\hline\hline
CFL & $\delta_{A}^{B}$ &  4\;(1--fold)
 & 1\;(8--fold) & $\delta_\alpha^i\delta_\beta^j/3$ & \;\;
$\delta_{\alpha\beta}\delta^{ij}-\delta_\alpha^i\delta_\beta^j/3$\;\;  \\ 
\hline  
2SC & \;\; $\delta_{A3}\delta^{B3}$\;\; & \;\; 1\;(4--fold)\;\; &\;\; 0\;(5--fold)\;\; 
    & \;\; $(\delta_{\alpha\beta}-
\delta_{\alpha 3}\delta_{\beta 3})(\delta^{ij}-
\delta^{i3}\delta^{j3})$\;\; &\;\; 
$\delta_{\alpha 3}\delta_{\beta 3}\delta^{i3}\delta^{j3}$ \;\; \\
\hline
\end{tabular}
\caption{Color-flavor structure of CFL and 2SC phases: Order 
parameters $\phi_{A}^{B}$, 
eigenvalues $\lambda_r$ of the matrix ${\cal M}{\cal M}^\dag$, and corresponding 
projectors ${\cal P}_r$, derived from Eq.\ (\ref{calM}). Color (flavor) indices are denoted $\alpha$, 
$\beta$, ($i$, $j$).  
}
\label{table2SCCFL}
\end{table*}

The formalism of this section can easily be applied to patterns of pairing 
in which ${\cal M}_{\bf k}$ depends on the direction of the quark momentum 
${\bf k}$. Such phases arise if the Cooper pairs carry total angular momentum 
$J=1$. This allows for pairing between quarks of the same flavor, as discussed 
in Sec.~\ref{subsec:single}. Depending on the specific structure of 
${\cal M}_{\bf k}$, the eigenvalues $\lambda_r$ may become momentum dependent 
and lead to nodes in the gap function along certain directions in momentum 
space. 

\subsection{Pressure and condensation energy}
\label{subsubsec:pressure}   

We can now return to our expression (\ref{pressure}) for the pressure 
$P$ (equivalently, the thermodynamic potential since $\Omega=-P$)
for a color superconductor and use the results of Sec.~\ref{sec_qp}
to evaluate it for a superconducting phase of the form (\ref{calM}).
We first substitute the expressions (\ref{prop}) and (\ref{anomalous}) for the fermion 
propagator (\ref{fullquark}) in the pressure (\ref{pressure}).  In order to obtain a result that is 
valid at both nonzero and zero temperature, it is then convenient to 
switch to Euclidean space, and perform the sum over Matsubara frequencies. 
Upon doing the trace over Nambu-Gorkov, color, flavor, and Dirac space we 
find
\bea 
\label{Tnonzero}
P &=& \sum_{e,r}\int\frac{d^3 k}{(2\pi)^3} \,
 \Tr[{\cal P}_r] \Bigg\{ \epsilon_{k,r}^{(e)} 
+ 2\,T\,\log\left(1 + e^{-\epsilon_{k,r}^{(e)}/T} \right)
 \nonumber   \\
 & & \;\; \mbox{}
  -  \frac{\lambda_r\, |\Delta|^2}{2\,\epsilon^{(e)}_{k,r}}
  \,\tanh\left(\frac{\epsilon_{k,r}^{(e)}}{2\,T}\right)  \Bigg\} 
  \, .
\eea
Including the effects of wave function renormalization would modify this 
expression at order $g$. In most contexts, we shall only consider the pressure 
(\ref{Tnonzero}) at zero temperature.  
In this case, with $\sum_r\Tr[{\cal P}_r] =N_cN_f$, 
\beq \label{totalpressure}
P = N_cN_f\frac{\mu^4}{12\pi^2} + \delta P \,\, .
\eeq
where we denote the pressure difference of the color-superconducting phase 
compared to the unpaired phase by $\delta P$.  If we make the simplifying
assumption (corrected in the next subsection) that the gap function
is a constant in momentum space in the vicinity of the Fermi surface,
we find the easily interpretable result
\beq \label{deltap}
\delta P =  
\frac{\mu^2}{4\pi^2}\,\sum_r {\rm Tr}[{\cal P}_r]\,\lambda_r\Delta^2
\, .
\eeq
At $T=0$ this quantity is the condensation energy density of the 
color-superconducting state. The fact that $\delta P>0$ implies that the 
superconducting state is favored relative to the normal phase. We observe 
that $\delta P$ is proportional to the sum of the energy gap squared of 
the $r$-th branch, multiplied by the corresponding degeneracy 
${\rm Tr}[{\cal P}_r]$. 

We can use the result (\ref{deltap}) to understand how to compare
the favorability of different patterns of color superconducting pairing:
the phase with lowest free energy (highest $\delta P$) is favored.  As
an example, in the CFL phase 
$\delta P = (\mu^2/(4\pi^2))(8\cdot 1+ 1\cdot 4)\Delta_{\rm CFL}^2$ 
while in the 2SC phase
$\delta P = (\mu^2/(4\pi^2))(4\cdot 1+0 \cdot 5)\Delta_{\rm 2SC}^2$ 
suggesting that the CFL phase is favored. (We shall make this conclusion
firm in Sec.~\ref{sec_qcdgaps}, where we shall find that $\Delta_{\rm CFL}$ is
smaller than $\Delta_{\rm 2SC}$ but only by a factor of $2^{1/3}$. 
This factor will also turn out to be determined entirely by 
the $\lambda_r$'s and ${\rm Tr}[{\cal P}_r]$'s.)

In principle, in order to generalize the conclusion that the CFL
phase is favored one has to compare the condensation energies of all 
possible phases described by the order parameter ${\cal M}$ in 
Eq.~(\ref{calM}). This is difficult because $\phi$ is an arbitrary 
complex $3\times 3$ matrix. At asymptotic densities, however, we can 
neglect the strange quark mass and treat the quarks as degenerate in 
mass. The resulting $SU(3)_c\times SU(3)_{f}$ 
symmetry simplifies the task. ($f$ is $L$ or $R$ for $\phi_L$ or $\phi_R$.)
The matrix $\phi$ transforms under color-flavor rotations as $\phi\to 
U^T\phi\,V$ with $U\in SU(3)_c$, $V\in SU(3)_{f}$. This means that two order 
parameters $\phi$ and $U^T\phi\,V$ describe the same physics. Now note that 
for any $\phi$ there exists a transformation $(U,V)$ such that $U^T\phi\,V$ 
is diagonal. Therefore, we need consider only diagonal matrices $\phi$. 
Choosing all diagonal elements to be nonzero corresponds to the maximum 
number of gapped quasiparticles. Hence, once we show (below) that 
$\Delta_{\rm 2SC}$ is not much larger than $\Delta_{\rm CFL}$ it is 
easy to understand that the CFL phase with $\phi={\bf 1}$, yielding an 
order parameter that is invariant under the largest possible subgroup
of the original symmetries, is the ground state at asymptotically large 
densities.

At lower densities, the flavor symmetry is explicitly broken by the mass 
of the strange quark (the symmetries are further broken by different 
chemical potentials due to neutrality constraints). In this case, the 
above argument fails and non-diagonal matrices $\phi$ become possible 
candidates for the ground state \cite{Rajagopal:2005dg,Malekzadeh:2006ik}.

\subsection{Weak coupling solution of the gap equation}
\label{sec_eliash}

We are now in a position to solve the QCD gap equation (\ref{gapeq}) for 
an order parameter with a given matrix structure ${\cal M}$. The matrix 
structure of the gap equation (\ref{gapeq}) is handled by multiplying both 
sides of the equation by ${\cal M}^\dag\Lambda_{\bf k}^{(+)}$ and taking the 
trace over color, flavor, and Dirac indices. 

The gap equation is sensitive to gluon modes with small momentum ($p \ll 
m_g$) and even smaller energy ($p_0\sim p^3/m_g^2 \ll p$), meaning that 
medium effects in the gluon propagator have to be taken into account. In 
the low momentum limit, the gluon propagator takes on the standard hard-dense 
loop approximation form \cite{Braaten:1991gm}, which we shall give below in 
Eqs.~(\ref{Fl}) and (\ref{Ft}) upon simplifying it as appropriate for $p_0
\ll p$.  In order to obtain $\log(\Delta/\mu)$ to order $g^0$, it suffices 
to keep only the leading terms in the propagator in the $p_0\ll p$ limit. 
We shall work in Coulomb gauge. Gauge independence of the gap in a generalized 
Coulomb gauge was established in \cite{Pisarski:2001af}, and a more formal 
proof of gauge invariance was given in \cite{Gerhold:2003js,Hou:2004bn}. The 
gap equation reads
\bea \label{gapeq1}
\Delta_{k,r} &=& \frac{g^2}{4}\,\int\frac{d^3q}{(2\pi)^3}\,
  \sum_s\,Z(\epsilon_{q,s})
   \frac{\Delta_{q,s}}{\epsilon_{q,s}}\,
    \tanh\left(\frac{\epsilon_{q,s}}{2T}\right) \non
&&\hspace{-1cm}\times\,\Big[D_\ell(p)\, {\cal T}_{00}^s({\bf k},{\bf q}) + 
 D_t(p,\epsilon_{q,s},\epsilon_{k,r})\, {\cal T}_t^s({\bf k},{\bf q})\Big] \, ,
\eea
where we have abbreviated $P\equiv K-Q$ and have denoted the gap function 
on the quasiparticle mass shell by $\Delta_{k,r}\equiv\Delta(\epsilon_{k,r},
{\bf k})$. We have denoted the traces over color, flavor, and Dirac space by
\beq \label{defT}
{\cal T}_{\mu\nu}^s({\bf k},{\bf q}) \equiv 
  -\frac{ {\rm Tr} \left[
           \ga_\mu T^T_a \ga_0 {\cal M}_{\bf q} \ga_0 {\cal P}_s
           \Lambda_{\bf q}^{(-)} \ga_\nu T_a {\cal M}_{\bf k}^\dag
           \Lambda_{\bf k}^{(+)} \right]}
        { {\rm Tr}\left[
           {\cal M}_{\bf k} {\cal M}_{\bf k}^\dag
           \Lambda_{\bf k}^{(+)}\right]} \, ,
\eeq
and ${\cal T}_t^s({\bf k},{\bf q})\equiv-(\delta^{ij}-\hat{p}^i\hat{p}^j)\,
{\cal T}_{ij}^s({\bf k},{\bf q})$. The two terms inside the square bracket in 
Eq.~(\ref{gapeq1}) correspond to the contributions from electric and magnetic
gluons. The dominant contribution comes from almost static gluons with $p_0\ll
p$. The static electric and almost static magnetic gluon propagator give
\bea
\label{Fl}
D_\ell(p) &\equiv & \frac{2}{p^2 + 3m_g^2} \\
\label{Ft}
D_t(p,\epsilon,\epsilon') &\equiv& 
 \frac{p^4}{p^6+M_g^4(\epsilon+\epsilon')^2} 
 + (\epsilon'\to-\epsilon')\, ,
\eea
where $M_g^2\equiv (3\pi/4) m_g^2$. 
With the gap equation now stated fully explicitly, all that remains is 
to solve it.  

We can solve (\ref{gapeq1}) for the zero temperature gap $\Delta$ on the Fermi surface. 
Or, we can solve for $T$ in the $\Delta
\rightarrow 0$ limit, thus obtaining the critical temperature $T_c$. Solving 
for $\Delta$, we find that it has a weak coupling expansion of the form
\beq\label{pQCD_exp}
 \log\left(\frac{\Delta}{\mu}\right) = 
 -\frac{b_{-1}}{g}- \bar{b}_0\log(g) - b_0 - \ldots . 
\eeq
In our treatment of the fermion propagator, the gluon propagator, and in our
truncation of the self-energy in Fig.~\ref{fig:sigma} to one loop (for example 
neglecting vertex renormalization) we have been careful to keep all effects 
that contribute to $b_0$, but we have neglected many that contribute at order 
$g \log g$ and $g$. The formalism that we have presented can be used to evaluate 
$b_{-1}$, $\bar{b}_0$ and $b_0$, and we shall describe the results in 
Sec.~\ref{sec_qcdgaps}.  

Before turning to quantitative results, it is worth highlighting the origin 
and the importance of the leading $-1/g$ behavior in (\ref{pQCD_exp}), namely 
the fact that $(\Delta/\mu) \sim \exp(-{\rm constant}/g)$.  If in the gap 
equation of Fig.~\ref{fig:sigma} we were to replace the exchanged gluon by 
a contact interaction, we would obtain a gap equation of the form
\beq
\Delta\propto g^2 \int d\xi \frac{\Delta}{\sqrt{\xi^2 + \Delta^2}}
\eeq
with $\xi\equiv k - \mu$. This always has the solution $\Delta=0$; to seek 
nonzero solutions, we cancel $\Delta$ from both sides of the equation. Then,
if $\Delta$ were $0$, the remaining integral would diverge logarithmically 
at small $\xi$. Therefore, we find a nonzero $\Delta$ for any positive 
nonzero $g$ no matter how small, with $\Delta \propto \exp(-{\rm constant}/g^2)$. 
This is the original BCS argument for superconductivity as a consequence 
of an attractive interaction at a Fermi surface. However, once we restore the 
gluon propagator the argument is modified.  The crucial point is that magnetic 
gluon exchange is an unscreened long-range interaction, meaning that the angular 
integral will diverge logarithmically at forward scattering in the absence of 
any mechanism that screens the magnetic interaction.  The gap equation 
therefore takes the form
\beq
\Delta\propto g^2 \int d\xi \frac{\Delta}{\sqrt{\xi^2 + \Delta^2}}
d \theta \frac{ \mu^2}{\theta \mu^2 + \delta^2}
\eeq
where $\theta$ is the angle between the external momentum ${\bf k}$
and the loop momentum ${\bf q}$ and where $\delta$ is some quantity with
the dimensions of mass that cuts off the logarithmic collinear divergence
of the angular integral. In the superconducting phase this divergence
will at the least be cut off by the Meissner effect, which screens
gluon modes with $p<\Delta$ (since the Cooper pairs have size $1/\Delta$)
giving $\delta\sim\Delta$. This yields $\Delta\sim g^2 \Delta (\log\Delta)^2$ 
and hence a nonzero gap $\Delta\sim \exp(-{\rm constant}/g)$. This consequence 
of the long-range nature of the magnetic gluon exchange was first discovered 
by Barrois \cite{Barrois:1979pv}. However, pursuing the argument as just 
stated yields the wrong value of the constant $b_{-1}$; it was Son who 
realized that the collinear divergence is cut off by Landau damping at 
a larger value of the angle $\theta$ than that at which the Meissner 
effect does so. Loosely speaking, Landau damping leads to $\delta \sim
(\Delta m_g^2)^{1/3}\gg \Delta$. Son was then able to calculate the 
coefficients of the $1/g$ term and of the logarithm in (\ref{pQCD_exp}) 
\cite{Son:1998uk}. The calculation of the constant $b_0$ was initiated in 
\cite{Schafer:1999jg,Hong:1999fh,Pisarski:1999bf,Pisarski:1999tv,Hsu:1999mp}
and completed in \cite{Brown:1999aq,Wang:2001aq}. Higher order terms are 
expected to be of order $g\log g$, order $g$, and at higher order still may
contain fractional powers and logarithms of $g$, see Sec.~\ref{sec_nfl}. 

The  $(\Delta/\mu)\propto \exp(-{\rm constant}/g)$ behavior means that the
color superconducting gap is parametrically larger at $\mu\rightarrow\infty$
than it would be for any four-fermion interaction.  Furthermore, asymptotic
freedom ensures that $1/g(\mu)^2$ increases logarithmically with $\mu$, which 
means that $\exp[-{\rm constant}/g(\mu)]$  decreases more slowly than $1/\mu$ 
at large $\mu$. We can therefore conclude that $\Delta$ increases with
increasing $\mu$ at asymptotically large $\mu$, although of course 
$\Delta/\mu$ decreases.  

We conclude this subsection with a derivation of the correct value of the
coefficient $b_{-1}$, namely the constant in $(\Delta/\mu)\propto \exp(
-{\rm constant}/g)$. This coefficient turns out to be independent of the 
spin-color-flavor structure ${\cal M}$, and it is therefore simplest to 
present its derivation in the 2SC phase, in which there is only one gap 
parameter $\Delta_k \equiv \Delta_{k,r=1}$, $\epsilon_k \equiv \epsilon_{k,r=1}$. 
The leading behavior of the gap is completely determined by magnetic gluon 
exchanges. We can also approximate the trace term by its value in the 
forward direction ${\cal T}_t({\bf k},{\bf q})\simeq {\cal T}_t({\bf k},
{\bf k})=2/3$ and set the wave function renormalization $Z(q_0)=1$ (in
the forward limit we also find  ${\cal T}_{00}({\bf k},{\bf q})\simeq 
{\cal T}_t({\bf k},{\bf q})$). Carrying out the angular integrals in 
the gap equation gives
\beq
\label{eliash}
 \Delta_k = \frac{g^2}{18\pi^2} \int dq \,
 \frac{\Delta_q}{\epsilon_q} \,\frac{1}{2}
  \log\left(\frac{\mu^2}{|\epsilon_q^2-\epsilon_k^2|}\right).
\eeq
Son observed that at this order we can replace the logarithm by 
$\max\{\log(\mu/\epsilon_k),\log(\mu/\epsilon_q)\}$. Introducing logarithmic 
variables $x=\log[2\mu/(\xi_{k}+\epsilon_{k})]$ with $\xi_k=
|k-\mu|$, the integral equation (\ref{eliash}) can be written as a 
differential equation 
\beq 
\Delta''(x) = -\frac{g^2}{18\pi^2}\Delta(x),
\eeq
with the boundary conditions $\Delta(0)=0$ and $\Delta'(x_0)=0$. Here, $x_0=
\log(2\mu/\Delta)$ determines the gap on the Fermi surface. The solution is 
\beq 
\Delta(x)=\Delta\sin\left(\frac{gx}{3\sqrt{2}\pi} \right),\hspace{0.15cm}
 \Delta=2\mu \exp\left(-\frac{3\pi^2}{\sqrt{2}g}\right),
\label{SimpleSinusoid}
\eeq
and thus $b_{-1}=3\pi^2/\sqrt{2}$. We conclude that in the weak-coupling 
limit the gap function is peaked near the
Fermi surface, with a width that is much smaller
than $\mu$ but much larger than $\Delta$.
Had we not set $Z(q_0)=1$, the $x$-dependence of $\Delta(x)$
would be more complicated than the simple sinusoid in (\ref{SimpleSinusoid}),
but the conclusion remains unchanged~\cite{Wang:2001aq}.

\subsection{Gap and critical temperature at weak coupling}
\label{sec_qcdgaps}

The gap on the Fermi surface of a color superconductor at zero 
temperature can be written as 
\beq
\label{gapgeneral0}
\Delta = \mu g^{-\bar{b}_0} e^{-b_0}
   \exp\left(-\frac{3\pi^2}{\sqrt{2}g}\right) \, , 
\eeq
to order $g^0$ in the weak-coupling expansion of $\log(\Delta/\mu)$.  We 
have derived the coefficient $b_{-1}$ in the exponent above, starting from 
a simplified version of the gap equation (\ref{gapeq1}), with no wave function 
renormalization and a simplified gluon propagator. Upon restoring these effects,
analysis of the gap equation (\ref{gapeq1}) yields
\beq  
\label{gapgeneral}
 g^{-\bar{b}_0}e^{-b_0}= g^{-5} 512\pi^4\left(\frac{2}{N_f}\right)^{5/2}\,
    e^{-b_0'}\,e^{-d}\,e^{-\zeta}\ .
\eeq
In the following we shall define and explain the origin of each of the terms 
in this equation; we shall not present a complete derivation.
\begin{itemize}
\item
The factor $g^{-5}$ and the numerical factor in Eq.~(\ref{gapgeneral}) 
are due to large angle magnetic as well as electric gluon exchanges and 
are independent of the pattern of pairing in the color superconducting phase, 
i.e. independent of ${\cal M}$. 
\item
The factor 
\beq
e^{-b_0'} =  \exp\left(-\frac{\pi^2 + 4}{8}\right)\simeq 0.177
\eeq
arises from the wave function renormalization factor $Z(q_0)$ in (\ref{sigmaapp})
\cite{Brown:1999aq,Wang:2001aq} and is also independent of ${\cal M}$ and hence 
the same for all color superconducting phases. 
\item
The factors that we have written as $e^{-d} e^{-\zeta}$ are different in 
different color superconducting phases. The factor $e^{-d}$ is due to the 
angular structure of the gap. For the $J=0$ condensates whose gap equation
we have derived, $e^{-d}=1$.  Upon redoing the angular integrals for spin-1 
condensates, we find that they are strongly suppressed
\cite{Schafer:2000tw,Schmitt:2002sc,Schmitt:2004et}. For spin-1 pairing 
patterns in which quarks of the same chirality form Cooper pairs, $d=6$. 
A smaller suppression occurs when quarks of opposite chirality pair, $d=4.5$. 
Superpositions of these states yield values of $d$ between these limits. 
Regardless, perturbative QCD predicts spin-1 gaps to be two to three orders 
of magnitude smaller than spin-0 gaps.

\item 
The factor $e^{-\zeta}$ depends on ${\cal M}$, the color-flavor-spin matrix
that describes the pattern of pairing in a particular color superconducting
phase. In a phase in which ${\cal M}{\cal M}^\dagger$ has two different 
eigenvalues $\lambda_1$ and $\lambda_2$, describing ${\rm Tr}[{\cal P}_1]$ and 
${\rm Tr}[{\cal P}_2]$ quasiparticles respectively, we find
\beq \label{zeta}
\zeta = \frac{1}{2}
\frac{\langle {\rm Tr}[{\cal P}_1] \lambda_1\log\lambda_1 
            + {\rm Tr}[{\cal P}_2] \lambda_2\log\lambda_2 \rangle}
  {\langle {\rm Tr}[{\cal P}_1] \lambda_1 +  {\rm Tr}[{\cal P}_2] 
   \lambda_2\rangle} \, ,
\eeq
where the angular brackets denote an angular average (trivial for $J=0$ 
phases).  In the CFL phase, $\lambda_1=1$ and ${\rm Tr}[{\cal P}_1]=8$ 
while $\lambda_2=4$ and ${\rm Tr}[{\cal P}_2]=1$, meaning that there are 
8 quasiparticles with gap $\Delta$ and 1 with gap $2\Delta$. Evaluating 
Eq.~(\ref{zeta}), we find $e^{-\zeta}=2^{-1/3}$ in the CFL phase 
\cite{Schafer:1999fe}. In the 2SC phase, $e^{-\zeta}=1$.  Note that the 
ratio $\Delta_{\rm CFL}/\Delta_{\rm 2SC}$ is also $2^{-1/3}$ in an NJL
model analysis \cite{Rajagopal:2000wf}; this result depends on the structure 
of the condensates not on the nature of the interaction. From
$\Delta_{\rm CFL}/\Delta_{\rm 2SC}=2^{-1/3}$ we can conclude the
discussion begun in Sec.~\ref{subsubsec:pressure}, noting now that the condensation
energy in the CFL phase is larger than that in the 2SC phase by
a factor $3\cdot 2^{-2/3}$.  

\item
We can also determine the admixture of a color symmetric condensate in 
the CFL phase. In this case we have to use a two-parameter ansatz
for the gap and solve two coupled gap equations. The color-symmetric
gap parameter $\Delta_6$ is parametrically small compared to the 
color-antisymmetric gap $\Delta_{\bar{3}}$, and the two gap equations
decouple. We find $\Delta_{6}/\Delta_{\bar{3}}=\sqrt{2}\log(2)g/
(36\pi)$ which is suppressed by both the coupling constant $g$ and 
a large numerical factor.
\end{itemize}

In evaluating (\ref{gapgeneral0}) and (\ref{gapgeneral}), it suffices at present
to evaluate $g$ at the scale $\mu$.  The effect of choosing $g(a\mu)$ with $a$
either some purely numerical constant or some constant proportional to $g$
or $\log\Delta$ is order $g$, meaning that we cannot and need not determine 
$a$ within our present calculation of $\log\Delta$ to order $g^0$.  For a 
numerical estimate of the effects of a running $g$ on $\Delta$, see 
\cite{Schafer:2004yx}.  The effects are not as large as envisioned in
\cite{Beane:2000hu}.

 We shall discuss a systematic approach to the calculation of corrections 
beyond ${\cal O}(g^0)$ in $\log(\Delta/\mu)$ in Sec.~\ref{sec_nfl}. There 
are a number of effects that have been considered, and were shown not 
to contribute at ${\cal O}(g^0)$, but for which the actual size of the ${\cal O}(g)$ 
(or higher) correction is not known. These include vertex corrections 
\cite{Brown:2000eh}, the imaginary part of the gap function 
\cite{Feng:2006yg,Reuter:2006rf}, and the modification of the gluon
propagator due to the Meissner effect \cite{Rischke:2001py}.


 It is instructive to extrapolate the perturbative results to lower baryon 
densities for which the running coupling constant is not small. Taking $\mu
\simeq 400-500$ MeV, and a strong coupling constant $g\simeq 3.5$ (note 
that $g=3.56$ at $\mu=400$ MeV according to the two-loop QCD beta function, 
which of course should not be taken seriously at these low densities) one 
obtains $\Delta\simeq 20$ MeV. This is comparable to (but on the small 
side of) the range of typical gaps $\Delta\sim (20-100)$ MeV 
\cite{Rajagopal:2000wf} obtained using models in which the interaction 
between quarks is described via a few model parameters whose values are 
chosen based upon consideration of zero-density physics, like the NJL models 
that we shall discuss in Sec.~\ref{sec:NJL} or numerical solutions
of the Dyson-Schwinger equations \cite{Nickel:2006vf,Marhauser:2006hy}.  
This qualitative agreement between two completely different approaches,
one based on using a model to extrapolate from $\mu=0$ to $\mu=400-500$ MeV
and the other based on applying a rigorous calculation that is valid
for $\mu>10^8$ MeV where the QCD coupling is weak at $\mu=400-500$ MeV
gives us confidence that we understand the magnitude of $\Delta$, the
fundamental energy scale that characterizes color superconductivity.  Furthermore,
the one nonperturbative interaction in QCD whose contribution to $\Delta$
has been evaluated reliably at high density, namely that due to the 't Hooft
interaction induced by instantons, serves to increase $\Delta$, bringing the
high density computation into even better agreement with the model-based
approaches \cite{Alford:1997zt,Rapp:1997zu,Rapp:1999qa,Berges:1998rc,Carter:1998ji,Schafer:2004yx}.

Finally, we can use the gap equation (\ref{gapeq1}) to extract the 
critical temperature $T_c$. The 
result is~\cite{Pisarski:1999bf,Pisarski:1999tv,Schmitt:2002sc,Wang:2001aq,Schmitt:2004et,Brown:1999yd,Brown:1999aq}
\beq
\frac{T_c}{\Delta}=\frac{e^\ga}{\pi} e^\zeta\, ,
\eeq
where $\ga\simeq 0.577$ is the Euler-Mascheroni constant. This should 
be compared to the BCS result $T_c/\Delta=e^\ga/\pi\simeq 0.57$. We 
observe that deviations from the BCS ratio occur in the case of two-gap 
structures and/or anisotropic gaps. Nevertheless, since $e^\zeta$ is 
of order one, the critical temperature is always of the same order of 
magnitude as the zero-temperature gap.  We see that for the 2SC phase 
$T_c/\Delta$ is as in BCS theory, whereas in the CFL phase this ratio 
is larger by a factor of $2^{1/3}$.  It therefore turns out that $T_c$ 
is the same in the CFL and 2SC phases.   

 These estimates of $T_c$ neglect gauge field fluctuations, making them 
valid only at asymptotic densities. We shall see in Sec.~\ref{sec_lg} 
that including the gauge field fluctuations turns the second order phase 
transition that we find by analyzing (\ref{gapeq1}) into a first order 
phase transition, and increases $T_c$ by a factor $1+{\cal O}(g)$, see Eq.\ (\ref{elevatedTc}).

\subsection{Color and electromagnetic Meissner effect}
\label{subsubsec:screening}

 One of the characteristic properties of a superconductor is the Meissner
effect, the fact that an external magnetic field does not penetrate
into the superconductor. The external field is shielded by supercurrents
near the interface between the normal phase and the superconducting phase.
The inverse penetration length defines a mass scale which can be viewed
as an effective magnetic gauge boson mass.

 This effect can also be described as the Anderson-Higgs phenomenon
\cite{Anderson:1963pc,Higgs:1964ia}.
The difermion condensate acts as a composite Higgs field which breaks all
or part of the gauge symmetry of the theory. The gauge fields acquire a
mass from the Higgs vacuum expectation value, and the would-be Goldstone
bosons become the longitudinal components of the gauge fields. A well known
example in particle physics is provided by the electroweak sector of the
standard model. The $SU(2)_L\times U(1)_Y$ gauge symmetry of the
electroweak standard model
is broken down to $U(1)_{Q}$ by the expectation value of an $SU(2)$ Higgs
doublet which carries hypercharge. There are three
massive gauge bosons, the $W^\pm$ and the $Z$ boson. The $Z$ is a linear
combination of the original $I_3$ and $Y$ gauge bosons. The orthogonal
linear combination is the photon, which remains massless because
the Higgs condensate is electrically neutral.

The gauge symmetry in QCD is $SU(3)_c\times U(1)_{Q}$. Different
color superconducting order parameters realize different Higgs phases.
The color gauge group may be partially or fully broken, and mixing
between diagonal gluons and photons can occur. In the following we
shall concentrate on the 2SC and CFL phases and briefly mention other
phases at the end of the section. Our starting point is the one-loop
gauge boson polarization tensor
\cite{Rischke:2000ra,Rischke:2000qz,Rischke:2002rz,Litim:2001mv,Son:1999cm},
\beq \label{poltensors}
\Pi_{ab}^{\mu\nu}(P)\equiv\frac{1}{2}\frac{T}{V}\sum_K{\rm Tr}
[\hat{\Gamma}_a^\mu\, S(K)\,\hat{\Gamma}_b^\nu\,S(K-P)] \,\, ,
\eeq
where
\beq \label{vertex}
\hat{\Gamma}_a^\mu\equiv \left\{\begin{array}{lll}
  {\rm diag}(g\,\ga^\mu T_a,-g\,\ga^\mu T_a^T) &
    \mbox{for} & \quad a=1,\ldots,8 \,\, , \\ \\
  {\rm diag}(e\, \ga^\mu Q,-e\,\ga^\mu Q) & \mbox{for} & \quad a=9 \,\, .
\end{array} \right.
\eeq
%
%
Here, $Q$ is the electric charge matrix $Q={\rm diag}(2/3,-1/3,-1/3)$, and
$e$ the electromagnetic coupling constant. The polarization function can
be defined as the second derivative of the thermodynamic potential with
respect to an external gauge field. This quantity is equal to the derivative
of the induced charge/current with respect to the gauge field. Electric
charge screening is governed by the zero momentum limit of the $\Pi_{00}$
component of the polarization tensor. The Meissner effect is related
to a non-vanishing zero momentum limit of the spatial components $\Pi_{ij}$.
We can define electric (Debye) and magnetic screening masses as
\begin{subequations} \label{defmasses}
\bea
m^2_{D,ab}&\equiv& -\lim\limits_{p\to 0}\Pi_{ab}^{00}(0,{\bf p}) \, ,\\
m^2_{M,ab}&\equiv& \frac{1}{2}\lim\limits_{p\to 0}
(\delta^{ij}-\hat{p}^i\hat{p}^j)\Pi_{ab}^{ij}(0,{\bf p})
 \, .
\eea
\end{subequations}
A calculation of the full momentum dependence of $\Pi_{\mu\nu}$ in
the 2SC and CFL phases can be found in \cite{Rischke:2001py}
and \cite{Malekzadeh:2006ud}, respectively. One result that
we shall need in Sec.~\ref{sec_gbs} is the electric screening mass for gluons
in the CFL phase, which is given by 
\bea \label{debyeCFL}
m_{D,aa}^2= \frac{21-8\log 2}{36}\frac{g^2 \mu^2}{\pi^2} \, .
\eea
The numerical factor $21-8\log 2$ can be written as
$15+(6-8\log 2)$, where the first term comes from
diagrams in which the gluon couples to two octet
quasiparticles and the second from coupling to one
octet and one singlet quasiparticle.
The $\log 2$ factor is the log of the ratio of the singlet
and octet gaps.

In the following,
we shall discuss the Meissner masses. Results for the CFL phase 
\cite{Casalbuoni:2000na,Rischke:2000ra,Zarembo:2000pj,Son:1999cm} and the 
2SC phase \cite{Rischke:2000qz,Casalbuoni:2001ha} are summarized in Table \ref{tablemeissner}, where we also
list the screening masses for the single-flavor CSL phase \cite{Schmitt:2003xq,Schmitt:2003aa}.

 We observe that the chromomagnetic screening masses are of order
$g\mu$. This means that the screening length is much shorter than
the coherence length $\xi=1/\Delta$, and color superconductivity
is type I, see Sec.~\ref{sec_lg}. The fact that the screening masses
are independent of the gap does not contradict the fact that there
is no magnetic screening in the normal phase. Magnetic screening disappears 
for energies and momenta larger than the gap. Therefore, if the
$\Delta\to 0$ limit is taken before the limit $p\to 0$ then the magnetic
screening vanishes, as expected. Of course, magnetic screening
masses also vanish as the temperature approaches $T_c$.

 We also note that $m^2_{M,ab}$ is a $9\times 9$ matrix, and the
physical masses are determined by the eigenvalues of this matrix.
In the CFL phase, all magnetic gluons
acquire the same nonzero mass, reflecting the residual $SU(3)_{c+L+R}$.
In the 2SC phase, the Meissner masses of the gluons 1 through 3 vanish,
reflecting the symmetry breaking 
pattern $SU(3)_c\to SU(2)_c$. The unscreened gluons correspond to the
generators of the unbroken $SU(2)_c$, as they
only see the first two colors, red and green. Cooper pairs 
are red-green singlets and 
so cannot screen these low momentum gluons.

 In both 2SC and CFL phases, the off-diagonal masses vanish except for
the eighth gluon and the photon, $m_{M,\ga 8}^2=m_{M,8\ga}^2\neq 0$.
The two-by-two part of the gauge boson mass matrices that describe the
eighth gluon and the photon
has one vanishing eigenvalue and one nonzero eigenvalue.
The eigenvectors are characterized by a mixing angle $\theta$, given
in the last column of Table \ref{tablemeissner}. This angle defines
the new gauge fields,
\begin{subequations}
\bea
\tilde{A}_\mu^8 = \cos\theta\,A_\mu^8 + \sin\theta\,A_\mu \, , \\
\tilde{A}_\mu = -\sin\theta\,A_\mu^8 + \cos\theta\,A_\mu \, ,
\eea
\end{subequations}
where $A_\mu^8$ and $A_\mu$ denote the fields for the eighth gluon and the
photon, respectively. The $\tilde{A}_\mu^8$ 
gauge boson feels a Meissner effect; it is the
analogue of the massive $Z$-boson in the electroweak standard model.
The $\tilde{A}_\mu$ gauge boson, on the other hand, experiences
no Meissner effect because the diquark condensate is $\tilde Q$-neutral.
This is the photon of the unbroken Abelian $U(1)_{\tilde Q}$ gauge symmetry,
consisting of simultaneous color and flavor (i.e. electromagnetic) rotations.
The $\tilde{A}_\mu$ field satisfies Maxwell's equations. Because $g\gg e$, 
the mixing angle is very small and the $\tilde{A}_\mu$ photon contains 
only a small admixture of the original eighth gluon.

In contrast, $J=1$ color superconductors show an
electromagnetic Meissner effect \cite{Schmitt:2003xq,Schmitt:2003aa}. For
example, in the CSL phase there is no mixing between
the gluons and the photon, as can be seen in the last row of Table
\ref{tablemeissner}.
The photon acquires a mass since the
electromagnetic group is spontaneously broken. Other candidate spin-1
phases, such as the polar, planar, or ${\it A}$ phase involve mixing
but also (except for a one-flavor system) exhibit an electromagnetic
Meissner effect. This difference in phenomenology of spin-0
vs.~spin-1 color superconductors may have consequences
in compact stars \cite{Aguilera:2006xv}.

\begin{table*}[ht]
\begin{tabular}{|c||cccccccc|cc|c||c|c|c|}
\hline
& \multicolumn{8}{c}{$m^2_{M,aa}$}\vline &
\multicolumn{2}{c}{$\quad m^2_{M,a\ga}=m^2_{M,\ga a}\quad$}\vline
&  $\quad m_{M,\ga\ga}^2\quad$ & $\tilde{m}_{M,88}^2$
&$\tilde{m}_{M,\ga\ga}^2$
&  $\cos^2\theta$\\
\hline\hline
$a$ & 1 & 2 & 3 \;&\; 4 & 5 & 6 & 7\; & 8 & 1-7 & 8 & 9 &
  \multicolumn{3}{c}{}\vline\\  \hline\hline
\;\;CFL\;\;  & \multicolumn{8}{c}{$\eta\,g^2$}\vline & 0  &
$-\frac{2}{\sqrt{3}}\,\eta\,eg$  &  $\frac{4}{3}\,\eta\,e^2$ &
$\left(\frac{4}{3}e^2+g^2\right)\,\eta$ & 0 & $3g^2/(3g^2+4e^2)$\\ \hline
\;\;2SC\;\;  & \multicolumn{3}{c}{0}\vline &
\multicolumn{4}{c}{$\frac{1}{2}g^2$}\vline
 & $\frac{1}{3}g^2$ & 0 &  $\frac{1}{3\sqrt{3}}eg $ &
$ \frac{1}{9}e^2$ & $\frac{1}{3}g^2+\frac{1}{9}e^2$ & 0
 & $ 3g^2/(3g^2+e^2)$\\
\hline
\;\;CSL\;\;  & $\beta g^2$ & $\alpha g^2$ & $\beta g^2$ &$\beta g^2$ &
$\alpha g^2$ & $\beta g^2$ &
$\alpha g^2$ &$\beta g^2$ & 0  &  0  &  $6q^2 e^2$ &
$\beta g^2$ & $6q^2 e^2$& 1 \\ \hline
\end{tabular}
\caption{Zero-temperature Meissner masses $m_M$, rotated Meissner masses
$\tilde{m}_M$, and
gluon/photon mixing angle $\theta$. The number $a$ labels the gluons
($a=1,\ldots,8$) and the photon
($a=9$). All masses are given in units of $N_f\mu^2/(6\pi^2)$,
where $N_f=3,2,1$ in the CFL, 2SC, CSL phases, respectively. We have
abbreviated
$\eta\equiv (21-8\log 2)/54$, $\alpha\equiv(3+4\log 2)/27$,
$\beta\equiv(6-4\log 2)/9$.
For the one-flavor CSL phase we denoted the quark electric charge by $q$.
While the rotated
photon in the CFL and 2SC phases is massless, the photon acquires a
Meissner mass in the CSL phase.
 }
\label{tablemeissner}
\end{table*}

\subsection{Chromomagnetic instability}
\label{subsubsec:chromo}

We have just seen in Sec.~\ref{subsubsec:screening} that color 
superconductors have nonzero Meissner masses for some gluons and/or 
the photon, indicating a color or electromagnetic Meissner effect.   
However, as we discussed previously, in Sec.~\ref{sec:gCFL}, if the CFL 
phase is stressed by a nonzero strange quark mass to the point that 
Cooper pairs break, the resulting gapless CFL (gCFL) phase found in analyses 
that presume a translationally invariant condensate exhibits 
{\it imaginary} Meissner masses \cite{Casalbuoni:2004tb,Fukushima:2005cm}.
This phenomenon was first discovered in the simpler gapless 2SC (g2SC) phase 
\cite{Huang:2004am,Huang:2004bg} and can be understood in either the 
gCFL or g2SC context via a simplified analysis involving two quark 
species only \cite{Alford:2005qw} that we introduced in Sec.~\ref{sec:BCSstress} 
and shall pursue here. The negative Meissner mass squared implies that 
these phases are unstable toward the spontaneous generation of currents, 
that break translation invariance. In this section we shall review the 
calculation of the Meissner mass in a gapless color superconductor.

We have seen in Sec.~\ref{sec:cfl} that the introduction of a nonzero
strange quark mass, combined with the requirement that matter
be neutral and in beta equilibrium, serve to exert a stress on 
the CFL pairing that is controlled by the parameter $m_s^2/(2\mu)$.
This stress seeks to separate the $bu$ and $rs$ Fermi surfaces (and
the $bd$ and $gs$ Fermi surfaces) but in the CFL phase they remain
locked together in order to gain pairing energy $\propto \Delta$ per
Cooper pair.  In the gCFL phase, on the other hand, there are 
unpaired $bu$ and $bd$ quarks in regions of momentum space in which
the corresponding $rs$ and $gs$ states are empty --- Cooper pairs
have been broken yielding gapless excitations.  We can describe the
resulting chromomagnetic instability generically by picking one
of these pairs, calling the quarks $1$ and $2$, and labelling their
effective chemical potentials $\mu_1$ and $\mu_2$. The quasiparticle 
dispersion relations are
\beq \label{gapless}
\epsilon_k \equiv \left|\sqrt{(k-\bar{\mu})^2+\Delta^2}\pm\delta\mu\right| \, ,
\eeq
with the average chemical potential $\bar{\mu}$ and the mismatch in chemical 
potentials $\delta\mu$ as in Eq.~(\ref{deltamu}). (Note that the leading 
effect of the strange mass, $\propto m_s^2/(2\mu)$, is included in the 
effective chemical potential, meaning that we may use the massless 
dispersion relation of Eq.~(\ref{gapless}).) 
For $\mu_1=\mu_2$ this yields identical dispersion relations 
for both degrees of freedom (and the same with a minus sign for the 
corresponding hole degrees of freedom which are omitted here). This is 
the usual situation of BCS superconductivity. For $\mu_1\neq\mu_2$, 
however, one obtains two different quasiparticle excitations. 
A qualitative change appears at $\delta\mu=\Delta$. For $\delta\mu>
\Delta$ the dispersion relations become gapless at the two momenta 
\beq \label{kpm}
k_\pm = \bar{\mu}\pm\sqrt{\delta\mu^2-\Delta^2} \, ,
\eeq
meaning that there are gapless quasiparticles on two spheres in momentum 
space.  In the region of momentum space between these two spheres, the 
states of species 1 are filled while those of species 2 are empty: the 
$1$-$2$ pairing has been ``breached'' \cite{Gubankova:2003uj}.  (We 
have taken $\delta\mu>0$.)  This seems a natural way for the system 
to respond to the stress $\delta\mu$, by reducing the number of $2$ 
particles relative to the number of $1$ particles, albeit at the
expense of lost pairing energy.  In the larger gCFL context, such
a response serves to alleviate the stress introduced by the 
requirements of neutrality and weak equilibrium.

 Gapless superconductivity \cite{Alford:1999xc} refers to the circumstance 
in which two species of fermions are paired in some regions of momentum 
space but in a shell (breach) in momentum space, bounded by two spherical effective
Fermi surfaces, one finds unpaired fermions of just one of the two species.  
The term does 
not refer to situations in which some fermion species pair throughout
momentum space while others do not pair at all, as for example in the 2SC 
phase.
Nor does it apply to anisotropic superconductors
in which the gap parameter vanishes in certain
directions on the Fermi surface, as for example
in some single-flavor color superconductors
or in the curCFL-$K^0$ and crystalline color superconducting
phases.
The g2SC and gCFL phases are gapless superconductors, 
in which the same quarks pair, yielding a nonzero order parameter $\Delta$,
while simultaneously featuring gapless excitations. Such phases turn out 
to suffer from the chromomagnetic instability as we now explain.

 The calculation of the Meissner mass  can be done starting from 
Eq.~(\ref{poltensors}). At zero temperature, in this simple context 
with two fermion species one finds
\beq \label{negmeissner}
m_M^2 = m_0^2\left[1-\frac{\delta\mu\,\Theta(\delta\mu-\Delta)}
                          {\sqrt{\delta\mu^2-\Delta^2}}\right] \, , 
\eeq
where $m_0$ is the Meissner mass obtained upon setting $\delta\mu=0$, 
removing the stress. This expression shows that the Meissner mass becomes 
imaginary if and only if the spectrum is gapless. 

In essence, this is also what happens in the CFL phase 
\cite{Casalbuoni:2004tb,Fukushima:2005cm}. In this case, of course, there 
are nine gauge bosons whose Meissner masses were discussed in Sec.~\ref{subsubsec:screening} 
for the case of pairing in the absence of stress. The Meissner masses 
for pairing with mismatched Fermi surfaces are complicated and have 
to be computed numerically in general. However, the reason for
the negativity of $m_M^2$ is the same as in Eq.~(\ref{negmeissner}): a 
negative term $\propto \delta\mu/\sqrt{\delta\mu^2-\Delta^2}$ appears for 
$\delta\mu > \Delta$. At the onset of gapless modes, $\delta\mu=\Delta$, this
term diverges and thus it dominates the Meissner masses at least for 
$\delta\mu$ close to, but larger than, $\Delta$. This is the story for 
the gluons $A_1$, $A_2$, but it turns out that the Meissner masses 
for the gluons $A_a$, $a=4,5,6,7$, at first remain well-behaved for 
values of $\delta\mu$ larger than $\Delta$ before eventually also becoming 
imaginary for sufficiently large mismatches. The gluons $A_3$, $A_8$ and 
the photon mix with each other. Two of the resulting new gauge bosons 
acquire an imaginary mass, just as the first two gluons. The third 
combination, $A_{\rm \tilde Q}$, remains massless, as expected from 
symmetry arguments. (The mixing between these gauge bosons is a 
function of the mismatch and cannot be described by the mixing 
angle given in Table \ref{tablemeissner}.) Although the details are 
clearly more complicated than in the simple two-species model, the 
conclusion remains that the chromomagnetic instability occurs if 
and only if there are gapless modes. 

 This statement is not always correct, as the analysis of the 
gapless 2SC phase demonstrates 
\cite{Huang:2004am,Huang:2004bg,Kiriyama:2006jp,Kiriyama:2006xw}.
In this phase, the gluons 1,2 and 3 are massless for arbitrary 
mismatches, reflecting the unbroken $SU(2)_c$. One combination of 
the eighth gluon with the photon behaves as in Eq.~(\ref{negmeissner}) 
while the other combination is massless. The Meissner masses for 
the gluons 4-7, however, are imaginary for $\delta\mu>\Delta$ as 
before but they are also imaginary in the parameter region 
$\Delta/\sqrt{2} < \delta\mu < \Delta$. Hence, the 2SC phase 
is unstable in a region where there are no gapless modes. Possible 
consequences of this interesting behavior are discussed in 
\cite{Gorbar:2006up} and have been related to 
gluon condensation 
\cite{Gorbar:2005rx}. 

 We also know of an example where gapless pairing need not be 
accompanied by an instability. This is a two-species system 
where the coupling is allowed to grow so large that the gap 
becomes of the order of $\bar{\mu}$ and even larger. In this 
case, a strong coupling regime has been identified where the 
gapless phase is free of the chromomagnetic instability 
\cite{Kitazawa:2006zp}. See \cite{Gubankova:2006gj} for 
a similar analysis in a non-relativistic system.  The scenario 
in these examples cannot arise in QCD, since before $\bar{\mu}$ 
drops so low that $\Delta \gtrsim \bar{\mu}$, quark matter is
replaced by nuclear matter. 

The chromomagnetic instability of the gCFL phase only demonstrates
that this phase is unstable; it does not determine the nature of
the stable phase.  However, the nature of the instability suggests
that the stable phase should feature currents, which must be
counterpropagating since in the ground state there can be no
net current.  Among the possible resolutions to the instability
that we have enumerated in Sec.~\ref{sec:non_cfl}, two stand out by this
argument. In the meson supercurrent phase of Sec.~\ref{sec_kcur}, that
we shall discuss further in Sec.~\ref{sec_gb_cur}, the phase of the CFL kaon
condensate varies in space, yielding a current \cite{Gerhold:2006dt}. 
Ungapped fermion modes carry a counter-propagating current such that 
the total current vanishes. In the crystalline color superconducting 
phases of Sec.~\ref{subsec:crystal}, that we shall discuss further in Sec.~\ref{sec:NJL}, the 
diquark condensate varies in space in some crystalline pattern 
constructed as the sum of multiple plane waves. 
If the total current carried by the condensate is nonzero, it is
cancelled by a counter-propagating current carried by the
ungapped fermion modes that are also found in the crystalline phases.
It is important to note that in both these phases, the ungapped 
modes have different Fermi surface topologies compared to that 
in the gCFL phase: they are anisotropic in momentum space, with 
unpaired fermions accommodated in one or many ``caps'' rather 
than in spherically symmetric shells. It turns out that in both 
these phases, the Meissner masses are real, meaning that these 
phases do not suffer from a chromomagnetic instability. This was 
shown in the meson supercurrent phase in \cite{Gerhold:2006np} 
and in the crystalline color superconducting phase in 
\cite{Ciminale:2006sm}. We compare the free 
energies of these phases in Fig.~\ref{fig:energy}. These two 
phases have to date been analyzed ``in isolation''.  It remains 
to be seen whether when they are analyzed in a sufficiently general 
framework that currents of either or both types are possible they 
are distinct possibilities or different limits of the same more 
general inhomogeneous phase.

\section{Effective theories of the CFL phase}
\label{sec_eft}

At energies below the gap the response of superconducting quark 
matter is carried by collective excitations of the superfluid 
condensate. The lightest of these excitations are Goldstone bosons
associated with broken global symmetries. Effective theories
for the Goldstone modes have a number of applications. They 
can be used to compute low temperature thermodynamic and transport 
properties, and to study the response to perturbations like nonzero 
quark masses and lepton chemical potentials. Other light degrees
of freedom appear near special points in the phase diagram. Fermion
modes become light near the CFL-gCFL transition, and fluctuations
of the magnitude of the gap become light near $T_c$. 

Effective field theories can be constructed ``top down'', by 
integrating out high energy degrees of freedom, or ``bottom up'',
by writing down the most general effective Lagrangian consistent
with the symmetries of a given phase. In QCD at moderate or low
density the microscopic theory is nonperturbative, and the top 
down approach is not feasible. In this case the parameters of the 
effective Lagrangian can be estimated using dimensional analysis 
or models of QCD. If the density is very large then effective 
theories can be derived using the top down method. However, even 
in this case it is often easier to follow the bottom up approach, 
and determine the coefficients of the effective Lagrangian using 
matching arguments. Matching expresses the condition that low energy 
Green functions in the effective and fundamental theory have to agree. 

 Quark matter at very high density is characterized by several energy 
scales. In the limit of massless quarks the most important scales 
are the chemical potential $\mu$, the screening scale $m_g$, and 
the pairing gap $\Delta$. In the weak coupling limit we have $\mu
\gg m_g\gg \Delta$. This hierarchy of scales can be exploited 
in order to simplify calculations of the properties of low 
energy degrees of freedom in the color superconducting phase. 
For this purpose we introduce an intermediate effective theory, 
the high density effective theory (HDET), which describes quark
and gluon degrees of freedom at energies below $m_g$. This theory
will be described in Secs.~\ref{sec_hdet}-\ref{sec_mhdet}. 
Secs.\ \ref{sec_lg}-\ref{sec_bar} are devoted to effective theories of
the CFL phase that allow us to determine the physics of the its low
energy excitations. As we shall see in Secs.~\ref{sec_trans} and \ref{sec:astro}, these
theories govern the phenomenology of the CFL phase even at densities
not high enough for the weak coupling calculation of the gap parameter
described in Sec.~\ref{sec:QCD} to be reliable. In \ref{sec_mefts} we
briefly mention effective field theories for some other color 
superconducting phases.

\subsection{High density effective theory}
\label{sec_hdet_intro}
 
 The formalism discussed in Sec.~\ref{sec:QCD} can be extended
to include higher orders in the coupling constants and the effects 
of nonzero quark masses. It can also be used to compute more 
complicated observables, like the dispersion relations of collective 
excitations. In practice these calculations are quite difficult, 
because the number of possible gap structures quickly proliferate,
and it is difficult to estimate the relative importance of corrections
due to the truncation of the Dyson-Schwinger equations, kinematic 
approximations, etc., a priori. 

 There are two, related, strategies for addressing these issues: 
effective field theories and the renormalization group. Within the 
effective field theory approach we try to derive an effective 
Lagrangian for quasi-quarks and gluons near the Fermi surface, 
together with a power counting scheme that can be used to determine 
the magnitude of diagrams constructed from the propagators and 
interaction terms of the theory. This is the strategy that we 
will describe in Secs.~\ref{sec_hdet}-\ref{sec_mhdet} below. 

 In the renormalization group approach we consider a general 
effective action for quarks and gluons at high baryon density, 
and study the evolution of the action as high energy degrees of 
freedom, energetic gluons and quarks far away from the Fermi 
surface, are integrated out \cite{Shankar:1993pf,Polchinski:1992ed}. 
This approach was applied to QCD with short range interactions 
in \cite{Evans:1998nf,Evans:1998ek,Schafer:1998na}. In this case
one can show that for typical initial conditions the color antisymmetric, 
flavor antisymmetric, $J=0$, BCS interaction does indeed grow faster 
than all other terms, confirming in another way the arguments of 
Secs.~\ref{sec:intro} and \ref{sec:cfl} that these are the channels in which the dominant
diquark condensation occurs. In order to use the renormalization
group approach more quantitatively, one has to deal with the unscreened 
long range gluon exchanges, which is more difficult. Son 
studied the evolution of the BCS interaction using the hard
dense loop gluon propagator as an input \cite{Son:1998uk}. The 
coupled evolution of static and dynamic screening and the BCS
interaction has not been solved yet. A general scheme constructing
effective actions by integrating out hard modes  
was proposed in \cite{Reuter:2004kk}.

\subsubsection{Effective Lagrangian}
\label{sec_hdet}
 
 Consider the equation of motion of a free quark with a chemical 
potential $\mu$
\beq
\label{dirac_mu}
\left(\bm{\alpha}\cdot{\bf p} - \mu\right) \psi_\pm 
  = E_\pm \psi_\pm\ ,
\eeq
where $\psi_\pm$ are eigenvectors of $(\bm{\alpha}\cdot\hat{\bf p})$
with eigenvalue $\pm 1$ and $E_\pm=-\mu\pm p$. If the quark momentum 
is near the Fermi momentum, $p \sim p_F=\mu$, then the solution $\psi_+$ 
describes a low energy excitation $E_+\sim 0$, whereas $E_-\sim -2\mu$ 
corresponds to a high energy excitation. In order to construct an 
effective field theory based on this observation we define low and 
high-energy components of the quark field \cite{Hong:1998tn,Hong:1999ru} 
\beq
\label{psi_pm}
\psi_{\pm} = e^{i p_F v_\mu x^\mu} 
   \left( \frac{1\pm \bm{\alpha}\cdot \hat{\bf v}_F}{2}\right) \psi\ ,
\eeq
where ${\bf v}_F$ is the Fermi velocity and $v_\mu = (1,{\bf v}_F)$.
The prefactor removes the rapid phase variation common to all 
fermions in some patch on the Fermi surface specified 
by $\hat{\bf v}_F$.
We can insert the decomposition Eq.~(\ref{psi_pm}) into the 
QCD Lagrangian and integrate out the $\psi_-$ field as well as 
hard gluon exchanges. This generates an expansion of 
the QCD Lagrangian in powers
of $1/p_F$. At tree level, integrating out the $\psi_-$ 
fields is equivalent to solving their equation of motion
\beq
\label{psi-eom}
\psi_{-,L} = \frac{1}{2p_F}\left(i \bm{\alpha}_\perp \cdot {\bf D}\,
    \psi_{+,L} + \gamma^0 M \psi_{+,R}\right)\ ,
\eeq
where $\bm{\gamma}_\parallel\equiv \hat{\bf v}_F(\hat{\bf v}_F\cdot\bm{\gamma})$, 
$\bm{\gamma}_\perp=\bm{\gamma}-\bm{\gamma}_\parallel$ and $M$ is 
the quark mass matrix. At ${\cal O}(1/p_F)$ the effective Lagrangian for
$\psi_+$ is 
\bea
\label{l_hdet}
{\cal L} &=& 
 \psi_{+,L}^\dagger (iv^\mu D_\mu) \psi_{+,L}
  - \frac{1}{2p_F} \psi_{+,L}^\dagger \left[  (D{\!\!\!\!/}_\perp)^2 
  + MM^\dagger \right]  \psi_{+,L}  \nonumber \\
 & & \mbox{} 
  + \left( L\leftrightarrow R, M\leftrightarrow M^\dagger\right)  
  + \ldots .
\eea
The low energy expansion was studied in more detail in \cite{Schafer:2003jn}. 
There are a number of physical effects that have to be included in order 
to obtain a well-defined expansion. First, four-fermion operators have to 
be included. These operators naturally appear at ${\cal O}(1/p_F^2)$ but their
effects are enhanced by the large density of states $N\sim p_F^2$ on the 
Fermi surface. The most important of these operators is the BCS interaction 
$[\psi({\bf v})\psi(-{\bf v})][\psi^\dagger({\bf v}')\psi^\dagger(-{\bf v'})]$.
The coefficient of the BCS operator was determined in \cite{Schafer:2003jn}.

 Because of the large density of states it is also necessary
to resum quark loop insertions in gluon n-point functions. There 
is a simple generating functional for these effects, known
as the hard dense loop (HDL) effective action \cite{Braaten:1991gm}
\beq
\label{S_hdl}
{\cal L}_{\rm HDL} = -\frac{m^2}{2}\sum_v \,G^a_{\mu\alpha}
  \frac{v^\alpha v^\beta}{(v^\lambda D_\lambda)^2} G^a_{\mu\beta}.
\eeq
This is a gauge invariant, but non-local, effective Lagrangian.
Expanding ${\cal L}_{\rm HDL}$ in powers of the gauge field 
produces $2,3,\ldots$ gluon vertices. The quadratic term
describes dielectric screening of electric modes and Landau 
damping of magnetic modes. Higher order terms contain 
corrections to the gluon self interaction in a dense medium.

\subsubsection{Non-Fermi liquid effects and the gap equation}
\label{sec_nfl}

 In this section we shall analyze the low energy expansion in the 
regime $\Delta<k_0<m_g$ \cite{Schafer:2005mc}. This energy range 
gives the dominant contribution to the pairing gap and other low
energy constants in the superconducting phase. Since electric fields 
are screened the interaction is dominated by the exchange of magnetic 
gluons. The transverse gauge boson propagator is
\beq
D_{ij}(K) = -\frac{i(\delta_{ij}-\hat{k}_i\hat{k}_j)}
      {k_0^2-k^2+iM_g^2 \frac{k_0}{k}} ,
\eeq
where $M_g^2=(3\pi/4) m_g^2$ and we have assumed that $|k_0|<k$. 
We observe that the propagator becomes large in the regime $|k_0|\sim 
k^3/m_g^2$. If the energy is small, $|k_0|\ll m_g$, then the 
typical energy is much smaller than the typical momentum,
\beq
\label{ld_kin}
 k \sim (m_g^2 |k_0|)^{1/3} \gg |k_0| .
\eeq
This implies that the gluon is very far off its energy shell and not 
a propagating state. We can compute loop diagrams containing quarks 
and transverse gluons by picking up the pole in the quark propagator, 
and then integrate over the cut in the gluon propagator using the 
kinematics dictated by Eq.~(\ref{ld_kin}). In order for a quark to 
absorb the large momentum carried by a gluon and stay close to the 
Fermi surface the gluon momentum has to be transverse to the momentum 
of the quark. This means that the term $k_\perp^2/(2\mu)$ in the quark 
propagator is relevant and has to be kept at leading order. 
Eq.~(\ref{ld_kin}) shows that $k_\perp^2/(2\mu)\gg k_0$ as $k_0\to 0$. 
This 
means that the pole of the quark propagator is governed by the condition 
$k_{||}\sim k_\perp^2/(2\mu)$. We find
\beq
\label{ld_reg}
 k_\perp \sim g^{2/3}\mu^{2/3}k_0^{1/3},\hspace{0.5cm}
 k_{||}  \sim g^{4/3}\mu^{1/3}k_0^{2/3}.
\eeq
In this regime propagators and vertices can be simplified even further. 
The quark and gluon propagators are
\beq
   S_{\alpha\beta}(K) = \frac{i\delta_{\alpha\beta}}
       {k_0-k_{||}-\frac{k_\perp^2}{2\mu}
              +i\epsilon {\rm sgn}(k_0)},
\eeq
\beq
   D_{ij}(K) = \frac{i\delta_{ij}}
       {k_\perp^2-iM_g^2\frac{k_0}{k_\perp}},
\eeq
and the quark gluon vertex is $gv_i(\lambda^a/2)$. Higher order 
corrections can be found by expanding the quark and gluon propagators 
as well as the HDL vertices in powers of the small parameter $\epsilon
\equiv (k_0/m)$.

 The regime characterized by Eq.~(\ref{ld_reg}) is completely perturbative, 
i.e.~graphs with extra loops are always suppressed by extra powers of 
$\epsilon^{1/3}$ \cite{Schafer:2005mc}. The power of $\epsilon$ can be 
found by using the fact that loop integrals scale as $(k_0k_{||}k_\perp^2)
\sim k_0^{7/3}$, fermion propagators scale as $1/k_{||}\sim 1/k_0^{2/3}$, 
gluon propagators scale as $1/k^2_\perp \sim 1/k_0^{2/3}$, and the 
quark-gluon vertex scales as a constant. Quark matter in the regime 
$\Delta<k_0<m$ is a non-Fermi liquid. The excitations are quasi-particles 
with the quantum numbers of quarks, but Green functions scale with 
fractional powers and logarithms of the energy and the coupling constant 
\cite{Ipp:2003cj,Schafer:2004zf,Gerhold:2005uu}. 

 The corrections to Fermi liquid theory do not upset the logic that 
underlies the argument that leads to the BCS instability. For quark 
pairs with back-to-back momenta the basic one gluon exchange
interaction has to be summed to all orders, but all other interactions
remain perturbative \cite{Schafer:2005mc}. The gap equation that sums 
the leading order transverse gluon exchange in the color-anti-symmetric 
channel is 
\bea
\label{eft_gap}
\Delta(p_0) &=& -i\frac{2g^2}{3} \int \frac{dk_0}{2\pi} 
     \int \frac{dk_\perp^2}{(2\pi)^2} \ 
          \frac{k_\perp}{k_\perp^3+iM_g^2 (k_0-p_0)}  \nonumber \\
 & & \hspace{0.5cm}\mbox{}\times
  \int \frac{dk_{||}}{2\pi} \  \frac{\Delta(k_0)}
        {k_0^2+k_{||}^2+\Delta(k_0)^2}.
\eea
This equation is exactly equivalent to Eq.~(\ref{eliash}). In
particular, all the kinematic approximations that were used to
derive Eq.~(\ref{eliash}), like the low energy approximation
to the HDL self energies and the forward approximation to the
Dirac traces, are built into the effective field theory vertices
and propagators. The effective theory can now be used to study
corrections to the leading order result. Higher order corrections
to the propagators and vertices of the effective theory modify
the kernel of the integral equation Eq.~(\ref{eft_gap}). The 
resulting correction to the gap function can be computed 
perturbatively, without having to solve the integral equation
again, using a method that is similar to Rayleigh-Schr\"odinger
perturbation theory \cite{Brown:2000eh,Schafer:2003jn}.

 The coefficients $b_0$ and $\bar{b}_0$ introduced in 
Sec.~\ref{sec_eliash} can be determined by matching the four-fermion 
operators in the effective theory \cite{Schafer:2003jn}. The $b_0$ 
term also receives contributions from the fermion wave function 
renormalization $Z\sim \log(k_0)$. All other terms give corrections 
beyond ${\cal O}(g^0)$ in $\log(\Delta_0/\mu)$. Vertex corrections scale
as $\Gamma\sim p_0^{1/3}$ and are suppressed compared to the
fermion wave function renormalization. The analogous statement
in the case of phonon-induced electronic superconductors is
known as Migdal's theorem. Gluon self energy insertions beyond
the $k_0/k_\perp$ term included in the leading order propagator
are also suppressed by fractional powers of the coupling and
the gluon energy.

\subsubsection{Mass terms}
\label{sec_mhdet}

\begin{figure}[t]
\bc\includegraphics[width=7cm]{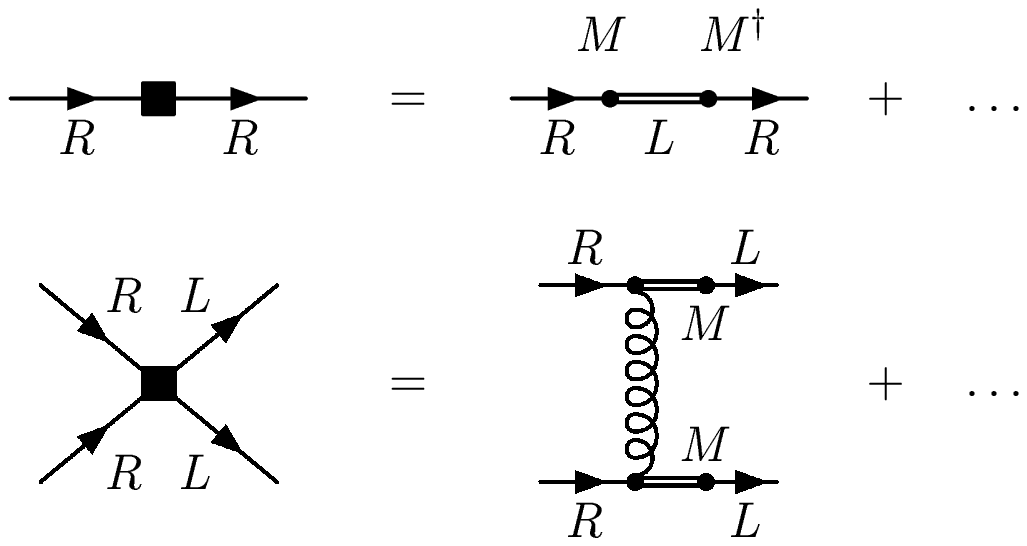}\ec
\caption{\label{fig_hdet_m}
Mass terms in the high density effective theory. The first 
diagram shows a ${\cal O}(MM^\dagger)$ term that arises from integrating 
out the $\psi_-$ field in the QCD Lagrangian. The second
diagram shows a ${\cal O}(M^2)$ four-fermion operator which arises from 
integrating out $\psi_-$ and hard gluon exchanges.}
\end{figure}

 A systematic determination of mass corrections to the high 
density effective theory is needed for calculations of the Goldstone 
boson masses in the CFL phase, and in order to understand the response
of the CFL ground state to nonzero quark masses. Mass terms affect
both the quark propagator and the quark-quark interaction. From
Eq.~(\ref{psi-eom}) and (\ref{l_hdet}) we see that integrating 
out the $\psi_-$ field gives a correction to the energy of the 
$\psi_+$ field of the form $MM^\dagger/(2p_F)$. This term can be
viewed as an effective flavor dependent chemical potential. We 
also note that this term is just the first in a tower of operators
that arise from expanding out the energy of a free massive quark, 
$E=(p^2+m^2)^{1/2}$, for momenta near the Fermi surface.  
Higher order terms correspond to additional corrections to the 
chemical potential, the Fermi velocity, and to non-linear terms
in the dispersion relation. 

 There are no mass corrections to the quark-gluon vertex at 
${\cal O}(1/p_F^2)$. There are, however, mass corrections to the 
quark-quark interaction. In connection with color superconductivity
we are mainly interested in the BCS interaction. The diagram 
shown in Fig.~\ref{fig_hdet_m} gives \cite{Schafer:2001za}
\bea
\label{hdet_m}
 {\mathcal L} &=& \frac{g^2}{32p_F^4}
     \left(\psi^{a\,\dagger}_{i,L} C\psi^{b\,\dagger}_{j,L}\right)
     \left(\psi^c_{k,R} C \psi^d_{l,R}\right)  \\
 & &  \mbox{}\times
    \left[(\bm{\lambda})^{ac}(\bm{\lambda})^{bd}(M)_{ik}(M)_{jl} \right]
  \; + \; \left(L\leftrightarrow R, M\leftrightarrow M^\dagger\right)\, .
  \nonumber
\eea
This is the leading interaction that couples the gap equations for
left and right handed fermions. We shall also see that the mass
correction to the BCS interaction gives the leading contribution to 
the mass shift in the condensation energy, and the masses of the 
Goldstone bosons.

\subsection{Ginzburg-Landau theory}
\label{sec_lg}

 At zero temperature fluctuations of the superconducting state are 
dominated by fluctuations of the phase of the order parameter. Near
the critical temperature the gap becomes small and fluctuations 
of the magnitude of the gap are important, too. This regime can be 
described using the Ginzburg-Landau theory. Ginzburg and Landau
argued that in the vicinity of a second order phase transition the
thermodynamic potential of the system can be expanded in powers 
of the order parameter and its derivatives. This method was used
very successfully in the study of superfluid phases of $^3$He. 

 The Ginzburg-Landau approach was first applied to color superconductivity
in \cite{Bailin:1979nh}. The problem was revisited by \cite{Iida:2000ha},
who included the effects of unscreened gluon exchanges and charge 
neutrality. Consider the $s$-wave color anti-triplet condensate in 
QCD with three massless flavors. The order parameter 
can be written as
\beq \label{nofluct}
\langle\psi^\alpha_i C\gamma_5\psi^\beta_j\rangle
= \epsilon^{\alpha\beta A}\epsilon_{ij B} \phi_A^B
\eeq
where $\phi_A^B$ is a matrix in color-flavor space. Note that here we have included the 
energy gap into $\phi_A^B$, in contrast to Eq.\ (\ref{calM}), where $\phi_A^B$ is dimensionless.
We have fixed the orientations of left and right-handed 
condensates. Fluctuations in the relative color-flavor orientation
of the left- and right-handed fermions correspond to the Goldstone 
modes related to chiral symmetry breaking, and will be considered 
in Sec.~\ref{sec_gbs}. Therefore, the ansatz (\ref{nofluct}) implies the assumption that chiral fluctuations 
near $T_c$ are small compared to non-chiral gap fluctuations and fluctuations of the 
gauge field. The thermodynamic potential can be expanded 
as
\bea
 \Omega &=& \Omega_0 + \alpha \,{\rm Tr}\left(\phi^\dagger\phi\right)
  + \beta_1 \left[ {\rm Tr}\left(\phi^\dagger\phi\right) \right]^2 \\
 & & \mbox{} 
  + \beta_2\,  {\rm Tr}\left(\left[\phi^\dagger\phi\right]^2\right)
  + \kappa\, {\rm Tr}\left(\nabla \phi\nabla\phi^\dagger\right) 
  + \ldots  \nonumber
\eea
The coefficients $\alpha,\beta_i,\kappa$ can be treated as 
unknown parameters, or determined in QCD at weak coupling. The 
weak coupling QCD result is \cite{Iida:2000ha}
\bea 
\label{LG_c}
\alpha &=& 4N \frac{T-T_c}{T}, 
\hspace{0.25cm}
\beta_1=\beta_2 = \frac{7\zeta(3)}{8(\pi T_c)^2} \, N ,  \\
 \kappa &=& \frac{7\zeta(3)}{8\pi^2T_c^2}\, N
\eea
where $N=\mu^2/(2\pi^2)$ is the density of states on the Fermi surface. 
This result agrees with the BCS result. Using Eq.~(\ref{LG_c}) we
can verify that the ground state is in the CFL phase $\phi_A^B\sim
\delta^B_A$. We can also study many other issues, like the gluon 
screening lengths, the structure of vortices, the effects of 
electric and color neutrality, and the effects of nonzero quark 
masses \cite{Iida:2002ev,Iida:2003cc,Iida:2004cj}.

 From the study of electronic superconductors, it is known that 
the nature of the finite temperature phase transition depends
on the ratio $\kappa=\lambda/\xi$ of the screening length $\lambda$
and the correlation length $\xi$. If $\kappa>1/\sqrt{2}$ the 
superconductor is type II, fluctuations of the order parameter 
are more important than fluctuations of the gauge field, and 
the transition is second order. In a type I superconductor
the situation is reversed, and fluctuations of the gauge field
drive the transition first order \cite{Halperin:1973jh}. 

 In the weak coupling limit, $\xi\sim1/\Delta \gg \lambda\sim 1/(g\mu)$ 
and color superconductivity is strongly type I. The role of gauge 
field fluctuations was studied in 
\cite{Bailin:1983bm,Giannakis:2003am,Matsuura:2003md,Noronha:2006cz,Giannakis:2004xt}. 
The 
contribution to the thermodynamic potential is 
\beq 
\Omega_{fl} = 8T\int \frac{d^3k}{(2\pi)^3} 
 \left\{ \log\left( 1+\frac{m_A^2(k)}{k^2} \right) 
    - \frac{m_A^2(k)}{k^2} \right\},
\eeq
where $m_A(k)$ is the gauge field screening mass. In QCD the 
momentum dependence of $m_A$ cannot be neglected. The 
contribution of the fluctuations $\Omega_{fl}$ induces a cubic term $\propto \phi^3$ 
in the thermodynamic potential which drives the transition first order.
The first order transition
occurs at a critical temperature $T_c^*$ \cite{Giannakis:2004xt}
\beq\label{elevatedTc}
 \frac{T_c^*-T_c}{T_c} = \frac{\pi^2}{12\sqrt{2}}\, g ,
\eeq
where $T_c$ is the critical temperature of the second order
transition obtained upon neglecting the cubic term.
Although the result (\ref{elevatedTc})
cannot be trusted quantitatively at accessible densities, say
$\mu\sim 400$~MeV where $g\sim 3.6$, it does make it clear that the
phase transition between the CFL (or 2SC) 
phase and the quark-gluon plasma will be strongly first order.
Noronha et al.~\cite{Noronha:2006cz} argue that
Eq.~(\ref{elevatedTc}) gives the complete $O(g)$ correction to
the critical temperature (see, however, \cite{Matsuura:2003md}).
This implies that the
transition to the color superconducting phase will occur at a
critical temperature that is significantly elevated relative
to the BCS estimate $T_c=0.57\Delta$ that we obtained 
in Sec.~\ref{sec_qcdgaps}.
The effects of gluon fluctuations are 
much more important here than those of photon fluctuations in
a conventional type I superconductor.

\subsection{Goldstone bosons in the CFL phase}
\label{sec_gbs}
\subsubsection{Effective Lagrangian}

 In the CFL phase the pattern of chiral symmetry breaking is identical 
to the one at $T=\mu=0$. This implies that the effective Lagrangian has 
the same structure as chiral perturbation theory. The main difference is 
that Lorentz-invariance is broken and only rotational invariance is a 
good symmetry. The effective Lagrangian for the Goldstone modes is given 
by \cite{Casalbuoni:1999wu}
\bea
\label{l_cheft}
{\mathcal L}_{\rm eff} &=& \frac{f_\pi^2}{4} {\rm Tr}\left[
 \partial_0\Sigma\partial_0\Sigma^\dagger - v_\pi^2
 \partial_i\Sigma\partial_i\Sigma^\dagger \right]   \\
 & & \hspace*{0cm}\mbox{} 
 +\left[ B {\rm Tr}(M\Sigma^\dagger) + h.c. \right] 
    \nonumber \\ 
 & & \hspace*{0cm}\mbox{} 
     +\left[ A_1{\rm Tr}(M\Sigma^\dagger)
                        {\rm Tr} (M\Sigma^\dagger) 
     + A_2{\rm Tr}(M\Sigma^\dagger M\Sigma^\dagger) \right.
 \nonumber \\[0.1cm] 
  & &   \hspace*{0.5cm}\mbox{}\left. 
     + A_3{\rm Tr}(M\Sigma^\dagger){\rm Tr} (M^\dagger\Sigma)
         + h.c. \right]+\ldots . \nonumber 
\eea
Here $\Sigma=\exp(i\phi^a\lambda^a/f_\pi)$ is the chiral field, $f_\pi$ 
is the pion decay constant and $M$ is a complex mass matrix. The fields
$\phi^a$ describe the octet of Goldstone bosons $(\pi^\pm,\pi^0,K^\pm, 
K^0,\bar{K}^0,\eta)$.  These Goldstone bosons are an octet under the 
unbroken $SU(3)_{c+L+R}$ symmetry of the CFL phase and their 
$\tilde Q$-charges under the unbroken gauge symmetry of the CFL phase 
are $\pm 1$ and $0$ as indicated by the superscripts, meaning that they 
have the same $\tilde Q$-charges
as the $Q$-charges of the vacuum pseudoscalar mesons. 
The chiral field and the 
mass matrix transform as $\Sigma\to L\Sigma R^\dagger$ and  $M\to 
LMR^\dagger$ under chiral transformations $(L,R)\in SU(3)_L\times 
SU(3)_R$. For the present, we have suppressed the singlet fields 
associated with the breaking of the exact $U(1)_B$ and approximate 
$U(1)_A$ symmetries.  We will give the effective Lagrangian for the 
massless Goldstone boson associated with superfluidity (i.e. from 
$U(1)_B$ breaking) below.

 The form of the effective Lagrangian follows from the symmetries of the 
CFL phase. It is nevertheless useful to understand how this Lagrangian 
arises upon integrating out high energy degrees of freedom. We start 
from the high density effective Lagrangian in the presence of a CFL 
gap term
\bea
\label{l_cfl1}
 {\cal L} &=& {\rm Tr}\left[\psi^\dagger_L (iv^\mu D_\mu) \psi_L \right] 
      \\
 & &  \mbox{} + \frac{\Delta}{2} \left\{ {\rm Tr} \left( X^\dagger \psi_L
                       X^\dagger \psi_L \right) 
  - \left[ {\rm Tr}\left( X^\dagger\psi_L\right)\right]^2 
    + h.c. \right\} \nonumber \\[0.2cm]
 & &  \mbox{} + \;\; \left(L\leftrightarrow R,
                           X\leftrightarrow Y\right).\nonumber 
\eea
Here, $\psi_{L,R}$ are left and right-handed quark fields which transform 
as $\psi_L\to L\psi_L U^T$ and $\psi_R\to R\psi_R U^T$ under chiral 
transformations $(L,R)\in SU(3)_L\times SU(3)_R$ and color transformations 
$U\in SU(3)_c$. We have suppressed the spinor indices and defined $\psi\psi
=\psi^\alpha C^{\alpha\beta}\psi^\beta$, where $C$ is the charge conjugation 
matrix. The traces run over color or flavor indices and $X,Y$ are fields 
that transform as $X\to L X U^T$ and $Y\to R Y U^T$. We will assume that 
the vacuum expectation value is $\langle X \rangle =\langle Y \rangle = 1$. 
This corresponds to the CFL gap term 
$\Delta\psi^\alpha_i\psi^\beta_j \,\epsilon_{\alpha\beta A}\epsilon^{ijA}$.
$X,Y$ parametrize fluctuations around 
the CFL ground state. Note that fluctuations of the type $X=Y$ correspond
to the field $\phi$ introduced in the previous section. 

 For simplicity we have assumed that the gap term is completely 
anti-symmetric in flavor. We will derive the effective Lagrangian 
in the chiral limit $M=M^\dagger=0$ and study mass terms later.
We can redefine the fermion fields according to 
\beq
\chi_L \equiv \psi_L X^\dagger , \hspace{1cm}
\chi_R \equiv \psi_R Y^\dagger.
\eeq
In terms of the new fields the Lagrangian takes the form 
\bea
{\cal L} &=& {\rm Tr}\left[ \chi_L^\dagger (iv^\mu\partial_\mu ) 
   \chi_L\right] \\
 & & \mbox{}
  -i\,  {\rm Tr}\left[ \chi_L^\dagger \chi_L
   X v^\mu \left( \partial_\mu -iA^T_\mu \right)X^\dagger \right]
   \nonumber \\ 
 & &  \mbox{} + \frac{\Delta}{2}  
     \left\{{\rm Tr}  \left(\chi_L\chi_L \right) 
   - \left[ {\rm Tr}\left(\chi_L\right)\right]^2 \right\} \nonumber\\
   && +\, \left(L\leftrightarrow R,X\leftrightarrow Y\right) .\nonumber
\eea
At energies below the gap we can integrate out the fermions. The
fermion determinant generates a kinetic term for the chiral fields 
$X$ and $Y$ \cite{Casalbuoni:1999wu}
\beq
\label{l_higgs}
{\cal L}=-\frac{f_\pi^2}{2}
  {\rm Tr}\left[ (X^\dagger D_0 X)^2-v_\pi^2(X^\dagger D_i X)^2\right]
+  (X\leftrightarrow Y)
\eeq
For simplicity we have ignored the flavor singlet components of $X$ 
and $Y$. 

The low energy constants $f_\pi$ and $v_\pi$ were calculated 
by matching the effective theory to weak coupling QCD calculations in 
\cite{Son:1999cm,Son:2000tu}, see also 
\cite{Bedaque:2001je,Beane:2000ms,Zarembo:2000pj,Rho:1999xf}.
The results are
\beq
\label{f_pi}
f_\pi^2=\frac{21-8\log 2}{18}
  \frac{\mu^2}{2\pi^2}, \hspace{1cm}
v_\pi^2= \frac{1}{3}.
\eeq
The simplest way to derive these results, given the results that we have 
already reviewed in Sec.~\ref{subsubsec:screening}, is to recall that 
the gluon field acquires a magnetic mass due to the Higgs mechanism
and an electric mass due to Debye screening, and then to notice that
Eq.~(\ref{l_higgs}) shows that the electric mass is $m_D^2=g^2f_\pi^2$,
while the magnetic mass is $m_M^2=v_\pi^2 m_D^2$.  This means that
$f_\pi$ and $v_\pi$ are determined by the Debye and Meissner masses
for the gluons in the CFL phase that we have presented in 
Sec.~\ref{subsubsec:screening}, see Eq.\ (\ref{debyeCFL}) and 
Table \ref{tablemeissner}.

Since the gluon is heavy, it can also be integrated out. Using 
Eq.~(\ref{l_higgs}) we get
\beq
 A_\mu^T = \frac{i}{2}\left( X^\dagger \partial_\mu X 
  + Y^\dagger \partial_\mu Y \right) + \ldots
\eeq
This result can be substituted back into the effective Lagrangian.
The result is 
\beq
{\mathcal L}_{\rm eff} = \frac{f_\pi^2}{4} {\rm Tr}\left[
  \partial_0\Sigma\partial_0\Sigma^\dagger - v_\pi^2
  \partial_i\Sigma\partial_i\Sigma^\dagger \right] \ ,
\eeq
where the Goldstone boson field is given by $\Sigma=XY^\dagger$. 
This shows that the light degrees of freedom correspond to fluctuations
of the color-flavor orientation of the left-handed CFL condensate 
relative to the right-handed one, as expected since these are the
fluctuations associated with the spontaneously broken global symmetry.

\subsubsection{$U(1)_B$ modes and superfluid hydrodynamics}
\label{sec_eft_u1}

 Finally, we quickly summarize the effective theory for the $U(1)_B$
Goldstone mode. At order ${\cal O}((\partial\varphi)^2)$ we 
get \cite{Son:1999cm,Son:2000tu}
\beq 
\label{u1_eft}
{\cal L} = \frac{f^2}{2}\left[ (\partial_0\varphi)^2 - v^2
  (\nabla\varphi)^2 \right] + \ldots ,
\eeq
where the low energy constants $f$ and $v$ are given by 
\beq \label{fv}
 f^2 = \frac{6\mu^2}{\pi^2}, \hspace{1cm} v^2=\frac{1}{3} .
\eeq
The field $\varphi$ transforms as $\varphi\to\varphi+\alpha$ under 
$U(1)_B$ transformation of the quark fields $\psi\to \exp(i\alpha)\psi$. 
Because $U(1)_B$ is an Abelian symmetry, the two-derivative terms do not 
contain any Goldstone boson self interactions. These terms are needed 
in order to compute transport properties of the CFL phase. Son noticed 
that self-interactions are constrained by Lorentz invariance (of 
the microscopic theory) and $U(1)_B$ invariance \cite{Son:2002zn}. 
The analogous argument for non-relativistic superfluids is described 
in \cite{Greiter:1989qb}. To leading order in $g$ the effective 
theory of the $U(1)_B$ Goldstone boson can be written as
\beq 
\label{u1_eft_2}
{\cal L} = \frac{3}{4\pi^2}\left[ (\partial_0\varphi-\mu)^2 -
  (\nabla\varphi)^2 \right]^2 + \ldots ,
\eeq
where the omitted terms are of the form $\partial^i\varphi^k$ 
with $i>k$. Expanding Eq.~(\ref{u1_eft_2}) to second order
in derivatives reproduces Eq.~(\ref{u1_eft}). In addition to 
that, Eq.~(\ref{u1_eft_2}) contains the leading three and four 
boson interactions. Using microscopic models one can obtain more 
detailed information on the properties of collective modes. A
calculation of the spectral properties of the $\varphi$
mode in an NJL model at $T=0$ and $T\neq 0$ can be found 
in \cite{Fukushima:2005gt}.

The spontaneous breaking of $U(1)_B$ is related to 
superfluidity, and the $U(1)_B$ effective theory can be 
interpreted as superfluid hydrodynamics \cite{Son:2002zn}. We 
can define the fluid velocity as 
\beq 
\label{def_u}
 u_\alpha = -\frac{1}{\mu_0} \, D_\alpha\varphi, 
\eeq
where $D_\alpha\varphi\equiv \partial_\alpha\varphi+(\mu,{\bf 0})$
and $\mu_0\equiv (D_\alpha\varphi D^\alpha\varphi)^{1/2}$. Note 
that this definition ensures that the flow is irrotational, 
${\bf\nabla}\times{\bf u}=0$. The identification (\ref{def_u}) is
motivated by the fact that the equation of motion for the $U(1)$
field $\varphi$ can be written as a continuity equation
\beq
 \partial^\alpha  (n_0 u_\alpha) = 0,
\eeq
where $n_0 =3\mu_0^3/\pi^2$ is the superfluid number density. At 
$T=0$ the superfluid density is equal to the total density of 
the system, $n=dP/d\mu|_{\mu=\mu_0}$. The energy-momentum tensor
has the ideal fluid form 
\beq
 T_{\alpha\beta} = (\epsilon+P)u_\alpha u_\beta -Pg_{\alpha\beta},
\eeq
and the conservation law $\partial^\alpha T_{\alpha\beta}=0$
corresponds to the relativistic Euler equation of ideal 
fluid dynamics. We conclude that the effective theory for the 
$U(1)_B$ Goldstone mode accounts for the defining characteristics 
of a superfluid: irrotational, non-dissipative hydrodynamic
flow.

\subsubsection{Mass terms}

\begin{figure}[t]
\bc\includegraphics[width=7cm]{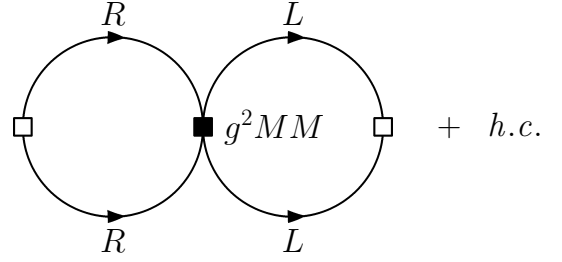}\ec
\caption{\label{fig_4fvac}
Contribution of the ${\cal O}(M^2)$ BCS four-fermion operator to the 
condensation energy in the CFL phase. The open squares correspond
to insertions of the anomalous self energy $\Delta$.}
\end{figure}

The structure of the mass terms in Eq.~(\ref{l_cheft}) is 
completely determined by chiral symmetry. The coefficients 
$B,A_i$ can be determined by repeating the steps discussed
in the previous section, but keeping the mass terms in the 
high density effective theory. In practice it is somewhat
easier to compute the coefficients of the chiral Lagrangian
using matching arguments. For example, we noticed that the easiest
way to determine $f_\pi$ is to compute the gluon screening mass
in the microscopic theory. 

In Sec.~\ref{sec_mhdet} we showed that $X_L\equiv MM^\dagger/(2p_F)$ 
and $X_R\equiv M^\dagger M/(2p_F)$ act as effective chemical potentials 
for left and right-handed fermions, respectively. Formally, the 
effective Lagrangian has an $SU(3)_L\times SU(3)_R$ gauge 
symmetry under which $X_{L,R}$ transform as the temporal components
of non-Abelian gauge fields. We can implement this approximate gauge 
symmetry in the CFL chiral theory by promoting time derivatives
to covariant derivatives \cite{Bedaque:2001je}, 
\beq
\label{mueff}
\partial_0 \Sigma\rightarrow \nabla_0\Sigma \equiv \partial_0 \Sigma 
 + i \left(\frac{M M^\dagger}{2p_F}\right)\Sigma
 - i \Sigma\left(\frac{ M^\dagger M}{2p_F}\right) .
\eeq
The mass dependent terms in the quark-quark interaction contribute
to the gap and to the condensation energy. In the chiral theory 
the shift in the condensation energy due to the quark masses is 
\bea
\label{dE_m}
 {\cal E} &=& - B {\rm Tr}(M) 
     - A_1\left[{\rm Tr}(M)\right]^2
     - A_2{\rm Tr}\left(M^2\right) \\
 & & \mbox{}
     - A_3{\rm Tr}(M){\rm Tr} (M^\dagger)
         + h.c. + \ldots . \nonumber 
\eea
The contribution to the condensation energy from the mass correction
to the BCS interaction is shown in Fig.~\ref{fig_4fvac}. The diagram 
is proportional to the square of the condensate 
\beq
\label{qq_cond}
\langle \psi^\alpha_{i,L} C\psi^\beta_{j,L}\rangle
 = \epsilon^{\alpha\beta A}\epsilon_{ijA}
     \Delta\frac{3\sqrt{2}\pi}{g}
     \left(\frac{\mu^2}{2\pi^2}\right) ,
\eeq
with the dependence on the mass matrix $M$ arising from the contraction 
of the BCS interaction with the CFL condensate. We get
\bea 
\epsilon^{\alpha\beta A}\epsilon_{ijA}
 \left(T^a\right)^{\alpha\gamma} \left(T^a\right)^{\beta\delta}  
     (M)_{ik}(M)_{jl}
\epsilon^{\gamma\delta B}\epsilon_{klB}
    \nonumber \\
  =-\frac{4}{3}\left\{  \Big( {\rm Tr}[M]\Big)^2 -{\rm Tr}\Big[ M^2\Big]
   \right\}\, , \hspace{1cm}
\eea
where $T^a=\lambda^a/2$. We note that the four-fermion operator 
is proportional to $g^2$ and the explicit dependence of the diagram 
on $g$ cancels. We find \cite{Son:1999cm,Son:2000tu,Schafer:2001za}
\beq
\label{E_MM}
{\cal E} = -\frac{3\Delta^2}{4\pi^2}
 \left\{  \Big( {\rm Tr}[M]\Big)^2 -{\rm Tr}\Big[ M^2\Big]
   \right\}
 + \Big(M\leftrightarrow M^\dagger \Big).
\eeq
This result can be matched against Eq.~(\ref{dE_m}). We find
$B=0$ and 
\beq
\label{Ai}
 A_1= -A_2 = \frac{3\Delta^2}{4\pi^2}\equiv A, 
\hspace{1cm} A_3 = 0.
\eeq
The result $A_1=-A_2$ reflects the fact that the CFL order parameter 
is anti-symmetric in flavor (pure $\bar{\bf 3}$) to leading order in $g$. 
Using Eqs.~(\ref{mueff}) and (\ref{Ai}) we can compute the energies of the 
flavored Goldstone bosons 
\bea
\label{mgb}
 E_{\pi^\pm} &=& \mu_{\pi^\pm} +
         \left[v_\pi^2 p^2+
      \frac{4A}{f_\pi^2}(m_u+m_d)m_s\right]^{1/2}\!\!\!\!, \nonumber \\
 E_{K^\pm}   &=& \mu_{K^\pm}  +
         \left[v_\pi^2 p^2+
      \frac{4A}{f_\pi^2}m_d (m_u+m_s)\right]^{1/2}\!\!\!\!, \\
 E_{K^0,\bar{K}^0} &=&  \mu_{K^0,\bar{K}^0} +
         \left[v_\pi^2 p^2+
      \frac{4A}{f_\pi^2}m_u (m_d+m_s)\right]^{1/2}\!\!\!\!, \nonumber
\eea
where 
\bea
\label{mu_mes}
\mu_{\pi^\pm} &=&  \mp\frac{m_d^2-m_u^2}{2\mu}, \hspace{0.5cm}
\mu_{K^\pm}    =  \mp \frac{m_s^2-m_u^2}{2\mu}, \nonumber \\
\mu_{K^0,\bar{K}^0}&=& \mp \frac{m_s^2-m_d^2}{2\mu}.
\eea
The mass matrix for the remaining
neutral Goldstone bosons, which mix, can be found in
\cite{Son:1999cm,Son:2000tu,Beane:2000ms}. 
We observe that the ${\cal O}(m)$ terms lead to an 
inverted mass spectrum with the kaons being lighter than the 
pions. This can be understood from the microscopic derivation 
of the chiral Lagrangian. The Goldstone boson field is $\Sigma
=XY^\dagger$, and a mode with the quantum number of the pion is 
given by $\pi^+\sim \epsilon^{abc}\epsilon^{ade}(\bar{d}^b_R
\bar{s}^c_R)(u^d_Ls^e_L)$. The structure of the field operators
suggests that the mass is controlled by $(m_u+m_d)m_s$. By the 
same argument the mass of the $K^+$ is governed by $(m_u+m_s)m_d$,
and $m_K<m_\pi$. We also note that the ${\cal O}(m^2)$ terms split the 
energies of different charge states. This can be understood from 
the fact that these terms act as an effective chemical potential 
for flavor. Explicit calculations in an NJL model reproduce $f_\pi$ in 
(\ref{f_pi}) and the results (\ref{mgb}), albeit with a different value of 
$A$~\cite{Ruggieri:2007pi,Kleinhaus:2007ve}.  This serves as a reminder that in the CFL phase 
at moderate densities, the effective theory is valid but the values of 
coefficients in it may not take on the values obtained by matching to high 
density calculations.

In perturbation theory the coefficient $B$ of the ${\rm Tr}(M
\Sigma)$ term is zero. $B$ receives non-perturbative contributions
from instantons. Instantons are semi-classical gauge configurations
in the Euclidean time functional integral 
that induce a fermion vertex of the form \cite{'tHooft:1976up}
\beq
\label{l_inst}
{\cal L} \sim G\det_f(\bar\psi_L\psi_R) + {\it h.c.},
\eeq
where $\det_f$ denotes a determinant in flavor space. The 't Hooft 
vertex (\ref{l_inst}) can be written as the product of the CFL 
condensate and its conjugate times $\langle \bar\psi\psi\rangle$, 
meaning that in the CFL phase Eq.~(\ref{l_inst}) induces a nonzero 
quark condensate $\langle\bar\psi\psi\rangle$, as well as Goldstone
boson masses $m_{GB}^2\sim m\langle\bar\psi\psi\rangle/f_\pi^2$.
The instanton has gauge field $A_\mu \sim 1/g $, so its
action is $S=8\pi^2/g^2$. The effective coupling $G$ is proportional
to $\exp(-S)\sim\exp(-8\pi^2/g^2)$, where $g$ is the running coupling
constant at a scale set by the instanton size $\rho$.

 In dense quark matter perturbative gauge field screening suppresses
instantons of size $\rho>1/\mu$, and the effective coupling $G$ can
be computed reliably \cite{Schafer:2002ty}. Combined with the weak
coupling result for $\langle \psi\psi\rangle$, see Eq.~(\ref{qq_cond}),
we get
\beq
\label{B}
 B = c
 \left[\frac{3\sqrt{2}\pi}{g}\Delta
     \left(\frac{\mu^2}{2\pi^2}\right)\right]^2
  \left(\frac{8\pi^2}{g^2}\right)^{6}
  \frac{\Lambda_{QCD}^9}{\mu^{12}} ,
\eeq
where $c=0.155$ and $\Lambda_{QCD}$ is the QCD scale factor. 
In terms of $B$, $\langle \bar{\psi} \psi \rangle = - 2 B$ and
the instanton contribution to the $K^0$ mass is 
$\delta m_{K^0}^2 = B (m_d+m_s)/(2 f_\pi^2)$ \cite{Manuel:2000wm}.
In the weak coupling limit, $\mu\gg\Lambda_{QCD}$, the instanton 
contribution is very small. However, because of the strong dependence 
on $\Lambda_{QCD}$ the numerical value of $B$ is quite uncertain. 
Using phenomenological constraints on the instanton size distribution 
\cite{Schafer:2002ty} concluded that the instanton contribution to the 
kaon mass at $\mu=500$ MeV is of order 10 MeV.

 Finally, we summarize the structure of the chiral expansion 
in the CFL phase. Ignoring non-perturbative effects the effective 
Lagrangian has the form 
\beq
\label{leff_exp}
{\mathcal L}\sim f_\pi^2\Delta^2 
 \left(\frac{\partial_0\Sigma}{\Delta}\right)^k
 \left(\frac{\vec{\partial}\Sigma}{\Delta}\right)^l
 \left(\frac{MM^\dagger}{\mu\Delta}\right)^m
 \left(\frac{MM}{\mu^2}\right)^n .
\eeq
Higher order vertices are suppressed by $\partial\Sigma/\Delta$ whereas
Goldstone boson loops are suppressed by powers of $\partial\Sigma/(4\pi 
f_\pi)$. Since the pion decay constant scales as $f_\pi\sim \mu$ the
effects of Goldstone boson loops can be neglected relative to higher
order contact interactions. This is different from chiral perturbation
theory at zero baryon density. We also note that the quark mass expansion 
contains two parameters, $m^2/\mu^2$ and $m^2/(\mu\Delta)$. Since 
$\Delta\ll\mu$ the chiral expansion breaks down if $m^2\sim \mu\Delta$. 
This is the same scale at which BCS calculations find a transition from 
the CFL phase to a less symmetric state. We also note that the result 
for the Goldstone boson energies given in Eq.~(\ref{mgb}) contains 
terms of ${\cal O}(m^2/\mu^2)$ and ${\cal O}([m^2/(\mu\Delta)]^2)$, but neglects 
corrections of ${\cal O}([m^2/\mu^2]^2)$. 

The effective Lagrangians (\ref{l_cheft}) and (\ref{u1_eft}) describe 
the physics of the low momentum pseudo-Goldstone and Goldstone bosons 
of the CFL phase at any density.  We have described the weak coupling 
computation of the coefficients $f_\pi$, $v_\pi$, $A_1$, $A_2$, $A_3$, 
$B$, $f$ and $v$ as well as the $\mu_{\rm eff}$'s in (\ref{mu_mes}). 
With the exception of $B$, all these results are expressed simply in 
terms of $\Delta$, $\mu$ and the quark masses, with $g$ not appearing 
anywhere. This suggests that the range of validity of these results, 
when viewed as a function of $\Delta$, is bigger than the range of 
validity of the weak coupling calculations on which they are based.
As we decrease the density down from the very large densities at which
the weak coupling calculation of $\Delta$ is under control, there is no 
indication that the relations between the effective theory coefficients 
and $\Delta$ and $\mu$ that we have derived in this section break down. 
The only sense in which we lose control of our understanding of the 
CFL phase is that we must treat $\Delta$ as a parameter, in terms of which all the other
effective theory coefficients are known.  Since $B$ introduces
$U(1)_A$-breaking physics that is not present at weak coupling and
that does not enter the effective theory through any other
coupling, it is not well constrained.

\begin{figure}[t]
\bc \includegraphics[width=7cm]{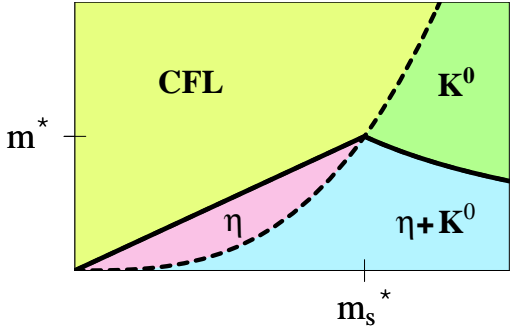}\ec
\caption{\label{fig_ph_ms_mu}
(Color online) Phase structure of CFL matter as a function of the light quark mass
$m$ and the strange quark mass $m_s$, from \cite{Kryjevski:2004cw}. 
CFL denotes pure CFL matter, while $K^0$ and $\eta$
denote CFL phases with $K^0$ and/or $\eta$ condensation. 
Solid lines are first order transitions, dashed lines
are second order. Instanton effects have been neglected.}
\end{figure}

\subsection{Kaon condensation}
\label{sec_kcond}

 If the effective chemical potential in Eq.~(\ref{mu_mes}) becomes
larger than the corresponding mass term in Eq.~(\ref{mgb}), then the energy of 
a Goldstone boson can become negative. In the physically relevant 
case $m_s\gg m_ u \sim m_d$ this applies in particular to the $K^0$ 
and the $K^+$. When the Goldstone boson energy becomes negative the 
CFL ground state is reorganized and a Goldstone boson condensate 
is formed. The physical reason is that a nonzero $m_s$ disfavors 
strange quarks relative to non-strange quarks. In normal quark 
matter the system responds to this stress by turning $s$ quarks 
into (mostly) $d$ quarks. In CFL matter this is difficult, since 
all quarks are gapped. Instead, the system can respond by populating 
mesons that contain $d$ quarks and $s$ holes. 

 The ground state can be determined from the effective potential 
\bea
\label{v_eff}
V_{\rm eff} &=& \frac{f_\pi^2}{4} {\rm Tr}\left[
 2X_L\Sigma X_R\Sigma^\dagger -X_L^2-X_R^2\right] \\
 & & \mbox{}
     - A_1\left\{ \left[{\rm Tr}(M\Sigma^\dagger)\right]^2 
     - {\rm Tr}\left[(M\Sigma^\dagger)^2\right] \right\},
  \nonumber
\eea
where $X_L=MM^\dagger/(2p_F)$,  $X_R=M^\dagger M/(2p_F)$
and $M={\rm diag}(m_u,m_d,m_s)={\rm diag}(m,m,m_s)$. Here we only
discuss the $T=0$ case. For nonzero temperature effects, in particular the calculation of the 
critical temperature of kaon condensation, see \cite{Alford:2007qa}. The first 
term on the right-hand side of Eq.\ (\ref{v_eff}) contains the effective 
chemical potential 
\beq
\mu_s\equiv -\mu_{K^0} \simeq\frac{m_s^2}{2 p_F}
\eeq
and favors states with a deficit of strange quarks. The second term 
favors the neutral ground state $\Sigma=1$. The lightest excitation with 
positive strangeness is the $K^0$~meson. We consider the ansatz $\Sigma 
= \exp(i\alpha\lambda_4)$ which allows the order parameter to rotate in 
the $K^0$ direction. The vacuum energy is 
\bea
\label{k0+_V}
 V(\alpha) &=& f_\pi^2 \Biggl[ -\frac{1}{2}\Big(\frac{m_s^2-m^2}{2p_F}
   \Big)^2\sin^2\alpha  \\
 & & \hspace{1.5cm}\mbox{}
    + m_{K^0}^2(1-\cos\alpha) \Biggr], \nonumber 
\eea
where $m_{K^0}^2= (4A_1 /f_\pi^2)m_u(m_d+m_s) + B (m_d+m_s)/(2 f_\pi^2)$. 
Minimizing the vacuum 
energy we obtain 
\beq
\cos(\alpha)= \left\{ \begin{array}{cl}
 1 & \mu_s<m_{K^0} \\
\;\frac{m_{K^0}^2}{\mu_s^2}\; & \mu_s >m_{K^0}\\
\end{array}\right.
\eeq
We conclude that there is a second order phase transition to 
a kaon condensed state at $\mu_s=m_{K^0}$. The strange quark mass 
breaks the $SU(3)$ flavor symmetry to $SU(2)_I\times U(1)_Y$. In 
the kaon condensed phase this symmetry is spontaneously broken to 
$U(1)_{\tilde{Q}}$. If $m_u=m_d$, isospin is an exact symmetry and there are two exact 
Goldstone modes \cite{Schafer:2001bq,Miransky:2001tw} with zero energy gap, the $K^0$ and 
the $K^+$. Isospin breaking leads to a small energy gap for the $K^+$.

Using the perturbative result for $A_1$, and neglecting instanton
effects by setting $B=0$, we can get an estimate of the critical 
strange quark mass. The critical strange quark mass scales as  
 $m_u^{1/3}\Delta^{2/3}$. Taking $\mu=500$~MeV, $\Delta=50$~MeV,
$m_u=4$~MeV and $m_d=7$~MeV, we find $m_s^{\rm crit}\simeq 68$ MeV, a 
result that corresponds to $m_{K^0}^{\rm crit}=5$~MeV. If  instanton 
contributions  increase $m_{K^0}$ by 10 MeV, this  would increase 
$m_s^{\rm crit}$ to 103~MeV, corresponding to the onset of kaon condensation 
depicted in Fig.~\ref{fig:energy}.

The difference in condensation 
energy between the CFL phase and the kaon condensed state is not 
necessarily small. In the limit $\mu_s\to \Delta$ we have 
$\sin \alpha \sim 1$ and $V(\alpha)\sim f_\pi^2\Delta^2/2$. Since $f_\pi^2$ 
is of order $\mu^2/(2\pi^2)$ this is an ${\cal O}(1)$ correction to the 
pairing energy in the CFL phase. Microscopic NJL model calculations 
of the condensation energy in the kaon condensed phase can be found 
in \cite{Buballa:2004sx,Forbes:2004ww,Warringa:2006dk,Kleinhaus:2007ve,Ruggieri:2007pi},
see also \cite{Ebert:2007bp,Ebert:2006tc}.

The CFL phase also contains a very light flavor neutral mode which 
can potentially become unstable. This mode is a linear combination 
of the $\eta$ and $\eta'$ and its mass is proportional to $m_u m_d$. 
Because this mode has zero strangeness it is not affected by the 
$\mu_s$ term in the effective potential. However, since $m_u,m_d
\ll m_s$ this state is sensitive to perturbative $\alpha_s m_s^2$ 
corrections \cite{Kryjevski:2004cw}. The resulting phase diagram 
is shown in Fig.~\ref{fig_ph_ms_mu}. The precise value of the 
tetra-critical point $(m^*,m_s^*)$ depends sensitively on the 
value of the coupling constant. At very high density $m^*$ is 
extremely small, but at moderate density $m^*$ can become as 
large as 5 MeV, comparable to the physical values of the up and 
down quark mass. 

\subsection{Fermions in the CFL phase}
\label{sec_bar}

 A single quark excitation with energy close to $\Delta$ is 
long-lived and interacts only weakly with the Goldstone modes in the CFL
phase. This means that it is possible to include quarks in the 
chiral Lagrangian. This Lagrangian not only controls the 
interaction of quarks with pions and kaons, but it also 
constrains the dependence of the gap in the
fermionic quasiparticle spectrum on the quark masses. 
This is of interest in connection with the existence and 
stability of the gapless CFL phase \cite{Alford:2003fq},
as we have discussed in Secs.~\ref{sec:BCSstress}, \ref{sec:CFLstress}, and \ref{sec:gCFL}. 

The effective Lagrangian for fermions in the CFL phase
is \cite{Kryjevski:2004jw,Kryjevski:2004kt}
\bea 
\label{l_bar}
{\mathcal L} &=&  
 {\rm Tr}\left(N^\dagger iv^\mu D_\mu N\right) 
 - D{\rm Tr} \left(N^\dagger v^\mu\gamma_5 
               \left\{ {\mathcal A}_\mu,N\right\}\right) \\[0.2cm]
 & & \mbox{}
 - F{\rm Tr} \left(N^\dagger v^\mu\gamma_5 
               \left[ {\mathcal A}_\mu,N\right]\right)
  \nonumber \\[0.2cm]
 & &  \mbox{} + \frac{\Delta}{2} \Big[
     \Big( {\rm Tr}\left(N_LN_L \right) 
   - \left[ {\rm Tr}\left(N_L\right)\right]^2 \Big) \nonumber \\
 & & \hspace{1cm}\mbox{}  
   - (L\leftrightarrow R) + h.c.  \Big]. \nonumber 
\eea
$N_{L,R}$ are left and right handed baryon fields in the adjoint 
representation of flavor $SU(3)$. The baryon fields originate from 
quark-hadron complementarity \cite{Schafer:1998ef,Alford:1999pa}. 
We can think of $N$ as describing a quark which is surrounded 
by a diquark cloud, $N_L \sim q_L\langle q_L q_L\rangle$. The 
covariant derivative of the nucleon field is given by $D_\mu N
=\partial_\mu N +i[{\mathcal V}_\mu,N]$. The vector and axial-vector 
currents are 
\beq
\label{v-av}
 {\mathcal V}_\mu = -\frac{i}{2}\left( 
  \xi \partial_\mu\xi^\dagger +  \xi^\dagger \partial_\mu \xi 
  \right), \hspace{0.2cm}
{\mathcal A}_\mu = -\frac{i}{2} \xi\left(\nabla_\mu 
    \Sigma^\dagger\right) \xi , 
\eeq
where $\xi$ is defined by $\xi^2=\Sigma$. It follows that $\xi$ 
transforms as $\xi\to L\xi U^\dagger=U\xi R^\dagger$ with 
$U\in SU(3)_V$. The fermion field transforms as $N\to UNU^\dagger$. 
For pure $SU(3)$ flavor transformations $L=R=V$ 
we have $U=V$. $F$ and $D$ are low energy constants that 
determine the baryon axial coupling. In  QCD at weak coupling, we
find $D=F=1/2$ \cite{Kryjevski:2004jw}.

\begin{figure}[t]
\bc\includegraphics[width=\hsize]{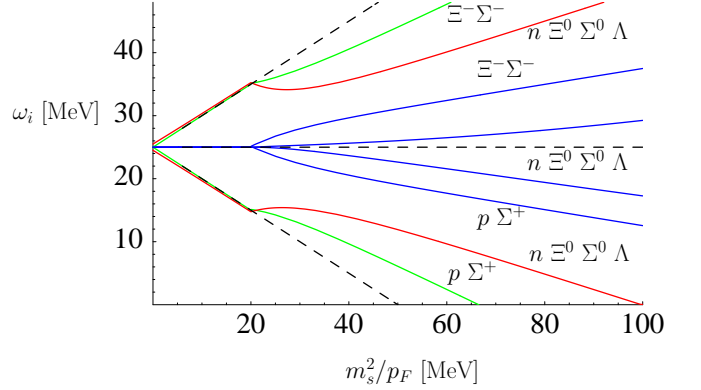}\ec
\caption{\label{fig_cfl_spec}
(Color online) This figure shows the fermion spectrum in the CFL phase. For 
$m_s=0$ there are eight fermions with gap $\Delta$ (set to 25 MeV as in Fig.\ \ref{fig:energy})
and one fermion with gap $2\Delta$ (not shown). As discussed in Sec.~\ref{sec:non_cfl},
the octet quasiparticles have the $SU(3)$ and $U(1)_{\rm \tilde Q}$ quantum
numbers of the octet baryons.
Without kaon condensation
gapless fermion modes appear at $\mu_s=\Delta$ (dashed lines).
With kaon condensation gapless modes appear at $\mu_s=4\Delta/3$.
(Note that the scale on the horizontal axis is $2\mu_s$.)}
\end{figure}

 The effective theory given in Eq.~(\ref{l_bar}) can be derived 
from QCD in the weak coupling limit. However, the structure of the 
theory is completely determined by chiral symmetry, even if the 
coupling is not weak. In particular, there are no free parameters
in the baryon coupling to the vector current. Mass terms are also
strongly constrained by chiral symmetry. The effective chemical 
potentials $(X_L,X_R)$ appear as left and right-handed gauge 
potentials in the covariant derivative of the nucleon field.
We have 
\bea
\label{V_X}
 D_0N     &=& \partial_0 N+i[\Gamma_0,N], \\
 \Gamma_0 &=& -\frac{i}{2}\left[
  \xi \left(\partial_0+ iX_R\right)\xi^\dagger + 
  \xi^\dagger \left(\partial_0+iX_L\right) \xi 
  \right], \nonumber 
\eea
where $X_L=MM^\dagger/(2p_F)$ and $X_R=M^\dagger M/(2p_F)$ as before.
$(X_L,X_R)$ covariant derivatives also appears in the axial vector 
current given in Eq.~(\ref{v-av}).

 We can now study how the fermion spectrum depends on the quark mass.
In the CFL state we have $\xi=1$. For $\mu_s=0$ the baryon octet
has an energy gap $\Delta$ and the singlet has gap $2\Delta$. The 
leading correction to this result comes from the commutator term
in Eq.~(\ref{V_X}). We find that the gap of the proton and neutron 
is lowered, $\Delta_{p,n}=\Delta-\mu_s$, while the gap of the cascade 
particles $\Xi^-,\Xi^0$ is increased, $\Delta_{\Xi}=\Delta+\mu_s$. 
As a consequence we find gapless $(p,n)$ excitations at $\mu_s=\Delta$. 
This result agrees with the spectrum discussed in Sec.~\ref{sec:gCFL}
if the identification $p\equiv (bu)$ and $n\equiv (bd)$ is made. 

  The situation is more complicated when kaon condensation is taken 
into account. In the kaon condensed phase there is mixing in the 
$(p,\Sigma^+,\Sigma^-,\Xi^-)$ and $(n,\Sigma^0,\Xi^0,\Lambda^8,
\Lambda^0)$ sector. For $m_{K^0}\ll\mu_s\ll \Delta$ the spectrum is 
given by
\beq
\label{KCFL_spec}
\omega_{p\Sigma^\pm\Xi^-}= \left\{
 \begin{array}{c}
 \Delta \pm \frac{3}{4}\mu_s, \\
 \Delta \pm \frac{1}{4}\mu_s,
\end{array}\right.  \hspace{0.75cm}
\omega_{n\Sigma^0\Xi^0\Lambda} = \left\{
 \begin{array}{c}
   \Delta \pm \frac{1}{2}\mu_s ,\\ 
   \Delta , \\
   2\Delta .
 \end{array} \right. 
\eeq
Numerical results for the eigenvalues are shown in Fig.~\ref{fig_cfl_spec}. 
We observe that mixing within the charged and neutral baryon sectors leads 
to level repulsion. There are two modes that become light in the CFL 
window $\mu_s\leq 2\Delta$. One mode is a charged mode which is a linear 
combination of the proton and the $\Sigma^+$, while the other mode is a 
linear combination of the neutral baryons $(n,\Sigma^0,\Xi^0,\Lambda^8,
\Lambda^0)$. The charged mode becomes gapless first, at $\mu_s=4\Delta/3$. 
Corrections to this result were studied in the NJL model calculation of 
\cite{Forbes:2004ww}, which includes various subleading condensates and 
obtains $\mu_s=1.22 \Delta$ at $\mu=500$ MeV. The neutral mode becomes 
gapless only at $\mu_s=2\Delta$. The most important difference as compared 
to the spectrum in the gapless CFL phase without kaon condensation is that 
for $\mu_s<2\Delta$ only the charged mode is gapless. 

\subsection{Goldstone boson currents}
\label{sec_gb_cur}

 In Sec.~\ref{subsubsec:chromo} we showed that gapless fermion modes lead to 
instabilities of the superfluid phase. Here we will discuss how these 
instabilities arise, and how they can be resolved, in the context 
of low energy theories of the CFL state, by the formation of the 
meson supercurrent state introduced in Sec.~\ref{sec_kcur}.  The chromomagnetic 
instability is an instability towards the spontaneous generation of 
currents, that is to say the spontaneous generation
of spatial variation in the phase of the diquark condensate. Consider 
a spatially varying $U(1)_Y$ rotation of the neutral kaon condensate
\beq
\label{xi_k}
\xi({\bf x}) = U({\bf x})\xi_{K} U^\dagger ({\bf x}) , 
\eeq
where $\xi_K=\exp(i\pi\lambda_4)$ and $U({\bf x})=\exp(i\phi_K({\bf x})\lambda_8)$.
This state is characterized by nonzero vector and axial-vector
currents, see Eq.~(\ref{v-av}). We shall study the dependence 
of the vacuum energy on the kaon current $\bm{\jmath}_K=\nabla
\phi_K$. The gradient term in the meson part of the effective 
Lagrangian gives a positive contribution
\beq
\label{e_kcur}
{\cal E}_m=\frac{1}{2}v_\pi^2f_\pi^2\jmath_K^2 .
\eeq
A negative contribution can arise from gapless fermions. In order 
to determine this contribution we have to calculate the fermion spectrum 
in the presence of a nonzero current. The relevant couplings are
obtained from the covariant derivative of the fermion field in 
Eq.~(\ref{V_X}) and the D and F-terms in Eq.~(\ref{l_bar}). The 
fermion spectrum is quite complicated. The dispersion relation of 
the lowest mode is approximately given by
\beq
\label{disp_ax}
\omega_l = \Delta +\frac{(l-l_0)^2}{2\Delta}-\frac{3}{4}
  \mu_s -\frac{1}{4}{\bf v}\cdot\bm{\jmath}_K,
\eeq
where $l={\bf v}\cdot{\bf p}-p_F$ and we have expanded $\omega_l$ 
near its minimum $l_0=(\mu_s+{\bf v}\cdot\bm{\jmath}_K)/4$.
Eq.~(\ref{disp_ax}) shows that there is a gapless mode if 
$\mu_s>4\Delta/3-\jmath_K/3 $. The contribution of the gapless mode 
to the vacuum energy is 
\beq
\label{e_fct} 
{\cal E}_q = \frac{\mu^2}{\pi^2}\int dl \int 
 \frac{d\Omega}{4\pi} \;\omega_l \theta(-\omega_l) ,
\eeq
where $d\Omega$ is an integral over the Fermi surface. The energy 
functional ${\cal E}_m+{\cal E}_q$ was analyzed in 
\cite{Schafer:2005ym,Kryjevski:2005qq}. 
There is an instability 
near the point $\mu_s=4\Delta/3$. The instability is resolved 
by the formation of a Goldstone boson current. If electric charge 
neutrality is enforced the magnitude of the current is very small,
and there is no tendency towards the generation of multiple 
currents. It was also shown that all gluonic screening masses 
are real \cite{Gerhold:2006np}. The situation is more complicated
if the neutral fermion mode becomes gapless, too. In this case 
the magnitude of the current is not small, and multiple currents
may appear. This
regime corresponds to the portion of the \mbox{curCFL-$K^0$} curve
in Fig.\ \ref{fig:energy} that is only slightly (invisibly) below the gCFL curve.

\subsection{Other effective theories}
\label{sec_mefts}

 Effective Lagrangians have also been been constructed for color 
superconducting phases other than the CFL phase. The effective 
theory for the light singlet axial mode in the 2SC phase can be 
found in \cite{Beane:2000ms}. The phonon effective theory in the 
crystalline color superconducting phase is discussed in Sec.~\ref{sec:rigid}.

 It is also interesting to study effective theories in QCD-like 
theories at large density. Some of these theories do not have 
a sign problem and can be studied on the lattice with algorithms
that are available today. Of particular interest are QCD with 
$N_c=2$ colors 
\cite{Nishida:2003uj,Kogut:1999iv,Kogut:2000ek,Fukushima:2007bj,Hands:1999md,Kogut:2001na,Hands:2006ve,Kogut:2002cm,Alles:2006ea} and QCD at finite 
isospin density \cite{Kogut:2002zg,Splittorff:2000mm,Son:2000xc}.

\section{NJL model comparisons among candidate phases below CFL densities}
\label{sec:NJL}
\newcommand{\ha}{\frac{1}{2}}
\newcommand{\rr}{{\bf r}}
\newcommand{\q}[2]{ {\bf q}_{#1}^{#2}}
\newcommand{\setq}[2]{\{{\bf q}_{#1}^{#2}\}}
\newcommand{\coleps}{\epsilon_{I\alpha\beta}}
\newcommand{\flaeps}{\epsilon_{Iij}}
\newcommand{\cross}[1]{#1\!\!\!/}
\newcommand{\cf}{\alpha i,\beta j}
\newcommand{\bp}{\bar{\psi}}
\newcommand{\qia}{ {\bf q}_{I}^{a}}
\newcommand{\setqia}{\{{\bf q}_{I}^{a}\}}
\newcommand{\unitq}[2]{\hat{{\bf q}}_{#1}^{#2}}
\newcommand{\unitqia}{\hat{{\bf q}}_{I}^{a}}
\newcommand{\hatq}[2]{\{\hat{{\bf q}}_{#1}^{#2}\}}
\newcommand{\hatqia}{\{\hat{{\bf q}}_{I}^{a}\}}
\newcommand{\vk}{{\bf{k}}}
\newcommand{\vl}{{\bf{l}}}
\newcommand{\vu}{{\bf u}}
\newcommand{\vv}{\hat{{\bf v}}}
\newcommand{\intspace}[1]{\int d^4 {#1}\,}
\newcommand{\mbyp}{\,\frac{\mu^2}{\pi^2}\,}

As we have explained in Sec.~\ref{sec:cfl}, 
at sufficiently high densities, where the up, down and
strange quarks can be treated on an equal footing and the disruptive
effects of the strange quark mass can be neglected, quark
matter is in the CFL phase.  
At asymptotic densities, the CFL gap
parameter $\Delta_{\rm CFL}$ 
and indeed any property of CFL
quark matter
can be calculated in full QCD, as described in Sec.~\ref{sec:QCD}. 
At any density at which the CFL phase arises,
its low energy excitations, and hence
its properties and phenomenology, can be described by the effective
field theory of Sec.~\ref{sec_eft}, whose form is known and whose parameters
can be systematically related to the CFL gap $\Delta_{\rm CFL}$.   
If we knew that the only form of color superconducting quark
matter that arises in the QCD phase diagram were CFL, there
would therefore be no need to resort to model analyses.
However, as we have discussed in 
Sec.~\ref{sec:non_cfl}, $M_s^2/(\mu\Delta_{\rm CFL})$ 
may not be small enough
(at $\mu=\mu_{\rm nuc}$ where the nuclear $\rightarrow$ quark matter 
transition occurs)
for the QCD phase diagram to be this simple.

Even at the very center of a neutron star, $\mu$ cannot be 
larger than about 500 MeV, meaning that the (density dependent) strange quark 
mass $M_s$ 
cannot be neglected.  In concert with the requirement that bulk matter
must be neutral and must be in weak equilibrium, a nonzero $M_s$ favors
separation of the Fermi momenta of the three different
flavors of quarks, and thus disfavors the cross-species BCS pairing
that characterizes the CFL phase.
If CFL pairing is disrupted by the heaviness
of the strange quark at a higher $\mu$ than that at which
color superconducting quark matter is superseded
by baryonic matter, 
the CFL phase must be replaced by some
phase of quark matter in which there is less, and less symmetric, pairing.

Within a spatially homogeneous ansatz, the next phase down in
density is the gapless CFL (gCFL)
phase
described in Sec.~\ref{sec:gCFL}.
However, as we have described in Sec.~\ref{subsubsec:chromo},
such  gapless paired states suffer
from a  chromomagnetic instability:  they can lower their energy by the
formation of counter-propagating
currents.
It seems likely,
therefore, that a ground state with counter-propagating currents is
required.  This could take the form of a crystalline color
superconductor,
that we have introduced in Sec.~\ref{subsec:crystal}. Or, given
that the CFL phase itself is likely augmented by kaon
condensation as described in Secs.~\ref{subsec:kaon} and
\ref{sec_kcond}, it could take
the form of the phase we have described in Sec.~\ref{sec_gb_cur}
in which a CFL kaon condensate carries a current
in one direction balanced by a counter-propagating current in the
opposite direction carried by gapless quark
quasiparticles.

Determining which phase or phases of quark matter occupy the regime of
density between hadronic matter and CFL quark
matter in the QCD phase diagram, if
there is such a regime, remains an outstanding challenge.  
Barring a major breakthrough that would allow
lattice QCD calculations to be brought to bear despite the fermion
sign problem,  a from-first-principles determination seems out of reach.
This leaves  two possible paths forward.  First, as we describe in this
section, we can analyze and compare many of the possible phases
within a simplified few parameter model, in so doing seeking qualitative
insight into what phase(s) are favorable.  Second, as we shall describe in
Sec.~\ref{sec:astro}, we can determine the observable consequences of the presence
of various possible color superconducting phases in neutron stars,
and then seek to use observational data to rule possibilities out or in.

\subsection{Model, pairing ansatz, and homogeneous phases}
\label{sec:NJLModel}

We shall employ a Nambu--Jona-Lasinio (NJL)
model in which the QCD interaction between
quarks is replaced by a point-like four-quark interaction, with the quantum
numbers of single-gluon exchange, analyzed in mean field theory.
This is not a controlled approximation.
However, it suffices for our purposes: because this model has  attraction
in the same channels as in QCD, its high density phase is the CFL phase; and, the
Fermi surface splitting effects whose
qualitative consequences we wish to study can be built
into the model.  
Note that we
shall assume throughout that $\Delta_{\rm CFL}\ll \mu$.  This weak coupling assumption
means that the pairing is dominated by modes near the Fermi surfaces.
Quantitatively,  this means that results for the gaps and condensation energies
of candidate phases are independent of the cutoff in the NJL model
when expressed in terms of the CFL gap $\Delta_{\rm CFL}$: if the 
cutoff is changed with
the NJL coupling constant adjusted so that $\Delta_{\rm CFL}$ stays fixed, 
the gaps and
condensation energies for the candidate crystalline phases also stay fixed.
This makes the NJL model valuable for making the comparisons
that are our goal.  
The NJL model has two parameters: the CFL gap $\Delta_{\rm CFL}$ 
which parametrizes the strength of the interaction and $M_s^2/(4\mu)$,
the splitting between Fermi surfaces in neutral quark matter in
the absence of pairing.  The free energy of candidate patterns of
pairing can be evaluated and compared as a function of these two
parameters.  

As a rather general pairing ansatz, we shall consider
\begin{eqnarray}
\langle ud \rangle &\sim& \Delta_3 \sum_{a} \exp\left(2i\q{3}{a} \cdot {\bf r}\right)\nonumber\\
\langle us \rangle &\sim& \Delta_2 \sum_{a} \exp\left(2i\q{2}{a}\cdot {\bf r} \right)\nonumber\\
\langle ds \rangle &\sim& \Delta_1 \sum_{a} \exp\left(2i\q{1}{a}\cdot {\bf r} \right)\ .
\label{udandusanddscondensates}
\end{eqnarray}
If we set all the wave vectors $\q{I}{a}$ to zero, we can use this ansatz to compare
spatially homogeneous phases including the CFL 
phase ($\Delta_1=\Delta_2=\Delta_3\equiv \Delta_{\rm CFL}$), the gCFL phase 
($\Delta_3>\Delta_2>\Delta_1>0$) and the 2SC phase 
($\Delta_3\equiv \Delta_{\rm 2SC}$; $\Delta_1=\Delta_2=0$).  Choosing different sets
of wave vectors will allow us to analyze and compare different crystalline
color superconducting phases of quark matter.

NJL models of varying degrees of 
complexity 
have been used for a variety of purposes
beyond the scope of this review.  
For example, whereas we treat $\Delta_{\rm CFL}$ and quark masses
as parameters and use the NJL model to compare
different patterns of pairing at fixed values of these 
parameters and $\mu$, it is possible instead to
fix the NJL coupling or couplings and then self-consistently
solve for the gap parameters and the 
$\langle \bar s s\rangle$ 
condensate as functions of 
$\mu$~\cite{Steiner:2002gx,Abuki:2004zk,Ruster:2005jc,Blaschke:2005uj,Abuki:2005ms,Ippolito:2007uz,Warringa:2006dk,Buballa:2001gj,Mishra:2006xh,Mishra:2003nr}.
Doing so reintroduces sensitivity to the cutoff in the NJL model and
so does not actually reduce the number of parameters.  Also, these models
tend to find rather larger values of $M_s$ than in analyses that go beyond
NJL models, for example the analysis using Dyson-Schwinger equations
in~\cite{Nickel:2006kc}.  
There have also been many investigations of the phase diagram
in the $\mu$-$T$ plane in
NJL models (either with $\Delta_{\rm CFL}$ and $M_s$ as parameters
or solved for self-consistently)~\cite{Berges:1998rc,Schwarz:1999dj,Kashiwa:2007pc,Ruester:2006aj,Warringa:2005jh,Ruster:2005jc,Ruster:2004eg,Fukushima:2004zq,Iida:2004cj,Abuki:2005ms,Barducci:2004tt,Mishra:2004gw,He:2006vr}. 
Although many of their features are sensitive to the cutoff as well
as the chosen couplings, these NJL phase diagrams indicate
how rich the QCD phase diagram may turn out to
be, as different condensates vanish at different temperatures.
One result that has been obtained using the Ginzburg-Landau approximation
as well as in NJL models
and so is of more general validity is
that upon heating the CFL phase at nonzero but
small $M_s^2/\mu$, as $T$ increases 
$\Delta_2$ vanishes first, then $\Delta_1$,
and then $\Delta_3$~\cite{Iida:2003cc,Fukushima:2004zq,Ruster:2004eg}.  
However, it remains to be seen how 
this conclusion is modified by the effects of
gauge-field 
fluctuations, which for $M_s=0$
turn the mean-field Ginzburg-Landau second order transition into
a strong first order phase transition at a significantly elevated
temperature, see Sec.~\ref{sec_lg} and Eq.~(\ref{elevatedTc}).

We shall analyze quark matter containing massless $u$ and $d$ quarks and $s$ quarks with
an effective mass $M_s$.
The Lagrangian density describing this system in the absence
of interactions is given by
\begin{equation}
{\cal L}_0=\bar{\psi}_{i\alpha}\,\left(i\,\partial\!\!\!
/\delta^{\alpha\beta}\delta_{ij} -M_{ij}^{\alpha\beta}+
\mu^{\alpha\beta}_{ij} \,\gamma_0\right)\,\psi_{\beta j}
\label{lagr1}\ \,,
\end{equation}
where $i,j=1,2,3$ are flavor indices and $\alpha,\beta=1,2,3$ are
color indices and we  have suppressed the Dirac indices,
where $M_{ij}^{\alpha\beta} =\delta^{\alpha\beta}\,{\rm
diag}(0,0,M_s)_{ij} $ is the mass matrix, 
and where
the quark chemical potential matrix is  given by
\begin{equation}\mu^{\alpha\beta}_{ij}=(\mu\delta_{ij}-\mu_e
Q_{ij})\delta^{\alpha\beta} + \delta_{ij} \left(\mu_3
T_3^{\alpha\beta}+\frac{2}{\sqrt 3}\mu_8 T_8^{\alpha\beta}\right) \,
, \label{mu}
\end{equation} with  $Q_{ij} = {\rm
diag}(2/3,-1/3,-1/3)_{ij} $ the quark electric-charge matrix and
$T_3$ and $T_8$ the diagonal color generators.
In QCD, $\mu_e$, $\mu_3$ and $\mu_8$ are the zeroth components of
electromagnetic and color gauge fields, and the gauge field dynamics
ensure that they take on values such that the matter is
neutral~\cite{Alford:2002kj,Gerhold:2003js,Kryjevski:2003cu,Dietrich:2003nu}, satisfying the
neutrality conditions (\ref{neutrality}).
In the NJL model, quarks interact
via four-fermion interactions and there are no gauge fields, so we introduce
$\mu_e$, $\mu_3$ and $\mu_8$ by hand, and choose them to satisfy
the neutrality constraints (\ref{neutrality}).  The assumption of weak equilibrium
is built into the calculation via the fact that the only flavor-dependent chemical
potential is $\mu_e$, ensuring for example that the chemical potentials of
$d$ and $s$ quarks with the same color must be equal.  
Because
the strange quarks have greater mass, the equality
of their chemical potentials implies that the $s$ quarks  have smaller
Fermi momenta than the $d$ quarks in the absence
of BCS pairing.  In the absence of pairing, then, because weak equilibrium
drives the massive strange quarks to be less numerous than
the down quarks, electrical neutrality
requires a $\mu_e>0$, which makes the up quarks less numerous than the
down quarks and introduces some electrons into the system.
In the absence of pairing, color neutrality is obtained with $\mu_3=\mu_8=0.$

The Fermi momenta of the quarks and electrons in quark matter
that is electrically and color neutral and in weak equilibrium
are given in the absence of pairing by
\begin{eqnarray}
p_F^d &=& \mu+\frac{\mu_e}{3}\nonumber\\
p_F^u &=& \mu-\frac{2 \mu_e}{3}\nonumber\\
p_F^s &=& \sqrt{\left(\mu+\frac{\mu_e}{3}\right)^2- M_s^2} \approx \mu + \frac{\mu_e}{3}
-\frac{M_s^2}{2\mu}\nonumber\\
p_F^e &=& \mu_e\ ,
\label{pF1}
\end{eqnarray}
where we have simplified $p_F^s$ 
by working to linear order in $\mu_e$ and $M_s^2$.
To this order, electric neutrality requires
$\mu_e={M_s^2}/(4\mu)$,
yielding
\begin{eqnarray}
p_F^d &=& \mu+\frac{M_s^2}{12\mu}=p_F^u+\frac{M_s^2}{4\mu}\nonumber\\
p_F^u &=& \mu-\frac{M_s^2}{6\mu}\nonumber\\
p_F^s &=& \mu-\frac{5 M_s^2}{12 \mu} =p_F^u-\frac{M_s^2}{4\mu}\nonumber\\
p_F^e &=& \frac{M_s^2}{4\mu}\ ,
\label{pF2}
\end{eqnarray}
as illustrated in Fig.~\ref{fig:splitting}.
We see from (\ref{pF1}) that to leading order in $M_s^2$ and $\mu_e$, the
effect of the strange quark mass on unpaired quark matter is as if instead
one reduced the strange quark chemical potential by $M_s^2/(2\mu)$.
We shall make this approximation throughout.
Upon making this assumption, we need no longer be careful about the
distinction between $p_F$'s and $\mu$'s, as we can simply think of the three
flavors of quarks as if they have chemical potentials
\begin{eqnarray}
\mu_d &=& \mu_u + 2 \delta\mu_3 \nonumber\\
\mu_u &=&p_F^u \nonumber\\
\mu_s &=& \mu_u - 2 \delta\mu_2
\label{pF3}
\end{eqnarray}
with
\beq
\delta\mu_3 = \delta\mu_2 = \frac{M_s^2}{8\mu}\equiv \delta\mu \ ,
\eeq
where the choice of subscripts indicates that
$2\delta\mu_2$ is the
splitting between the Fermi surfaces for quarks 1 and 3 and
$2\delta\mu_3$ is that between the Fermi surfaces for quarks 1 and 2,
identifying $u,d,s$ with $1,2,3$.  



As described in \cite{Rajagopal:2000ff,Alford:2002kj,Steiner:2002gx,Alford:2003fq},
BCS pairing introduces qualitative changes into the analysis of neutrality.  For example,
in the CFL phase $\mu_e=0$ and $\mu_8$ is nonzero and of order $M_s^2/\mu$.
This arises because wherever BCS pairing occurs between
fermions whose Fermi surface would be split in the absence of pairing,
the Fermi momenta of these fermions are locked together. This maximizes
the pairing energy gain while at the same time exacting a kinetic energy
price and changing
the relation between the chemical potentials and the particle numbers. This
means that the $\mu$'s required for neutrality can change qualitatively as
happens in the CFL example. 

The NJL interaction term with
the quantum numbers of single-gluon exchange that we add to the
Lagrangian (\ref{lagr1}) is
\begin{equation}
{\cal L}_{\rm interaction} = 
-\frac{3}{8}\lambda(\bar{\psi}\Gamma^{A\nu}\psi)(\bar{\psi}\Gamma_{A\nu}\psi)
\label{interactionlagrangian}\ ,
\end{equation}
where we have suppressed the color and flavor indices that we showed
explicitly in (\ref{lagr1}), and have continued to suppress the Dirac indices.
The full expression for $\Gamma^{A\nu}$ is 
$(\Gamma^{A\nu})_{\alpha i,\beta j} = \gamma^\nu (T^A)_{\alpha \beta}\delta_{i j}$. 
The NJL coupling constant $\lambda$ has dimension -2,
meaning that an ultraviolet cutoff $\Lambda$ 
must be introduced as a second parameter in order
to fully specify the interaction.    We shall define $\Lambda$ as 
restricting
the momentum
integrals 
to a shell around the Fermi surface, 
$\mu-\Lambda < |{\bf p}| < \mu + \Lambda$.

In the mean-field approximation, the interaction Lagrangian (\ref{interactionlagrangian})
takes on the form
\begin{equation}
{\cal L}_{\rm interaction}=
\ha\bar{\psi}\Delta(x)\bar{\psi}^T + \ha\psi^T\bar{\Delta}(x)\psi,
\label{meanfieldapprox}
\end{equation}
where $\Delta(x)$ is related to the diquark condensate by the
relations
\begin{equation}
\begin{split}
\Delta(x) &= \frac{3}{4}\lambda\Gamma^{A\nu}\langle\psi\psi^T\rangle(\Gamma_{A\nu})^T \\
\bar{\Delta}(x) &= \frac{3}{4}\lambda
   (\Gamma^{A\nu})^T\langle\bp^T\bp\rangle\Gamma_{A\nu} \\
   &=\gamma^0\Delta^{\dagger}(x)\gamma^0\label{deltaislambdacondensate}\;.
\end{split}
\end{equation}
The ansatz (\ref{udandusanddscondensates}) can now be made precise: we take
\begin{equation}
\Delta(x)=\Delta_{CF}(x)\otimes C\gamma^5\label{spinstructure}\;,
\end{equation}
with the color-flavor part
\begin{equation}
\Delta_{CF}(x)_{\cf} = \sum_{I=1}^3\sum_{\q{I}{a}}
 \Delta(\q{I}{a}) e^{2i\q{I}{a}\cdot \rr}\coleps\flaeps\label{precisecondensate}\ .
\end{equation}
We have introduced notation that 
allows for the possibility of gap parameters $\Delta(\q{I}{a})$
with different magnitudes for different $I$ {\it and for different} $a$.  In fact, we shall
only consider circumstances in which $\Delta(\q{I}{a})=\Delta_I$, as in 
(\ref{udandusanddscondensates}).

The full Lagrangian, given by the sum of (\ref{lagr1}) and (\ref{meanfieldapprox}),
is then quadratic and can be written very simply upon introducing the two component
Nambu-Gorkov spinors (\ref{NambuGorkov})
in terms of which 
\begin{equation}
{\cal L} = \ha \bar{\Psi} \left( \begin{array}{cc}
i\cross{\partial}+\cross{\mu} & \Delta(x)    \\
\bar{\Delta}(x) &   (i\cross{\partial}-\cross{\mu})^T
\end{array} \right) \Psi \;.
\label{fulllagrangian}
\end{equation}
Here, $\cross\mu\equiv \mu\gamma_0$ and $\mu$ is the matrix (\ref{mu}).

The propagator corresponding to the Lagrangian (\ref{fulllagrangian}) is given by
\begin{equation}
\begin{split}
\langle\Psi(x)\bar{\Psi}(x')\rangle
 &=\Bigl( \begin{array}{cc}
   \langle\psi(x)\bp(x')\rangle & \langle\psi(x)\psi^T(x')\rangle    \\
   \langle\bp^T(x)\bp(x')\rangle & \langle\bp^T(x)\psi^T(x')\rangle
\end{array} \Bigr) \\
 &=\Bigl( \begin{array}{cc}
   iG(x,x') & iF(x,x')    \\
   i\bar{F}(x,x') & i\bar{G}(x,x')
\end{array} \Bigr) \;,
\label{propagator}
\end{split}
\end{equation}
where $G$ and $\bar{G}$ are the ``normal'' components of the propagator and $F$
and $\bar{F}$ are the ``anomalous'' components. They satisfy the coupled
differential equations
\begin{equation}
\begin{split}
\Bigl( \begin{array}{cc}
i\cross{\partial}+\cross{\mu} & \Delta(x)    \\
\bar{\Delta}(x) &   (i\cross{\partial}-\cross{\mu})^T
\end{array} \Bigr) &
\Bigl( \begin{array}{cc}
   G(x,x') & F(x,x')    \\
   \bar{F}(x,x') & \bar{G}(x,x')
\end{array} \Bigr)\\
&=
\Bigl( \begin{array}{cc}
  1  & 0     \\
  0  & 1
\end{array} \Bigr)\delta^{(4)}(x-x')\ .
\label{Green's equation}
\end{split}
\end{equation}
We can now rewrite (\ref{deltaislambdacondensate}) as
\begin{equation}
\begin{split}
\Delta(x) &= \frac{3i}{4}\lambda\Gamma^{A\nu} F(x,x) (\Gamma_{A\nu})^T \\
\bar{\Delta}(x) &= \frac{3i}{4}\lambda
   (\Gamma^{A\nu})^T \bar{F}(x,x) \Gamma_{A\nu}\ , 
\label{formalgapequation}
\end{split}
\end{equation}
either one of which is the self-consistency equation, or gap equation, that we
must solve.  

Without further approximation, (\ref{formalgapequation}) is not tractable.
It yields an infinite set of coupled gap equations, one for each $\Delta(\q{I}{a})$, because
without further approximation it is not consistent to choose finite sets $\{{\bf q}_I\}$. 
When several plane waves are present in the condensate, they induce an infinite
tower of higher momentum condensates~\cite{Bowers:2002xr}.   In the next subsection,
we shall make a Ginzburg-Landau (i.e. small-$\Delta$) approximation which eliminates
these higher harmonics.

Of course, an even more dramatic simplification is obtained if we set all
the wave vectors $\q{I}{a}$ to zero.  Still, even in this case obtaining the general
solution with $M_s\neq 0$ and $\Delta_1 \neq \Delta_2 \neq \Delta_3$ is
somewhat involved~\cite{Alford:2003fq,Alford:2004hz,Fukushima:2004zq}.  
We shall not present the resulting analysis of the CFL$\rightarrow$gCFL transition
and the gCFL phase here.  The free energies of these phases are depicted in
Fig.~\ref{fig:energy}, and their gap parameters are depicted below in Fig.~\ref{fig:deltavsx}.

If we simplify even further, by setting $M_s=0$
and $\Delta_1=\Delta_2=\Delta_3\equiv\Delta_{\rm CFL}$, 
the gap equation determining the CFL gap parameter $\Delta_{\rm CFL}$
can then be evaluated analytically, yielding~\cite{Bowers:2002xr}
\beq
\Delta_{\rm CFL}=2^{\frac{2}{3}}\Lambda \exp\left[-\frac{\pi^2}{2\mu^2\lambda}\right]\ .
\end{equation}
We shall see below that in the limit in which
$\Delta\ll \Delta_{\rm CFL},\delta\mu\ll\mu$, all results for the myriad possible crystalline phases 
can be expressed in terms of $\Delta_{\rm CFL}$;
neither $\lambda$ nor $\Lambda$ shall appear.  This reflects the fact that in this
limit the physics of interest is dominated by quarks near the Fermi surfaces,  not near 
$\Lambda$, and so once $\Delta_{\rm CFL}$ is used
as the parameter describing the strength of the attraction between quarks, $\Lambda$ 
is no longer visible; the cutoff $\Lambda$ only appears in the relation between
$\Delta_{\rm CFL}$ and $\lambda$, not in any comparison among different possible
paired phases.  We are using the NJL model in a specific, limited, fashion in which
it serves as a two parameter model allowing the comparison among different possible
paired phases at a given $\Delta_{\rm CFL}$ and $M_s$.  NJL models have also
been employed to estimate the value of $\Delta_{\rm CFL}$ at
a given 
$\mu$ \cite{Alford:1997zt,Alford:1998mk,Rapp:1997zu,Rajagopal:2000wf,Berges:1998rc,Carter:1998ji}; 
doing so requires normalizing the four-fermion interaction 
by calculating some zero density quantity like the vacuum chiral condensate, and
in so doing introduces a dependence on the cutoff $\Lambda$.  
Such mean-field NJL analyses
are important complements to extrapolation down from an
analysis that is rigorous at high density and hence weak coupling, described
in Sec.~\ref{sec:QCD}, and give us
confidence that we understand the magnitude of 
$\Delta_{\rm CFL}\sim 10-100$~MeV.  
This estimate receives further support from the lattice-NJL calculation
of \cite{Hands:2004uv} which finds diquark condensation and a $\sim 60$ MeV gap 
in an NJL model whose parameters are normalized via calculation of $f_\pi$, $m_\pi$
and a constituent quark mass in vacuum.  With these as inputs, $\Delta$ is then
calculated on the lattice,
i.e. without making
a mean-field approximation.
With an understanding of its magnitude in hand, we shall treat $\Delta_{\rm CFL}$ as a parameter,
thus making our results insensitive to $\Lambda$.


We shall focus below on the use of the NJL model that we have
introduced to analyze and compare different possible crystalline phases,
comparing their free energies to that of the CFL phase as a benchmark.
The free energy of the 2SC phase is easily calculable in the same model,
and the free energies of the unstable gapless CFL and gapless 2SC phases
can also be obtained~\cite{Alford:2004hz}.  These free energies are all shown
in Fig.~\ref{fig:energy}.  The free energies of phases with various 
patterns of single-flavor pairing have also been calculated in the
same model~\cite{Alford:2002rz}.
The NJL model is not a natural starting point
for an analysis of the kaon condensate in the CFL-$K^0$ phase, but with considerable 
effort this has been accomplished 
in \cite{Buballa:2004sx,Forbes:2004ww,Warringa:2006dk,Kleinhaus:2007ve}.
The
curCFL-$K^0$ phase of Secs.~\ref{sec_kcur} and \ref{sec_gb_cur}, 
in which the $K^0$-condensate carries a current,
has not been analyzed in an NJL model.  But,
because both the CFL-$K^0$ and curCFL-$K^0$ phases 
are continuously connected to the CFL phase, they can both be analyzed in 
a model-independent fashion using the effective field theory described in
Sec.~\ref{sec_eft}.  The CFL-$K^0$ and curCFL-$K^0$ curves in Fig.~\ref{fig:energy}
were obtained as described in Sec.~\ref{sec_eft}.  It remains a challenge for future
work to do a calculation in which both curCFL-$K^0$ and crystalline phases
are possible, allowing a direct comparison of their free energies within a single
calculation and a study of whether they are distinct as current
results seem to suggest or are instead different limits of
some more general inhomogeneous color superconducting phase.

\subsection{Crystalline phases}
\label{subsec:NJLCrystalline}

Crystalline color superconductivity~\cite{Alford:2000ze,Bowers:2001ip,Casalbuoni:2001gt,Leibovich:2001xr,Kundu:2001tt,Bowers:2002xr,Casalbuoni:2003wh,Casalbuoni:2003sa,Casalbuoni:2004wm,Casalbuoni:2005zp,Ciminale:2006sm,Mannarelli:2006fy,Casalbuoni:2002hr,Casalbuoni:2002my,Casalbuoni:2006zs,Giannakis:2002jh}
naturally permits pairing between quarks living at
split Fermi surfaces by allowing Cooper pairs with nonzero net momentum.
In three-flavor quark matter, this allows pairing to occur even with the
Fermi surfaces split in the free-energetically optimal way as in the absence
of pairing, meaning that neutral crystalline phases are obtained
in three-flavor quark matter with the chemical potential matrix
(\ref{mu})
simplified to 
$\mu = \delta^{\alpha\beta}\otimes {\rm diag}\left(\mu_u,\mu_d,\mu_s\right)$
with the flavor chemical potentials given simply by 
(\ref{pF3})~\cite{Casalbuoni:2005zp,Mannarelli:2006fy,Rajagopal:2006ig},
up to higher order corrections that have been investigated in
\cite{Casalbuoni:2006zs}.  
This is the origin of the advantage that crystalline color superconducting
phases have over the CFL and gCFL phases at large values 
of the splitting $\delta\mu$.
For example, by
allowing $u$ quarks with momentum ${\bf p+q}_3$ to pair with $d$
quarks with momentum ${\bf -p+q}_3$, for any ${\bf p}$, 
we can pair $u$ and $d$ quarks
along rings on their respective Fermi surfaces. In coordinate
space, this corresponds to a condensate of the form 
$\langle ud \rangle\sim \Delta_3 \exp\bigl({2i{\bf q_3}\cdot{\bf r}}\bigr)$. 
The net free energy gained due to pairing is then a balance between increasing $|{\bf q}_3|$
yielding pairing on larger rings while exacting a greater kinetic energy cost. The optimum
choice turns out to be $|{\bf q}_3|=\eta \delta\mu_3$ with $\eta=1.1997$, corresponding
to pairing rings on the Fermi surfaces with opening angle $67.1^\circ$~\cite{Alford:2000ze}.
Pairing with only a single ${\bf q}_3$ is disadvantaged because
the only quarks on each Fermi surface that can then pair 
are those lying on a single ring.
This disadvantage can be overcome in two ways.
First, increasing $\Delta$ widens
the pairing rings on the Fermi surfaces into pairing bands which fill in,
forming pairing caps, at large enough $\Delta$~\cite{Mannarelli:2006fy}.
Second,
it is possible to cover larger areas of the Fermi surfaces by allowing Cooper pairs with 
the same $|{\bf q}_3|$ but various $\hat{\bf q}_3$, yielding 
$\langle ud \rangle \sim \Delta_3 \sum_{\q{3}{a}}\exp\bigl(2\, i\, \q{3}{a} \cdot {\bf r}\bigr)$
with the $\q{3}{a}$ chosen from some specified 
set $\{\q{3}{1},\q{3}{2},\q{3}{3},\ldots\}\equiv\setq{3}{}$.  This is a condensate
modulated in position space in some crystalline pattern, with the crystal structure defined
by $\setq{3}{}$.  
In this two-flavor context, a Ginzburg-Landau
analysis reveals that 
the best $\setq{3}{}$ contains eight vectors pointing at the corners of a cube,
say in the $(\pm 1,\pm 1,\pm 1)$ directions in momentum space, yielding a face-centered
cubic structure in position space~\cite{Bowers:2002xr}.

This subsection describes the analysis of three-flavor crystalline
phases in \cite{Rajagopal:2006dp}. 
We use the ansatz given by (\ref{spinstructure}) and (\ref{precisecondensate})
for the three-flavor crystalline color superconducting 
condensate.
This is antisymmetric in color ($\alpha,\beta$), spin, 
and flavor ($i,j$) indices and is a generalization
of the CFL condensate to crystalline color superconductivity. 
We set $\Delta_1=0$, neglecting
$\langle ds \rangle$ pairing because the $d$ and $s$ Fermi
surfaces  are twice as far apart from each other as each is from
the intervening $u$ Fermi surface. Hence, $I$ can be taken to run over $2$ and $3$ only.
$\setq{2}{}$ and $\setq{3}{}$ define the crystal structures of the $\langle us \rangle$ and
$\langle ud \rangle$ condensates respectively.   We only consider 
crystal structures
in which all the vectors in $\setq{2}{}$ are equivalent to each other
in the sense that any one can be transformed into any other by a symmetry
operation of $\setq{2}{}$
and same for
$\setq{3}{}$. This justifies our simplifying assumption that the $\langle us\rangle$
and $\langle ud \rangle$ condensates are each specified by a single gap parameter
($\Delta_2$ and $\Delta_3$ respectively), avoiding having to introduce one gap
parameter per ${\bf q}$.  We furthermore only consider crystal structures which
are exchange symmetric, meaning that $\setq{2}{}$ and $\setq{3}{}$ can be exchanged
by some combination of rigid rotations and reflections applied simultaneously to
all the vectors in both sets.  This simplification, together with 
$\delta\mu_2=\delta\mu_3$ (an approximation corrected only at order $M_s^4/\mu^3$),
guarantees that we find solutions with $\Delta_2=\Delta_3$.

We analyze and compare candidate crystal structures
by evaluating the free energy $\Omega(\Delta_2,\Delta_3)$ for each
crystal structure in
a Ginzburg-Landau expansion in powers of the $\Delta$'s. 
This approximation
is controlled
if $\Delta_2,\Delta_3 \ll \Delta_{\rm CFL},\delta\mu$, with $\Delta_{\rm CFL}$ the gap parameter in the
CFL phase at $M_s^2/\mu=0$.   
The terms in the Ginzburg-Landau expansion must respect the global $U(1)$ symmetry 
for each flavor,
meaning that each $\Delta_I$ can only
appear in the combination $|\Delta_I|^2$. (The $U(1)$ symmetries are spontaneously
broken by the condensate, but not explicitly broken.) Therefore,
$\Omega(\Delta_2,\Delta_3)$ is given to sextic order by
\begin{eqnarray}
\nonumber
\Omega(\Delta_2,\Delta_3)&=&\frac{2\mu^2}{\pi^2}\Biggl[P_2
\alpha_2 |\Delta_2|^2 + P_3 \alpha_3 |\Delta_3|^2\\
&+&\ha\Bigl( \beta_2|\Delta_2|^4 + \beta_3|\Delta_3|^4
+ \beta_{32} |\Delta_2|^2|\Delta_3|^2\Bigr)\nonumber\\
&+&\frac{1}{3}\Bigl( \gamma_2|\Delta_2|^6+\gamma_3|\Delta_3|^6\nonumber\\
&+&\gamma_{322}|\Delta_3|^2|\Delta_2|^4+\gamma_{233}|\Delta_3|^4|\Delta_2|^2\Bigr)\Biggr]
\label{GLexpansion}\;,
\end{eqnarray}
where we have chosen notation consistent with that  used in
the two flavor study of \cite{Bowers:2002xr}, which arises as a special
case of (\ref{GLexpansion}) if we take $\Delta_2$ or $\Delta_3$ to be zero. 
$P_I$ is the number of vectors in the set $\setq{I}{}$. 
The form of the
Ginzburg-Landau expansion (\ref{GLexpansion}) is model-independent, whereas
the expressions for the coefficients $\alpha_I$, $\beta_I$, $\beta_{IJ}$, $\gamma_I$,
and $\gamma_{IJJ}$ for a specific crystal structure are model-dependent.
We calculate them in the NJL model described in Sec.~\ref{sec:NJLModel}.
For exchange symmetric
crystal structures, $\alpha_2=\alpha_3\equiv\alpha$, $\beta_2=\beta_3\equiv\beta$, 
$\gamma_2=\gamma_3\equiv\gamma$ and $\gamma_{233}=\gamma_{322}$.

Because setting one of the $\Delta_I$ to zero reduces the problem to one with two-flavor
pairing only, we can obtain $\alpha$, $\beta$ and $\gamma$
via applying the two-flavor analysis described in \cite{Bowers:2002xr} to
either $\setq{2}{}$ or $\setq{3}{}$ separately.  
Using $\alpha$ as an example, we learn 
that
\begin{eqnarray}
\alpha_I =
\alpha(q_I,\delta\mu_I) &=& -1
 +\frac{\delta\mu_I}{2 q_I}
 \log\left(\frac{q_I+\delta\mu_I}{q_I-\delta\mu_I}\right)\nonumber\\
 &\quad&\quad - \ha\log\left(\frac{\Delta_{\rm 2SC}^2}{4(q_I^2-\delta\mu_I^2)}\right).
\label{AlphaEqn}
\end{eqnarray}
Here, $q_I\equiv |{\bf q}_I|$ and 
$\Delta_{\rm 2SC}$ is the gap parameter
for the 2SC (2-flavor, 2-color) BCS pairing 
obtained with $\delta\mu_I=0$ and $\Delta_I$ nonzero with the
other two gap parameters set to zero.    Assuming that $\Delta_{\rm CFL}\ll \mu$, the
2SC gap parameter
is given by
$\Delta_{\rm 2SC}= 2^{\frac{1}{3}}\Delta_{\rm CFL}$~\cite{Schafer:1999fe},
see Sec.~\ref{sec:QCD}.
In the Ginzburg-Landau approximation, in which the $\Delta_I$ are assumed
small, we must  first minimize the quadratic contribution to the
free energy, and only then  investigate the 
quartic and sextic contributions.  
Minimizing $\alpha_I$ fixes the length
of all the vectors in the set $\{{\bf q}_I\}$, and eliminates the possibility
of waves at higher harmonics,
yielding
$q_I = \eta\, \delta\mu_I$ with $\eta=1.1997$ the solution to
$\frac{1}{2\eta}\log\left[(\eta+1)/(\eta-1)\right]=1$~\cite{Alford:2000ze}. 
Upon setting $q_I=\eta\,\delta\mu_I$, (\ref{AlphaEqn}) becomes
\begin{equation}
\alpha_I(\delta\mu_I) 
=-\frac{1}{2}
\log\left(\frac{\Delta_{\rm 2SC}^2}{4 \delta\mu_I^2 (\eta^2-1)}\right)\ .
\label{AlphaEqn2}
\end{equation}
Once the $q_I$ have been fixed, the only dimensionful
quantities on which the quartic and sextic coefficients can depend are 
the $\delta\mu_I$~\cite{Bowers:2002xr,Rajagopal:2006ig},
meaning that for exchange symmetric crystal structures and with $\delta\mu_2=\delta\mu_3=\delta\mu$
we have 
$\beta=\bar\beta/\delta\mu^2$, $\beta_{32}=\bar\beta_{32}/\delta\mu^2$, 
$\gamma=\bar\gamma/\delta\mu^4$ and $\gamma_{322}=\bar\gamma_{322}/\delta\mu^4$
where the barred quantities are dimensionless numbers which depend only on
$\{\hat{\bf q}_2\}$ and $\{\hat{\bf q}_3\}$ that must  be evaluated for each crystal structure.
Doing so
requires
evaluating one-loop Feynman diagrams with 4 or 6 insertions of $\Delta_I$'s. Each insertion
of $\Delta_I$ ($\Delta_I^*$) adds (subtracts) momentum $2\q{I}{a}$ for some $a$. 
The vector sum of all these external momenta inserted into a given one-loop 
diagram must vanish,
meaning that
the calculation consists of a bookkeeping task (determining which combinations of
4 or 6 $\q{I}{a}$'s selected from the sets $\setq{I}{}$ satisfy this 
momentum-conservation constraint) 
that grows rapidly in complexity with the complexity of the crystal structure,
and a loop integration that is nontrivial because the momentum in the propagator
changes after each insertion. In~\cite{Rajagopal:2006ig}, this calculation
is carried out explicitly for 11 crystal structures in the mean-field NJL model 
of Sec.~\ref{sec:NJLModel} upon
making the weak coupling ($\Delta_{\rm CFL}$ and $\delta\mu$ both much less than $\mu$)
approximation. Note that in this approximation
neither the NJL cutoff nor the NJL coupling constant appear in any quartic
or higher Ginzburg-Landau coefficient, and as we have seen
above they appear in $\alpha$ only 
within  $\Delta_{\rm CFL}$.
Hence, the details of the model do not matter as long as one thinks of 
$\Delta_{\rm CFL}$ as a parameter, kept $\ll \mu$.

It is easy to show that  for exchange symmetric crystal structures any
extrema of $\Omega(\Delta_2,\Delta_3)$ in $(\Delta_2,\Delta_3)$-space
must either have $\Delta_2=\Delta_3=\Delta$,
or have one of $\Delta_2$ and $\Delta_3$ vanishing~\cite{Rajagopal:2006ig}.  
It is also possible to show that 
the three-flavor crystalline phases with 
$\Delta_2=\Delta_3=\Delta$ are electrically neutral whereas two-flavor solutions
in which only one of the $\Delta$'s is nonzero are not~\cite{Rajagopal:2006ig}.  
We therefore analyze only solutions with $\Delta_2=\Delta_3=\Delta$.
We find that $\Omega(\Delta,\Delta)$ is positive for large $\Delta$
for all the crystal structures that have been  investigated to 
date~\cite{Rajagopal:2006ig}.\footnote{This is
in marked contrast with what happens with only two flavors (and upon
ignoring the requirement of neutrality.)  in that context,
many crystal structures have negative $\gamma$ and hence
sextic order free energies that are unbounded from below~\cite{Bowers:2002xr}.
} 
This allows us to  minimize $\Omega(\Delta,\Delta)$ with respect to $\Delta$,
thus evaluating $\Delta$ and $\Omega$.

We begin with the simplest three-flavor  ``crystal'' structure in which
$\setq{2}{}$ and $\setq{3}{}$ each contain only a single vector, making
the $\langle us\rangle$ and $\langle ud\rangle$ condensates 
each a single plane wave~\cite{Casalbuoni:2005zp}.  We call this the 2PW phase.
Unlike in the more realistic
crystalline phases we describe below, in this ``crystal''
the magnitude of the $\langle ud\rangle$ and $\langle us \rangle$
condensates are unmodulated.
This simple condensate nevertheless
yields a qualitative lesson which proves helpful in winnowing the space
of multiple plane wave crystal structures~\cite{Rajagopal:2006ig}.
For this simple ``crystal'' structure, 
all the coefficients in the Ginzburg-Landau
free energy can be evaluated analytically~\cite{Casalbuoni:2005zp,Mannarelli:2006fy,Rajagopal:2006ig}. The 
terms that occur in the three-flavor case but not in the two-flavor case,
namely $\bar\beta_{32}$ and $\bar\gamma_{322}$, describe the
interaction between the two condensates, and depend on the angle
$\phi$ between $\q{2}{}$ and $\q{3}{}$. 
For any angle $\phi$, both $\bar{\beta}_{32}$ and
$\bar{\gamma}_{322}$ are
positive.
And, both increase monotonically with $\phi$ and
diverge as $\phi\rightarrow\pi$
This divergence 
tells us that choosing $\q{2}{}$ and
$\q{3}{}$ precisely antiparallel exacts an infinite free energy price
in the combined Ginzburg-Landau and weak-coupling
limit in which $\Delta\ll \delta\mu,\Delta_{\rm CFL}\ll \mu$,
meaning that in this limit if we chose $\phi=\pi$ we find $\Delta=0$.
Away from the Ginzburg-Landau limit, when the pairing rings
on the Fermi surfaces widen into bands, choosing $\phi=\pi$ exacts a finite
price meaning that $\Delta$  is nonzero but smaller than that for any other
choice of $\phi$.  
The high cost of choosing $\q{2}{}$ and
$\q{3}{}$ precisely antiparallel can be understood 
qualitatively as arising from the fact that in this case the
ring of states on the $u$-quark Fermi surface that ``want to'' pair
with $d$-quarks coincides precisely with the ring that ``wants to''
pair with $s$-quarks~\cite{Mannarelli:2006fy}.  
This simple two plane wave ansatz has been
analyzed upon making
the weak-coupling approximation but without
making the Ginzburg-Landau approximation~\cite{Mannarelli:2006fy}.
All the qualitative lessons learned from the Ginzburg-Landau approximation
remain valid
and we learn further that 
the Ginzburg-Landau 
approximation always  underestimates $\Delta$~\cite{Mannarelli:2006fy}.

The analysis of the simple two plane wave ``crystal'' structure, together with
the observation that in more complicated crystal structures with more than
one vector in $\setq{2}{}$ and $\setq{3}{}$ the Ginzburg-Landau coefficient
$\beta_{32}$ ($\gamma_{322}$) is given in whole (in part) by a sum of many 
two plane wave contributions,
yields one of two
rules for constructing favorable crystal structures for three-flavor
crystalline color superconductivity~\cite{Rajagopal:2006ig}:
$\setq{2}{}$ and $\setq{3}{}$ should be rotated with respect to each other
in a way that best keeps vectors in one set away from the antipodes
of vectors in the other set.  
The second rule is that 
the sets $\setq{2}{}$ and $\setq{3}{}$ should each
be chosen to yield crystal structures which, seen
as separate two-flavor crystalline phases, are as favorable as possible.
The 11 crystal structures analyzed in~\cite{Rajagopal:2006ig}
allow one to
make several pairwise
comparisons that test these two rules.
There are instances of two structures which differ only in the
relative orientation of $\setq{2}{}$ and $\setq{3}{}$ and in these cases the
structure in which vectors in $\setq{2}{}$ get closer to the antipodes
of vectors in $\setq{3}{}$ are disfavored. And, there are instances
where the smallest angle between a vector in $\setq{2}{}$ and the antipodes
of a vector in $\setq{3}{}$ are the same for two different crystal structures,  and
in these cases the one with the more favorable two-flavor structure is more favorable.
These considerations, together with explicit calculations, indicate that
two structures, which we denote ``2Cube45z'' and ``CubeX'', are particularly favorable.

In the 2Cube45z crystal, $\setq{2}{}$ and $\setq{3}{}$ each contain eight
vectors pointing at the corners of a cube. If we orient $\setq{2}{}$ so that its
vectors point in the $(\pm 1,\pm 1,\pm 1)$ directions in momentum space, then $\setq{3}{}$ is rotated
relative to $\setq{2}{}$ by $45^\circ$ about the $z$-axis.   In this crystal structure,
the $\langle ud \rangle$ and $\langle us \rangle$ condensates are each
given by the most favored 
two-flavor crystal structure~\cite{Bowers:2002xr}.  The relative rotation
maximizes the separation between any vector in
$\setq{2}{}$ and the nearest antipodes of a vector in $\setq{3}{}$.

\begin{figure}[t]
 \begin{center}
 \includegraphics[width=\hsize,angle=0]{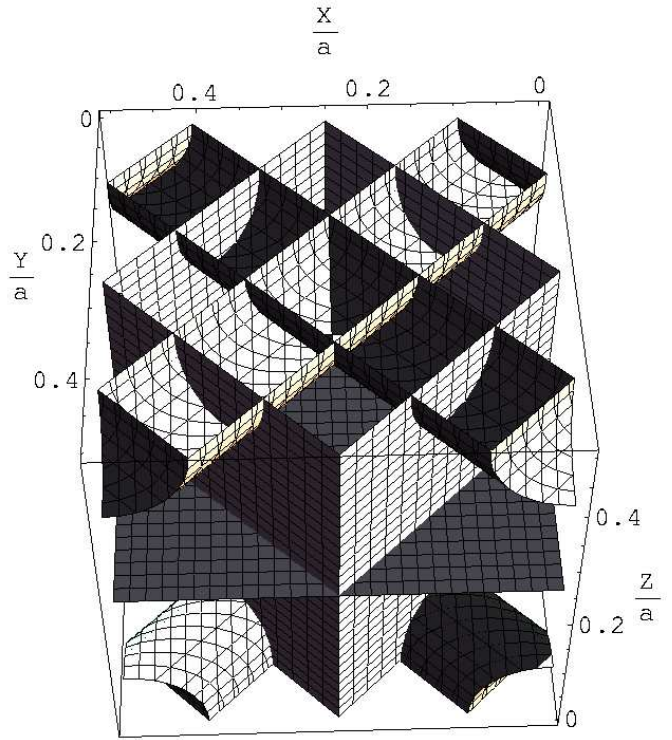}     
\end{center}
\caption{{ The CubeX crystal structure of Eq.~(\ref{CubeXStructure}). 
The figure extends from 0 to $a/2$ in the $x$, $y$ and $z$ directions.
Both $\Delta_2(\rr)$ and $\Delta_3(\rr)$
vanish at the horizontal plane.  $\Delta_2(\rr)$ vanishes on the darker vertical planes,
and $\Delta_3(\rr)$ vanishes on the lighter vertical planes.   
On the upper
(lower) dark cylinders and the lower (upper) two small corners of dark cylinders,
$\Delta_2(\rr)=  +3.3 \Delta$ ($\Delta_2(\rr)=  -3.3 \Delta$).
On the upper
(lower) lighter cylinders and the lower (upper) two small corners of lighter cylinders,
$\Delta_3(\rr)=  -3.3 \Delta$ ($\Delta_3(\rr)=  +3.3 \Delta$).  
The largest value of $|\Delta_I(\rr)|$ is $4\Delta$, occurring along lines 
at the centers of the cylinders.
The lattice spacing is $a$ when one takes into
account the signs of the condensates; 
if one looks only at $|\Delta_I(\rr)|$, 
the lattice spacing is $a/2$. 
}
\label{fig:CubeXContours}}
\end{figure}

We arrive at the CubeX structure by reducing the number
of vectors in $\setq{2}{}$ and $\setq{3}{}$.  This worsens the two-flavor
free energy of each condensate separately, but allows vectors in 
$\setq{2}{}$ to be kept farther away from the antipodes of vectors
in $\setq{3}{}$.  We have not analyzed all structures
obtainable in this way, but we have found one and only one which
has a condensation energy comparable to that of the 2Cube45z structure.
In the CubeX structure,
$\setq{2}{}$ and $\setq{3}{}$ each contain four vectors forming a rectangle. The eight 
vectors together point toward the corners of a cube. The 2 rectangles intersect to look like an ``X'' 
if viewed end-on.  
The color, flavor and position space dependence of 
the CubeX condensate is given by
\begin{eqnarray}
\epsilon_{2\alpha\beta}\epsilon_{2ij} 
\Biggl[ 
\cos \frac{2\pi}{a} \left( x+y+z\right) &+& \cos \frac{2\pi}{a}\left(-x-y+z\right) \Biggr]+\nonumber\\
 \epsilon_{3\alpha\beta}\epsilon_{3ij} 
 \Biggl[
\cos \frac{2\pi}{a} \left( -x+y+z\right) &+& \cos \frac{2\pi}{a}\left(x-y+z\right) \Biggr],
\label{CubeXStructure}
\end{eqnarray}
where
$a = \sqrt{3}\pi/q = 4.536/\delta\mu = 36.29 \mu /M_s^2$
is the lattice spacing.  For
example, with $M_s^2/\mu=100, 150, 200$~MeV the lattice
spacing is  $a=72, 48, 36$~fm. 
We depict this condensate in Fig.~\ref{fig:CubeXContours}.

\begin{figure}[t]
\begin{center}
\includegraphics[width=\hsize,angle=0]{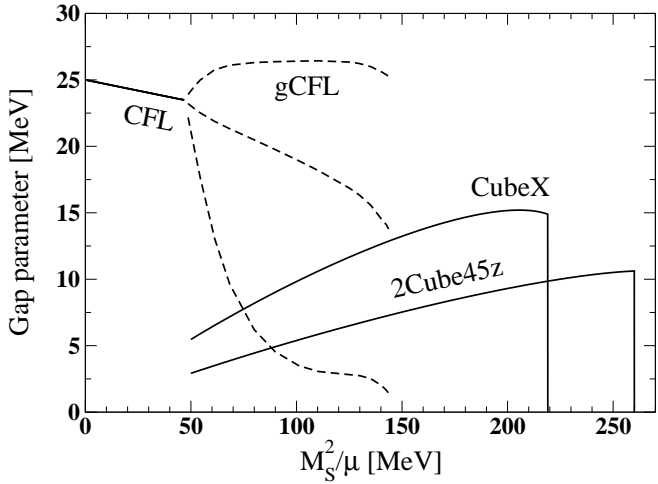}
\vspace{-0.2in}
\end{center}
\caption{Gap parameter $\Delta$ versus $M_s^2/\mu$ for: the CFL gap parameter
(set to 25~MeV at $M_s^2/\mu=0$), the three gap
parameters $\Delta_1<\Delta_2<\Delta_3$ describing $\langle ds\rangle$,
$\langle us\rangle$ and $\langle ud\rangle$ pairing in the gCFL phase, and the
gap parameters in the
crystalline color superconducting phases with CubeX and 2Cube45z  crystal
structures. 
Increasing $M_s^2/\mu$ corresponds to decreasing density.
}
\label{fig:deltavsx}
\end{figure}

In Figs.~\ref{fig:deltavsx} and \ref{fig:energy},
we plot $\Delta$
and $\Omega$ versus $M_s^2/\mu$ 
for the most favorable crystal structures that we have found, namely the
CubeX and 2Cube45z structures described above.
We have taken the CFL gap parameter $\Delta_{\rm CFL}=25$~MeV in these figures,
but they can easily be rescaled to any value of $\Delta_{\rm CFL}\ll \mu$~\cite{Rajagopal:2006ig}:
if the $\Delta$ and $M_s^2/\mu$ axes are rescaled by $\Delta_{\rm CFL}$ and the energy
axis is rescaled by $\Delta_{\rm CFL}^2$.
Fig.~\ref{fig:deltavsx} shows that the gap parameters 
are large enough that the Ginzburg-Landau approximation is
at the edge of its domain of reliability.
However, results obtained for the simpler 2PW crystal structures
suggest that the Ginzburg-Landau calculation underestimates $\Delta$ and the
condensation energy and that, even when
it breaks down, it is a good qualitative guide
to the favorable structure~\cite{Mannarelli:2006fy}.  We therefore trust the
result, evident in Fig.~\ref{fig:energy}, that these crystalline phases are both
impressively robust, with one or other of them favored over a wide swath of
$M_s^2/\mu$ and hence density. We do not trust the Ginzburg-Landau calculation
to discriminate between these two 
structures, particularly given that although we have a qualitative understanding
of why these two are favorable we have no qualitative argument for why one should
be favored over the other. We are confident that 2Cube45z
is the most favorable structure obtained by rotating one 
cube relative to another. We are not as confident that CubeX is the best possible
structure with fewer than 8+8 vectors. Regardless, the 2Cube45z and CubeX crystalline
phases together make the case
that three-flavor crystalline color superconducting phases are the ground
state of cold quark matter over a wide range of  densities.  If 
even better crystal structures can be found, this will only further strengthen this case.

Fig.~\ref{fig:energy} shows that over most of the range of $M_s^2/\mu$ where it was once considered
a possibility, the gCFL phase can be replaced by a {\it much} more favorable three-flavor
crystalline color superconducting phase.  
We find that the two most favorable crystal structures
have large condensation energies, easily 1/3 to 1/2 of that in the CFL phase
with $M_s=0$, which is  $3\Delta_{\rm CFL}^2\mu^2/\pi^2$.  This is at first surprising, given
that the only quarks that pair are those lying on rings on
the Fermi surfaces, whereas in the CFL phase with $M_s=0$ pairing
occurs over the entire $u$, $d$ and $s$ Fermi surfaces.  It can to a degree
be understood qualitatively once we recall that there are in fact many rings,
and note that as $\Delta$ increases, the pairing rings spread into bands
on the Fermi surfaces, and for $\Delta$ as large as that we find to be favored
these bands have expanded and filled in, becoming many ``polar caps" on 
the Fermi surfaces~\cite{Mannarelli:2006fy}. 
In addition to
being free-energetically favorable, these crystalline phases are, as far as we know, stable: they
do not suffer from the chromomagnetic 
instability~\cite{Giannakis:2005vw,Giannakis:2005sa,Ciminale:2006sm,Gatto:2007ja} 
and they are also stable
with respect to kaon condensation~\cite{Anglani:2007aa}. 
In simplified analogue contexts, it has even been possible to trace the path
in configuration space from the unstable gapless phase (analogue of gCFL)
downward in free energy  to the stable crystalline phase~\cite{Fukushima:2006su,Fukushima:2007bj}.

Fig.~\ref{fig:energy} also
shows that it is hard to find a crystalline
phase with lower free energy than the gCFL phase at the lower values
of $M_s^2/\mu$ (highest densities) within the ``gCFL window''.  
At these densities, however, the calculations described in Sec.~\ref{sec_eft}
demonstrate that the gCFL phase is superseded by the stable CFL-$K^0$
and curCFL-$K^0$ phases, as shown in Fig.~\ref{fig:energy}.

The three-flavor crystalline color superconducting phases with 
CubeX and 2Cube45z crystal structures are
the lowest free energy phases that we know of, and hence
candidates for the ground state of QCD, over a wide range
of densities.   Within the Ginzburg-Landau approximation
to the NJL model that
we have employed, one or other is favored over the CFL, gCFL and
unpaired phases for $2.9 \Delta_{\rm CFL} < {M_s^2}/{\mu} < 10.4 \Delta_{\rm CFL}$,
as shown in Fig.~\ref{fig:energy}.
For $\Delta_{\rm CFL}=25$~MeV and $M_s=250$~MeV,
this translates to $240 {\rm MeV} < \mu < 847 {\rm MeV}$.
With these choices of parameters, the 
lower part of this range of $\mu$ (higher part of the range of $M_s^2/\mu$) is certainly
superseded by nuclear matter.  And, the high end of this range extends 
beyond the $\mu\sim 500$~MeV characteristic of the quark matter at the densities
expected 
at the center of neutron stars. This qualitative feature persists
in the analysis of \cite{Ippolito:2007uz} in which $M_s$ is solved for
rather than taken as a parameter.
If  neutron stars
do have quark matter cores, then, it is reasonable to 
include 
the possibility that 
the {\it entire}
quark matter core could be in a crystalline color superconducting phase
on the menu of options
that must ultimately be winnowed by confrontation with astrophysical observations.
(Recall, that if $\Delta_{\rm CFL}$ is larger, say $\sim 100$~MeV, the entire quark matter
core could be in the CFL phase.)    
As we shall see in the next subsection, 
crystalline color superconducting quark
matter is rigid, with a very large shear modulus, while at the same time being
superfluid.  
This provides  a possible origin
for pulsar glitches, as we shall discuss in Sec.~\ref{sec:astro_glitches}

\subsection{Rigidity of crystalline color superconducting quark matter}
\label{sec:rigid}

The crystalline phases of color superconducting quark matter that we 
have described in the previous subsection
are unique among all forms of dense matter that may arise within
neutron star cores in one respect: they are rigid~\cite{Mannarelli:2007bs}.  
They are not solids in the usual sense: the quarks
are not fixed in place at the vertices of some crystal structure.
Instead, in fact, these phases are superfluid since the
condensates all spontaneously break the $U(1)_B$ symmetry
corresponding to quark number.
We shall always write the condensates as real.  This choice
of overall phase breaks $U(1)_B$, and spatial
gradients in this phase correspond to supercurrents.
And yet, we shall see that
crystalline color superconductors
are rigid solids with large shear moduli. 
The diquark condensate,
although spatially inhomogeneous, can carry 
supercurrents~\cite{Alford:2000ze,Mannarelli:2007bs}.
It is the spatial modulation of the gap parameter that breaks translation
invariance, as depicted for the CubeX phase in Fig. \ref{fig:CubeXContours}, 
and it is this pattern
of modulation that is rigid.\footnote{  
Supersolids~\cite{Andreev:1969,Chester:1970,Leggett:1970,Chan:2004,Chan2:2004,Son:2005ak} 
are another example of rigid superfluids,
but  they differ from crystalline color superconductors 
in that they
are rigid by virtue of the
presence of an underlying lattice of atoms.
}
This novel form of rigidity may sound tenuous
upon first hearing, but we shall present
the effective Lagrangian
that describes the phonons in the CubeX and 2Cube45z crystalline
phases, whose lowest order coefficients have  
been calculated in the NJL model that we are employing~\cite{Mannarelli:2007bs}.  
We shall then extract the shear moduli from the phonon
effective action, quantifying the rigidity and
indicating the presence of transverse phonons.
The fact that the crystalline phases are simultaneously rigid and superfluid
means that their presence within neutron star
cores has potentially observable consequences, as we shall describe
in Sec.~\ref{sec:astro_glitches}.

The shear moduli of a crystal may be extracted from the effective Lagrangian
that describes phonons in the crystal, namely space- and time-varying displacements
of the crystalline pattern.  Phonons in two-flavor crystalline
phases were first investigated in \cite{Casalbuoni:2002hr,Casalbuoni:2002my}.
In the present context, we introduce displacement fields
for 
the $\langle ud \rangle$,
$\langle us \rangle$ and $\langle ds \rangle$  condensates 
by making the replacement
\begin{equation}
\Delta_I \sum_{\q{I}{a}}e^{2i\q{I}{a}\cdot\rr} \rightarrow
\Delta_I \sum_{\q{I}{a}}e^{2i\q{I}{a}\cdot(\rr - \vu_I(\rr))}
\label{displacementfields}
\end{equation}
in (\ref{precisecondensate}).  
One way to obtain the effective action describing the dynamics of the
displacement fields $\vu_I(\rr)$, 
including both its form and the values of its coefficients
within the NJL model that we are employing, is to begin with the NJL model
of Sec.~\ref{sec:NJLModel} but with (\ref{displacementfields}) 
and integrate out the fermion fields.  
After a lengthy calculation~\cite{Mannarelli:2007bs}, this yields
\begin{equation}
\begin{split}
&S[{\bf u}]=
\ha\intspace{x}\sum_I \kappa_I \\
&\times\Biggl[
  \left(
    \sum_{\qia}(\hat{q}_I^a)^m(\hat{q}_I^a)^n \right)(\partial_0 u_I^m)(\partial_0 u_I^n) \\
 &-\left(
    \sum_{\qia}(\hat{q}_I^a)^m(\hat{q}_I^a)^v(\hat{q}_I^a)^n(\hat{q}_I^a)^w\right)
    (\partial_v u_I^m)(\partial_w u_I^n)
 \Biggr]\label{Seff4}\;
\end{split}
\end{equation}
where $m$, $n$, $v$ and $w$ are spatial indices running over $x$, $y$ and $z$
and where we have defined
\begin{equation}
\kappa_I\equiv
\frac{2\mu^2|\Delta_I|^2\eta^2}{\pi^2(\eta^2-1)}\;.
\label{lambdapw}
\end{equation}
Upon setting $\Delta_1=0$ and $\Delta_2=\Delta_3=\Delta$,
\begin{equation}
\kappa_2=\kappa_3\equiv\kappa=\frac{2\mu^2|\Delta|^2\eta^2}{\pi^2(\eta^2-1)}
\simeq 0.664\,\mu^2|\Delta^2|\;.\label{kappavalue}
\end{equation}
$S[{\bf u}]$  is the low energy effective action for phonons in any crystalline color
superconducting phase, valid to second order in derivatives, to
second order in the gap parameters $\Delta_I$ and to second order in
the phonon fields $\vu_I$.    Because we are interested in long wavelength,
small amplitude, phonon excitations, expanding to second order in derivatives
and in the phonon fields is satisfactory. More
complicated terms will arise at higher order, for example
terms that couple the different $\vu_I$'s, but it is legitimate
to neglect these complications~\cite{Mannarelli:2007bs}.  Extending this
calculation to higher order in the Ginzburg-Landau approximation would 
be worthwhile, however, since as we saw in Sec.~\ref{subsec:NJLCrystalline} this
approximation is at the edge of its domain of reliability.  

In order to extract the shear moduli, we need to compare the phonon
effective action to the general theory of elastic media~\cite{Landau:elastic},
which requires introducing the strain tensor
\begin{equation}
s_I^{mv}\equiv\ha\Bigl(\frac{\partial  u_I^m}{\partial
x^v}+\frac{\partial  u_I^v}{\partial x^m}\Bigr).\label{strain}
\end{equation}
We then wish to compare the action (\ref{Seff4}) to
\begin{eqnarray}
{\cal S}[{\bf u}]&=& \ha\intspace{x}\Biggl(
     \sum_I\sum_m \rho_I^m (\partial_0  u_I^m)(\partial_0  u_I^m)\nonumber\\
    &\quad& \quad\qquad\qquad -\sum_{I}\sum_{{mn}\atop{vw}}\lambda_I^{mvnw}
     s_I^{mv}s_I^{nw}\Biggr)
\label{full action},
\end{eqnarray}
which is the general form of the action in the case in which the effective
action is quadratic in displacements and 
which defines the elastic modulus tensor $\lambda_I^{mvnw}$ for this case.
In this case, the stress tensor 
(in general the derivative of the potential energy with respect to $s_I^{mv}$) is given
by
\begin{equation}
\sigma_I^{mv}
=
\lambda_I^{mvnw}s_I^{nw}\label{stress2}\; .
\end{equation}
The diagonal components of $\sigma$ are proportional to the
compression exerted on the system and are therefore related to  the
bulk  modulus of the crystalline color superconducting
quark matter. Since unpaired quark
matter  has a pressure $\sim \mu^4$, it gives a contribution to the
bulk modulus that  completely overwhelms the contribution from the
condensation into a crystalline phase, which is of order
$\mu^2\Delta^2$.  We shall therefore not calculate the
bulk modulus.  On the other hand, the response to shear
stress arises only because of the presence  of the crystalline condensate.
The shear modulus is defined as follows. Imagine exerting a
static external stress $\sigma_I$ having only an off-diagonal
component, meaning  $\sigma^{mv}_I\neq 0$ for a pair of space
directions $m\neq v$, and all the other components of $\sigma$ are
zero. The system will respond with a strain $s_I^{nw}$. 
The shear modulus in the $mv$ plane is then
\begin{equation}
\nu_I^{mv} \equiv \frac{\sigma_I^{mv}}{2s_I^{mv}} 
= \ha\lambda_I^{mvmv}
\label{defineshearmodulus}\;,
\end{equation}
where the indices $m$ and $v$ are not summed.  
For a general quadratic potential
with $\sigma_I^{mv}$ given by (\ref{stress2}), $\nu_I^{mv}$ simplifies
partially but the full simplification given by the last equality in  
(\ref{defineshearmodulus}) only arises for special cases in which 
the only nonzero entries in $\lambda^{mvnw}$ with $m\neq v$ are
the $\lambda^{mvmv}$ entries, as is the case for all the crystal
structures that we consider.



For a given crystal structure, upon evaluating the sums in (\ref{Seff4}) and
then using the definition (\ref{strain}) to compare (\ref{Seff4}) to
(\ref{full action}), we can extract expressions for the $\lambda$ tensor
and thence for the shear moduli.  
This analysis, described in detail in \cite{Mannarelli:2007bs},
shows that in the CubeX phase
\begin{equation}
\nu_2=\frac{16}{9}\kappa\left( \begin{array}{ccc}
0 & 0 & 1\\
0 & 0 & 0\\
1 & 0 & 0
\end{array}
\right)\,,\hspace{.3cm}  \nu_3=\frac{16}{9}\kappa\left(
\begin{array}{ccc}
0 & 0 & 0\\
0 & 0 & 1\\
0 & 1 & 0
\end{array}
\right)\label{nu2 and nu3}\;,
\end{equation}
while in the 2Cube45z phase
\begin{equation}
\nu_{2}=\frac{16}{9}\kappa\left( \begin{array}{ccc}
0 & 1 & 1\\
1 & 0 & 1\\
1 & 1 & 0
\end{array}
\right)\,,\ \
\nu_{3}=\frac{16}{9}\kappa\left( \begin{array}{ccc}
0 & 0 & 1\\
0 & 0 & 1\\
1 & 1 & 0
\end{array}
\right)\label{nu 2Cube45z}\;.
\end{equation}
We shall see in Sec.~\ref{sec:astro_glitches} that it is relevant to check
that both these crystals have enough 
nonzero entries in their shear moduli
$\nu_I$ that if there are rotational vortices are pinned within them, 
a force seeking to 
move such a vortex is opposed by the rigidity 
of the crystal structure described
by one or more of the nonzero entries in the $\nu_I$.  
This is demonstrated
in \cite{Mannarelli:2007bs}.

We see that all the nonzero shear moduli of both the CubeX and 2Cube45z
crystalline color superconducting phases turn out to take on the same value,
\begin{equation}
\nu_{\rm CQM} = \frac{16}{9}\kappa
\end{equation}
with $\kappa$ defined by (\ref{kappavalue}).
Evaluating $\kappa$ yields
\begin{eqnarray}
\nu_{\rm CQM} 
&=& 1.18\, \mu^2 \Delta^2\nonumber\\
&=&  2.47\, \frac{{\rm MeV}}{{\rm fm}^3}
\left(\frac{\Delta}{10~{\rm MeV}}\right)^2 \left(\frac{\mu}{400~\rm{MeV}}\right)^2.
\label{shearmodulus}
\end{eqnarray}
From (\ref{shearmodulus}) we first of all see that the shear modulus
is in no way suppressed relative to the scale  $\mu^2\Delta^2$ that could have
been guessed on dimensional grounds.  And, second, we discover that
a quark matter core in a crystalline color superconducting phase is 20 to 1000
times more rigid than the crust of a conventional neutron 
star~\cite{Strohmayer:1991,Mannarelli:2007bs}.
Finally, see \cite{Mannarelli:2007bs} for the 
extraction of the phonon dispersion relations from the effective 
action~(\ref{Seff4}).
The transverse phonons, whose restoring force is provided by the shear 
modulus and which correspond to propagating ripples in
a condensation pattern like that in Fig.~\ref{fig:CubeXContours},
turn out to have direction-dependent velocities that are 
typically a substantial
fraction of the speed of light, in the specific instances evaluated in \cite{Mannarelli:2007bs}
being given by $\sqrt{1/3}$ and $\sqrt{2/3}$.
This is yet a third way of seeing that this 
superfluid phase of matter is rigid indeed.

\section{Transport properties and neutrino processes}
\label{sec_trans}

In Sec.~\ref{sec:astro} we shall discuss how the observation of 
neutron star properties constrains the phase structure of dense quark 
matter. A crucial ingredient in these analyses are the transport 
properties as well as neutrino emissivities and opacities of different 
phases of quark matter. 

Using the methods introduced in Sec.~\ref{sec_eft} 
it is possible to perform rigorous calculations of transport properties 
of the CFL phase. The results are parameter free predictions of QCD at 
asymptotically large density, and rigorous consequences of QCD expressed 
in terms of a few phenomenological parameters ($f_\pi$, $m_\pi$, $\ldots$) 
at lower density. 

In the case of other color superconducting phases we
perform calculations using perturbative QCD or models of QCD. For many
quantities the results depend mainly on the spectrum of quark modes, 
and not on details of the quark-quark interaction. 

\subsection{Viscosity and thermal conductivity}
\label{sec_visc}

Viscosity and thermal conductivity determine the dissipated energy $\dot{E}$
in a fluid with nonzero gradients of the velocity ${\bf v}$ and the 
temperature $T$,  
\bea
\label{edot}
\dot{E} &=& -\frac{\eta}{2} \int d^3x\,  
  \left(\partial_iv_j+\partial_jv_i-\frac{2}{3}\delta_{ij}
      \partial_k v_k \right)^2  \nonumber \\
 & & 
   - \zeta \int d^3x \, \big( \partial_iv_i\big)^2 
   - \frac{\kappa}{T} \int d^3x \, \big( \partial_i T\big)^2 
  \, . 
\eea
The transport coefficients $\eta$, $\zeta$ and $\kappa$ are the 
shear and bulk viscosity and the thermal conductivity, respectively. 
Eq.\ (\ref{edot}) is strictly 
valid only for non-relativistic fluids. In the 
case of relativistic fluids there is an extra
contribution to the dissipated energy which is 
proportional to $\kappa$ and the gradient of 
$\mu$ \cite{Landau:Fluid}. In
terms of its hydrodynamic properties a superfluid can be 
viewed as a mixture of a normal and a superfluid component
characterized by separate flow velocities. The shear viscosity 
is entirely due to the normal component, but there are 
contributions to the bulk viscosity which are related to 
stresses in the superfluid flow relative to the normal one \cite{Gusakov:2007px,Andersson:2006nr,khala}. 
In the following we shall neglect these effects and interpret
$v_i$ in Eq.~(\ref{edot}) as the normal fluid velocity. 

 In neutron stars an important contribution to the bulk viscosity 
arises from electroweak effects. In a bulk compression mode the 
density changes periodically and electroweak interactions may 
not be sufficiently fast to reestablish weak equilibrium. Weak 
effects occur on the same time scale as the oscillation period
of the neutron star and the frequency dependence of the bulk
viscosity is important. We define
\beq 
\label{defbulk}
\zeta(\omega) =  2\left\langle\dot{E}\right\rangle
 \left(\frac{V_0}{\delta V_0}\right)^2\frac{1}{\omega^2} \, ,
\eeq
where $\omega$ is the oscillation frequency, $\langle \ldots \rangle$ 
is a time average, and $\delta V_0/V_0$ is the fractional change 
in the volume. The coefficient $\zeta$ in Eq.\ (\ref{edot}) is the $\omega\to 0$
limit of $\zeta(\omega)$. If a single weak process is responsible for reestablishing
chemical equilibrium, the frequency dependent 
bulk viscosity can be written in the form  
\beq 
\label{bulkfinal}
\zeta(\omega) = C \frac{\gamma}{\gamma^2+\omega^2} \, .
\eeq
The prefactor $C$ accounts for the dependence of the equilibrium densities 
(e.g., the net difference between the 
density of strange and non-strange quarks if the weak process changes 
strangeness) on 
the respective chemical potentials, and $\gamma$ is the characteristic inverse time
scale of the flavor changing process. Eq.~(\ref{bulkfinal}) shows that, 
for a given $\omega$, $\zeta$ has a maximum at $\gamma = \omega$.
At this point the time scale of the microscopic process matches 
the one of the external oscillation. If more than one weak process contributes to reequilibration,
Eq.~(\ref{bulkfinal}) becomes more 
complicated \cite{Haensel:2000vz,Alford:2006gy,Sa'd:2007ud}

\subsubsection{CFL phase}
\label{sec_visc_cfl}

\begin{figure}[t]
\begin{center} 
\includegraphics[width=\hsize]{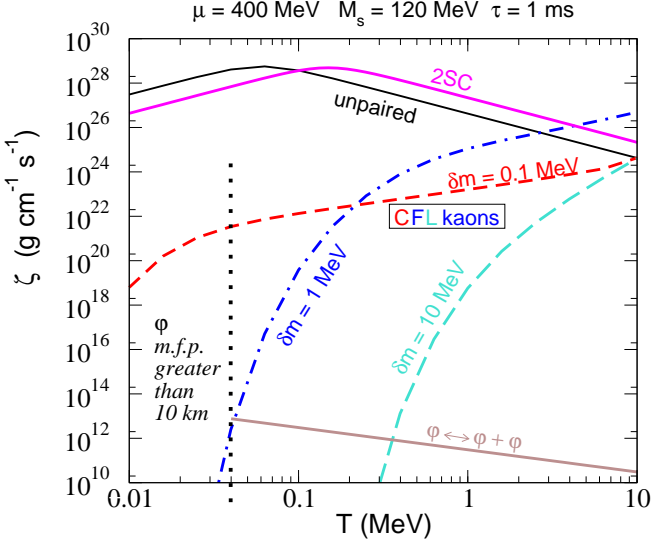}
\end{center}
\caption{(Color online) 
Bulk viscosities as functions of temperature for an oscillation period $\tau=2\pi/\omega=1$ ms.
CFL phase: contribution
from the process $K^0\leftrightarrow \varphi+\varphi$ for different values of 
$\delta m\equiv m_{K^0}-\mu_{K^0}$ and contribution 
from $\varphi\leftrightarrow\varphi+\varphi$, see Eq.\ (\ref{zeta_cfl}). 
2SC phase and unpaired quark matter:
contribution from the process $u+d\leftrightarrow u+s$.}
\label{fig:bulkCFL}
\end{figure}

 The normal fluid is composed of quasi-particle excitations. 
In the CFL phase all quark modes are gapped and the relevant 
excitations are Goldstone bosons. At very low temperature, transport 
properties are dominated by the massless Goldstone boson $\varphi$ 
associated with the breaking of the $U(1)_B$ symmetry. Using the 
results in Sec.~\ref{sec_eft_u1}, 
we can compute the mean free path $l_\varphi$ of 
the $\varphi$ due to $\varphi\leftrightarrow\varphi+\varphi$ and 
$\varphi+\varphi \leftrightarrow\varphi+\varphi$ scattering.  
Small angle scattering contributions give rise to  
$l_\varphi\propto\mu^4/T^5$ \cite{Manuel:2004iv} and $l_\varphi\simeq 1$ km at $T=0.1$ MeV, 
while large angle scattering contributions yield an even longer
$l_\varphi\propto\mu^8/T^9$  \cite{Shovkovy:2002kv}.
The thermal conductivity $\kappa$ due to the $\varphi$ is given by 
\cite{Shovkovy:2002kv}  
\beq 
\label{kappa_cfl}
\kappa = \frac{2\pi^2T^3}{45v^2}\, l_\varphi \, ,
\eeq
where $\ell_\varphi$ is the $\varphi$ mean-free path between
large angle scatterings and $v$ is the $\varphi$ velocity from Eqs.\ (\ref{u1_eft}) and
(\ref{fv}).
For temperatures below $\sim 1$ MeV the thermal conductivity 
is very large and macroscopic amounts of CFL matter are expected 
to be isothermal. The electric conductivity in CFL matter is dominated 
by thermal electrons and positrons and was estimated in \cite{Shovkovy:2002sg}.

At low temperatures, the shear viscosity of the CFL phase is
dominated by the $\varphi$ contribution, which
was computed in \cite{Manuel:2004iv} and is given by  
\beq
\label{eta_cfl}
\eta = 1.3\times 10^{-4} \, \frac{\mu^8}{T^5} \, . 
\eeq

The bulk viscosity 
$\zeta$ vanishes
in an exactly scale invariant system. For realistic quark masses
the dominant source of scale breaking is the strange quark mass. 
The contribution from the process $\varphi\leftrightarrow \varphi+\varphi$ 
is \cite{Manuel:2007pz}
\beq
\label{zeta_cfl}
\zeta=0.011 \frac{M_s^4}{T}  \, .
\eeq
We show this contribution in Fig.~\ref{fig:bulkCFL}. The other contribution to the CFL bulk viscosity
presented in the figure comes from the process $K^0\leftrightarrow \varphi+\varphi$ and 
was studied for arbitrary $\omega$ in \cite{Alford:2007rw}. We observe that 
at $T\simeq (1-10)$ MeV the bulk viscosity of CFL matter is 
comparable to that of unpaired quark matter. For $T<1$~MeV, $\zeta$
is strongly suppressed. Depending on the poorly known value for $\delta m\equiv m_{K^0}-\mu_{K^0}$
(here assumed to be positive, a negative value corresponds to kaon condensation), 
the pure $\varphi$ contribution
given in Eq.~(\ref{zeta_cfl}) may dominate over the $K^0\leftrightarrow 
\varphi+\varphi$ reaction at low enough temperatures. However, for $T<0.1$ MeV the $\varphi$ mean free 
path is on the order of the size of the star, i.e., the system 
is in the collisionless rather than in the hydrodynamic regime, and the 
result ceases to be meaningful. 
 
Thermal conductivity and viscosities for the CFL-$K^0$ phase have not yet 
been computed. The existence of a gapless $K^0$ Goldstone mode in this phase will introduce
new contributions. However, since the CFL results for
$\kappa$ and $\eta$ are already dominated by a gapless mode, namely
the $\varphi$, the modifications to these quantities are not expected
to be significant.  The modification to $\zeta$ will be more
significant, since the kaon contribution to this quantity
is already important in the CFL phase.

\subsubsection{Other phases}
\label{sec_visc_other}

\begin{figure}[t]
\begin{center} 
\includegraphics[width=\hsize]{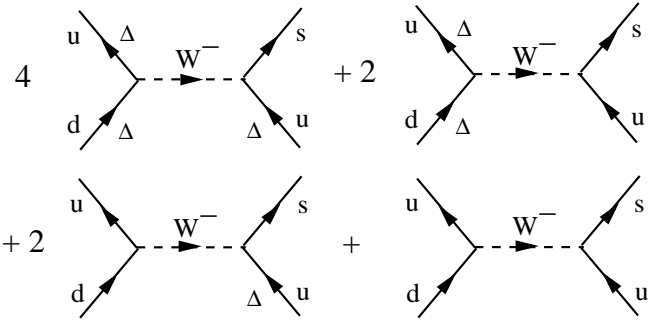}
\end{center}
\caption{Contributions to the process $u+d\to u+s$ in the 2SC phase. A 
gapped fermion is marked with the gap $\Delta$ at the respective line. 
(We have omitted (small) contributions from anomalous propagators.)}
\label{fig:diagrams}
\end{figure}

For unpaired, ultrarelativistic three-flavor quark matter, thermal and electric conductivity 
as well as shear viscosity have been computed in 
\cite{Heiselberg:1993cr}. In the low-temperature limit (in particular $T\ll m_D$ with the electric 
screening mass $m_D^2=N_fg^2\mu^2/(2\pi^2)$) they are 
\bea
\kappa &\simeq& 0.5\frac{m_D^2}{\alpha_s^2} \, , \qquad \sigma_{\rm el}\simeq 0.01
\frac{e^2\mu^2 m_D^{2/3}}{\alpha_s^2 T^{5/3}} \, ,
\eea
and
\bea
\eta &\simeq& 4.4\times 10^{-3}\,\frac{\mu^4m_D^{2/3}}{\alpha_s^2T^{5/3}} \, . 
\eea
These quantities have not yet been computed for partially gapped color superconductors
such as the 2SC phase. The presence of ungapped modes, however, suggests that the 
results only differ by a numerical factor from the unpaired phase results.

The dominant flavor changing process that contributes to the bulk viscosity in unpaired 
quark matter is the reaction \cite{Anand:1999bj,Madsen:1992sx}
\beq
\label{udus}
u+d \leftrightarrow u +s \, .
\eeq
Other relevant processes are the semi-leptonic processes $u+e \leftrightarrow 
d +\nu_e$ and $u+e \leftrightarrow s +\nu_e$ \cite{Sa'd:2007ud,Dong:2007mb}.

In a partially gapped phase the bulk viscosity is also dominated by the process (\ref{udus}). 
In the 2SC phase of three flavor quark 
matter, the number of $d$-quarks  produced per unit time and volume, $\Gamma$,
due to (\ref{udus}) can be computed from the diagrams shown in 
Fig.\ \ref{fig:diagrams}.
The combinatorical factors 
in front of the diagrams are obtained upon counting color degrees of freedom: one can 
attach one of three colors to each of the two weak vertices, giving rise to 9 possibilities.
In the 2SC phase all blue quarks and all strange 
quarks are unpaired while all other modes are paired, see Table \ref{table2SCCFL}. Consequently, 4 of the 9
possibilities contain three gapped modes (red or green for both vertices), 2 contain two gapped modes 
(red or green for one, blue for the other vertex), 2 contain one gapped
mode (blue for one, red or green for the other vertex), and one contains only unpaired modes (blue for both 
vertices). 
Therefore, at very low temperature, $T\ll\Delta$,   
where the contributions of gapped quarks are exponentially 
suppressed, 
$\Gamma$ is to a very good approximation given by \cite{Madsen:1999ci}
\beq 
\label{factor9}
\Gamma^{\rm 2SC} = \frac{1}{9}\Gamma^{\rm unp} \qquad 
   {\rm for} \;\; T\ll \Delta \, , 
\eeq
since only the one reaction containing only unpaired modes
contributes. The rate $\Gamma^{\rm unp}$ was computed in \cite{Madsen:1993xx}. 

For larger temperatures, the contribution from gapped modes cannot be 
neglected. Each diagram yields a contribution which (for one direction of 
the process) schematically reads
\bea
\Gamma &\propto&\sum_{\{e_i\}}\int_{\{p_i\}}{\cal F}\,
 \delta(e_1\varepsilon_1 +e_2\varepsilon_2-e_3\varepsilon_3-
    e_4\varepsilon_4+\delta\mu) \non
&&\times\; f(e_1\varepsilon_1)f(e_2\varepsilon_2) f(-e_3\varepsilon_3) 
    f(-e_4\varepsilon_4) \, .
\eea 
Here, $\varepsilon_i$ are the quasiparticle energies, $\delta\mu=\mu_s-\mu_d$, 
and $f$ is the Fermi distribution function. ${\cal F}$ is a function of the momenta 
$p_i$ and the signs $e_i=\pm 1$. The sum over the signs $e_i$ is 
very important in a paired system: the process $u+d\to u+s$ not 
only receives contributions from $2\to 2$ processes, but also
from $3\to 1$ and $1\to 3$ reactions involving pairs created or 
absorbed by the condensate. 

{}From the net production rate of $d$ quarks $\Gamma$ one obtains the characteristic inverse time
scale $\gamma$ needed for the bulk viscosity in Eq.~(\ref{bulkfinal}). For small external 
volume oscillations $\delta V_0/V_0$, 
$\Gamma$ is linear in the resulting oscillation in chemical potentials, 
$\Gamma=\lambda\,\delta\mu$. Then, $\gamma \equiv B \lambda$, where $B$ depends only  
on the equilibrium flavor densities. The resulting bulk viscosity as a function 
of temperature for a typical oscillation frequency 
$\omega/(2\pi)=1{\rm ms}^{-1}$ is shown in Fig.\ \ref{fig:bulkCFL}. A critical temperature
of $T_c=30$ MeV is assumed. For low temperatures, the time scale of the 
nonleptonic process is much smaller than the oscillation frequency
$\gamma\ll\omega$, implying $\zeta\propto \gamma$. Consequently, from 
Eq.\ (\ref{factor9}) we conclude $\zeta_{\rm 2SC}=\zeta_{\rm unp}/9$. 
For large temperatures, however, we have $\gamma\gg\omega$ and thus 
$\zeta\propto 1/\gamma$. Consequently, the superconducting phase, which 
has the slower rate, has the larger bulk viscosity. 

The bulk viscosity has also been computed for two-flavor quark matter 
with single-flavor pairing \cite{Sa'd:2006qv}. In this case there are also ungapped 
modes and thus the result is similar to the one of the 2SC phase. 
The main difference is the lower critical 
temperature for single-flavor pairing. As a consequence, these phases are unlikely 
to exist for temperatures larger than that at which the bulk viscosity of the unpaired phase
is maximal. Therefore, the bulk viscosity cannot be larger than that of the unpaired phase. 


\subsection{Neutrino emissivity and specific heat}
\label{sec_emis}
 
Neutrino emissivity determines the rate at which quark 
matter can loose heat via neutrino emission. For the purpose of studying 
how neutron stars
with ages ranging from tens of seconds to millions of
years cool, as we shall discuss in Sec.~\ref{sec:cooling}, it is appropriate
to treat the matter as completely transparent to the neutrinos
that it emits.

\subsubsection{CFL phase}
\label{sec_emis_cfl}
 
In CFL quark matter, all quasifermion modes are gapped and 
neutrino emissivity is dominated 
by reactions involving (pseudo)-Goldstone modes such as
\begin{subequations}
\bea
\pi^\pm,K^\pm &\to& e^\pm +\bar{\nu}_e \, , \\
\pi^0 &\to& \nu_e +\bar{\nu}_e \, , \\
\varphi + \varphi &\to& \varphi + \nu_e +\bar{\nu}_e \, .
\eea
\end{subequations}
These processes were studied in \cite{Jaikumar:2002vg,Reddy:2002xc}. 
The decay rates of the massive mesons $\pi^\pm$, $K^\pm$, and $\pi^0$ 
are proportional to their number densities and are suppressed by
Boltzmann factors $\exp(-E/T)$, where $E$ is the energy gap of the 
meson. Since the pseudo-Goldstone boson energy gaps are on the order 
of a few MeV, the emissivities are strongly suppressed as compared 
to unpaired quark matter for temperatures below this scale. Neutrino
emission from processes involving the $\varphi$ is not exponentially
suppressed, but it involves a very large power of $T$, 
\beq
\epsilon_\nu \sim  \frac{G_F^2 T^{15}}{f^2\mu^4}\, , 
\eeq
and is numerically very small. 
Reddy et al.~also studied the neutrino mean free path $l_\nu$. For 
$T\sim 30$ MeV the mean free path is on the order of 1 m, but 
for $T<1$~MeV, $l_\nu>10$ km \cite{Reddy:2002xc}. In the CFL-$K^0$ phase,
$l_\nu$ is almost the same as in the CFL phase, while the neutrino 
emissivity is larger \cite{Reddy:2003ap}.

The specific heat of CFL matter is also dominated by the $\varphi$, yielding  
\beq
\label{cv_cfl}
c_V = \frac{2\pi^2}{15 v^3} \, T^3 \, .
\eeq
This is much smaller than the specific heat of
any phase containing unpaired quarks, as we shall see below.

\subsubsection{Other phases}
\label{sec_emis_oth}

The density of thermally excited ungapped fermions is proportional to 
$\mu^2 T$ while that of ungapped bosons is $T^3$. This means that 
in any degenerate system ($T\ll \mu$) ungapped fermion modes, 
if they exist, will dominate the neutrino rates. In 
unpaired quark matter neutrino emissivity is dominated by the 
direct Urca processes
\begin{subequations} 
\label{directurca}
\bea 
u + e &\to & d + \nu_e \qquad \mbox{(electron capture)} \, , \\
d &\to & u + e + \bar{\nu}_e \qquad \mbox{($\beta$-decay)} \, .
\eea
\end{subequations}
The radiated energy per unit of time and volume is \cite{Iwamoto:1980eb}
\beq 
\label{emissivity}
\epsilon_\nu \simeq \frac{457}{630}\alpha_sG_F^2 T^6
   \mu_e\mu_u\mu_d  \, . 
\eeq
Note that this result is proportional to the strong coupling constant 
$\alpha_s$. The tree-level processes for massless quarks are 
approximately collinear and the weak matrix element vanishes in 
the forward direction. A nonzero emissivity arises from 
strong interaction corrections, which depress quark Fermi momenta
relative to their chemical potentials. Because they do not at
the same time depress the electron Fermi momentum, this opens
up phase space for the reactions (\ref{directurca}).
A nonzero emissivity can also arise from quark mass effects, or higher 
order corrections in $T/\mu$. Since strange quark decays are Cabbibo 
suppressed and $T/\mu$ is small the dominant contribution is 
likely to be that proportional to $\alpha_s$, namely (\ref{emissivity}). 
Note that we have not
included non-Fermi liquid corrections of $O(\alpha_s\log(T))$
\cite{Schafer:2004jp}. 

In order to determine the rate at 
which neutron stars cool we also need to know the specific heat. 
In unpaired quark matter 
\beq
\label{cv_unp} 
 c_V = \frac{N_cN_f}{3}\mu^2 T \, ,
\eeq
where we have again neglected terms of $O(\alpha_s\log(T))$
\cite{Ipp:2003cj} and assumed the flavor chemical potentials to be equal. 
We see that the specific heat (\ref{cv_cfl})
in the CFL phase, whose excitations are bosonic,
is much smaller than that in unpaired quark matter. 

In the case of 2SC matter, the neutrino emissivity at low temperature
is 1/3 of that of unpaired quark matter. The 2SC emissivity for 
arbitrary temperatures can be found in \cite{Jaikumar:2005hy}. 
In addition to the direct Urca process neutrino pair production
\beq
q + q \to q + q + \nu_\ell + \bar{\nu}_\ell 
\eeq
($q$ is any quark flavor and $\ell$ denotes neutrino flavor) has 
also been studied  \cite{Jaikumar:2001hq}. The rate of this process 
is parametrically smaller than the direct Urca process for very small 
temperatures ($\exp(-2\Delta/T)$ vs.\ $\exp(-\Delta/T)$), but it may 
play a significant role for temperatures close to the superconducting 
phase transition temperature $T_c$.
 
 For the LOFF phase similar arguments apply. The presence of
ungapped modes renders its specific heat ~\cite{Casalbuoni:2003sa}
and its neutrino emissivity due to direct Urca processes
virtually indistinguishable from the unpaired phase \cite{Anglani:2006br}.
However, interesting effects of crystalline structures may be expected 
for other cooling mechanisms. This is not unlike effects in the crust 
of a conventional neutron star, where for instance electron-phonon 
scattering as well as Bragg diffraction of electrons lead to neutrino 
emission via bremsstrahlung processes, see \cite{Yakovlev:2000jp} 
and references therein. 

 The direct Urca processes have also been considered for the gapless CFL 
phase. A distinctive feature of this phase is the fact that the energy 
of one of the quark modes is approximately quadratic in momentum. This 
implies a strong enhancement in the specific heat, which leads to 
very slow cooling at very small temperatures when photon emission 
from the surface dominates the energy loss \cite{Alford:2004zr}. 
However, the instability of this phase at small temperatures 
suggests that this result is most probably of no relevance 
for astrophysics. 

\begin{table*}[t]
\begin{tabular}{|c||c|c|c|c|} 
\hline
 phase  & gap structure &  
 ${\cal G}(\phi)\propto$ & ${\cal K}(\phi)\propto$  
     \\ \hline\hline
CSL & isotropic (no nodes) & \;\; $\phi\exp(-\sqrt{2}\phi)$\;\;  
  & \;\; $\phi^{5/2} \exp(-\sqrt{2}\phi)$\;\;    \\ \hline  
\;\;planar \;\;& anisotropic (no nodes)& $\phi^{1/2}\exp(-\phi)$ & 
$\phi^2\exp(-\phi)$  \\ \hline  
polar & point nodes (linear) & $\phi^{-2}$ & $\phi^{-2}$ 
   \\ \hline
{\it A} & \;\; point nodes (quadratic)\;\;  & $\phi^{-1}$ & 
   $\phi^{-1}$    \\ \hline
\end{tabular}
\caption{Suppression function ${\cal G}(\phi)$ for neutrino emissivity 
in direct Urca processes and suppression function ${\cal K}(\phi)$ for 
specific heat 
for four spin-one color-superconducting phases (abbreviating $\phi\equiv
\Delta/T$, and everything in the limit $\phi\to\infty$). While fully 
gapped modes yield exponential suppression, nodes in the gap yield power 
law suppressions. The gap functions in the polar and {\it A} phases differ 
in the angular direction in the vicinity of the point nodes. A linear 
behavior leads to a stronger suppression than a quadratic behavior.}
\label{tableemissivity}
\end{table*}

Finally, it is interesting to consider neutrino emission from 
single flavor paired matter. Single flavor spin-one
pairing involves small gaps, as well as nodes in the gap parameter, 
and the emissivity is expected to be larger than that of matter
with spin zero pairing. The emissivity of two-flavor quark matter 
with $\langle uu\rangle$ and $\langle dd\rangle$ pairing  
was studied for different spin-one order parameters 
in \cite{Schmitt:2005wg,Wang:2006tg}. The result can be written as
\beq
\label{emissresult}
\epsilon_\nu 
 = \frac{457}{630}\alpha_sG_F^2 T^6\mu_e\mu_u\mu_d\,\left[
\frac{1}{3} + \frac{2}{3}\,{\cal G}(\Delta/T)\right] \, ,
\eeq
where the $u$-quark and $d$-quark gaps are assumed to be identical. All spin-one phases analyzed in 
\cite{Schmitt:2005wg,Wang:2006tg} and described by Eq.\ (\ref{emissresult}) contain ungapped modes 
similar to the 2SC phase. 
Therefore, the
emissivity at low temperatures is simply one third of that of unpaired quark matter. 
(In the case of color-spin locking, all excitations become gapped if one takes into account
nonzero quark masses \cite{Aguilera:2005tg,Schmitt:2005wg} and/or more complicated structures 
of the order parameter \cite{Marhauser:2006hy}.)
The contribution to (\ref{emissresult}) that arises from the paired
quarks, is described
by the nontrivial function ${\cal G}(\Delta/T)$; see \cite{Schmitt:2005wg} for the explicit form 
and numerical evaluation of this function for arbitrary temperatures.
In Table \ref{tableemissivity}, we present the 
behavior of this function for temperatures much smaller than the gap, $T\ll 
\Delta$, for various single-flavor spin-one
color superconducting phases. Although this contribution is small compared to the 
contribution of the ungapped modes, we can use it to show the effect of different 
(anisotropic) gap structures on the parametric behavior of the neutrino emissivity.
We see that, while fully gapped modes lead to an exponential 
suppression of the emissivity, nodes in the gap weaken this 
suppression to a power law. The power law depends on the behavior of the gap in the
vicinity of the nodes.

The specific heat can be written as
\beq
c_V =  T(\mu_u^2+\mu_d^2) \, \left[\frac 13+\frac 23 
   {\cal K} (\Delta/T)\right] \,\, .
\label{s-heat}
\eeq
We show the suppression function ${\cal K} (\Delta/T)$ for the specific 
heat in Table \ref{tableemissivity}. We see that an exponential suppression 
of the emissivity goes along with an exponential suppression of the specific 
heat. 



\section{Color superconductivity in neutron stars}
\label{sec:astro}

Neutron stars are the densest material objects in the universe, with masses of order
that of the sun ($M_\odot$) and radii of order ten km.    Depending on their mass and on
the stiffness of the equation
of state of the material of which they are composed, their central density lies
between $\sim 3$ and $\sim 12$ times nuclear saturation density 
($n_0=0.16$ nucleons$/$fm$^3$)~\cite{Lattimer:2000nx,Lattimer:2006xb}. 
Neutron stars consist of an outer
crust made of a rigid lattice of positive ions embedded within a fluid
of electrons and (in the inner layer of the crust) superfluid neutrons~\cite{Negele:1971vb}.
Inside this crust, one finds a fluid ``mantle'' consisting of neutrons and protons,
both likely superfluid, and electrons.   Determining the composition of
neutron star cores, namely of the densest matter in the universe, remains
an outstanding challenge.\footnote{For review articles on neutron stars as 
laboratories for understanding dense matter, see for instance 
\cite{Weber:2006iw,Page:2006ud,weber,Prakash:2000jr,Lattimer:2004pg,Lattimer:2006xb,Yakovlev:2004iq}.}
If the nuclear equation of state is stiff enough,
neutron stars are made of neutrons, protons and electrons all the way
down to their centers.  If higher densities are reached,  other phases of
baryonic matter (including either a pion 
condensate~\cite{BahcallWolf,Migdal:1971cu,Sawyer:1972cq,Scalapino:1972fu,Baym:1978sz}, a kaon 
condensate~\cite{Kaplan:1986yq,Brown:1995mn},
or a nonzero density of one or several 
hyperons~\cite{Glendenning:1984jr}) may result. Or, neutron star
cores may be made of color superconducting quark matter.  

The density 
at which the transition from baryonic matter to quark matter occurs is
not known; 
this depends on a comparison between the equations of state
for both, which is not well-determined for either.  Very roughly, we expect
this transition to occur when the density exceeds one nucleon per nucleon
volume, a criterion which suggests a transition to quark matter
at densities $\gtrsim 3 n_0$.   The question we shall
pose in this section is how astrophysical observation of neutron stars
could determine whether they do or do not contain quark matter within
their cores.  We have seen throughout the earlier sections of this review
that quark matter at potentially accessible densities may be in the CFL phase,
with all quarks paired, or may be in one of a number of possible phases 
in which there are some unpaired quarks, some of which
are spatially inhomogeneous.   
If quark matter does exist within neutron stars, with their temperatures
far below the critical temperatures for these paired phases,
it will be in some color superconducting phase.
We shall see in this section that these
different phases have different observational consequences, making it
possible for a combination of different types of observational data
to cast light upon the question of which phase of color superconducting
quark matter is favored in the QCD phase diagram, if in fact neutron stars
do feature quark matter cores.

Before turning to the signatures of quark matter in
neutron star cores, we mention here the more radical
possibility that nuclear matter in bulk is metastable at zero pressure,
with the true ground state of strongly interacting matter in the infinite volume limit being 
color superconducting three-flavor quark matter.  According to this
``strange quark matter hypothesis"~\cite{Bodmer:1971we,Witten:1984rs,Farhi:1984qu},
ordinary nuclei are either stabilized by virtue of their small size or are metastable
with lifetimes vastly exceeding the age of the universe.
If this hypothesis is correct, some of the stars that we think are neutron stars
may be strange stars, made entirely of quark matter
\cite{Farhi:1984qu,Alcock:1986hz,Haensel:1986qb,Alcock:1988re}.
Strange stars may have a thin crust (of order 100 meters
thick) of positive ions suspended above the quark matter
surface by an electric field~\cite{Alcock:1986hz}, or they may have a comparably
thin crust of positive ions embedded within the (negatively charged) outer layer
of the quark matter itself~\cite{Jaikumar:2005ne,Alford:2006bx}.  
They cannot, however, have a conventional, km thick, crust.
And, there are many indications that
neutron stars in fact do have conventional crusts.   For example,
the rich phenomenology of X-ray bursts is well-understood only within this 
setting.  More recent evidence comes from the analysis of the quasi-periodic
oscillations with frequencies in the tens of Hz detected in the aftermath of
magnetar 
superbursts~\cite{Israel:2005av,Strohmayer:2005ks,Watts:2005ue,Strohmayer:2006py,Watts:2006ew}, which can be understood as seismic oscillations
of a conventional neutron star 
crust~\cite{Strohmayer:2005ks,Strohmayer:2006py,Watts:2006ew} whereas the thin crusts of
a strange star  would oscillate at  much higher frequencies~\cite{Watts:2006hk}.  
Even if most compact stars are neutron stars not strange stars, 
it remains a logical possibility that some strange stars exist, meaning that
all ordinary neutron stars are metastable. Although possible this scenario is unlikely, 
given that merger events in which strange stars in an inspiralling binary 
are tidally disrupted would litter the universe with small chunks
of quark matter (``strangelets'') and one must then understand why these
have not catalyzed the conversion of all neutron stars to strange 
stars~\cite{Friedman:1990qz}.  We shall devote the remainder of
this section to the more challenging task of using observational
data to constrain the more conservative scenario that quark
matter exists only
above some nonzero transition pressure, namely within
the cores of conventional neutron stars.

\subsection{Mass-radius relation}
\label{subsec:massradius}

It has long been a central goal of neutron star astrophysics 
to measure the masses $M$ and radii $R$ of many neutron stars to a reasonable
accuracy.  Mapping out the curve in the mass-radius plane along which
neutron stars are found
would yield a strong constraint on the equation of state of dense matter.
As this program represents such a large fraction of the effort to
use observations of neutron stars to constrain dense matter physics,
we begin by considering its implications for the presence of quark
matter within neutron star cores.

The larger the maximum mass that can be attained by a neutron star,
the stiffer the equation of state of dense matter, 
and if
stars with masses close to $2~M_\odot$ are found then
the existence of phases with a soft equation of state,
such as baryonic matter with kaon or pion condensation, can be ruled out.
However, although the quark matter equation of state is not known from first principles,
it may easily be as stiff as the stiffer equations of state posited for ordinary
nuclear matter, and neutron  stars with quark matter cores can in fact reach masses 
of order $2~M_\odot$~\cite{Baldo:2002ju,Fraga:2001id,Ruster:2003zh,Alford:2004pf,Blaschke:2007ri}.

The equation of state for CFL quark matter can be parametrized to a good
approximation as~\cite{Alford:2004pf}
\beq
\Omega= -P =  -\frac{3}{4\pi^2}(1-c)\mu^4 
+ \frac{3}{4\pi^2} (M_s^2 - 4 \Delta^2)\mu^2 + B_{\rm eff}\ .
\label{simpleEOS}
\eeq
If $c$ were zero, the $\mu^4$ term would be that for noninteracting quarks; $c$ 
parametrizes the leading effect of interactions, modifying the relation
between $p_F$ and $\mu$.   At high densities, $c=2\alpha_s/\pi$ to leading order in
the strong coupling constant \cite{Freedman:1976ub,Baym:1976yu}.  Analysis of higher order corrections suggests
that $c\gtrsim 0.3$ at accessible densities \cite{Fraga:2001id}.
$B_{\rm eff}$ can be thought of as parametrizing our ignorance of the $\mu$
at which the nuclear matter to quark matter transition occurs.
The $M_s^2\mu^2$ term is the leading effect of the strange quark mass, and
is common to all quark matter phases.   The pressure of 
a color superconducting phase
with less  pairing than in the CFL phase would have a smaller coefficient of
the $\Delta^2\mu^2$ term, and would also differ at order $M_s^4$, here lumped into
a change in $B_{\rm eff}$. 
Because pairing is a Fermi surface phenomenon, it only modifies
the $\mu^2$ term, leaving the larger $\mu^4$ term untouched. However, 
it can nevertheless be important 
because at accessible densities the $\mu^4$ term is largely cancelled
by $B_{\rm eff}$, enhancing the importance of the 
$\mu^2$ term ~\cite{Alford:2002rj,Lugones:2002va}.  
Remarkably, and perhaps coincidentally, if we make the 
(reasonable) parameter choices $c=0.3$, $M_s=275$~MeV 
and $\Delta=100$~MeV and choose $B_{\rm eff}$ such that nuclear matter gives
way to CFL quark matter at the relatively low density $1.5\; n_0$, then
over the entire range of higher densities relevant to neutron stars  the
quark matter equation
of state (\ref{simpleEOS}) is almost indistinguishable 
from the nuclear equation of state due
to Akmal, Pandharipande and Ravenhall (APR) \cite{Akmal:1998cf} that is one of the
stiffest nuclear equations of state in the compendium found 
in \cite{Lattimer:2004pg,Lattimer:2006xb}.  
Neutron stars made entirely of nuclear matter with the APR equation of state
and neutron stars with a quark matter core with 
the equation of state (\ref{simpleEOS})
with the parameters just described
fall along almost indistinguishable curves on a mass vs. radius plot,
with the most significant difference being that
the APR
equation of state admits neutron stars with maximum mass $2.3 M_\odot$, whereas
the introduction of a quark matter core reduces the maximum
mass slightly, to $2.0 M_\odot$~\cite{Alford:2004pf}.

The similarity between a representative
quark matter equation of state and a representative nuclear
equation of state makes clear that it will be very hard to use a future
determination of the  equation
of state to discern the presence of quark matter.  
However, although the numbers in the above paragraph should be taken as indicative
rather than definitive, they do suggest that 
the existence of  a neutron
star whose mass was reliably determined to be 
$>2 M_\odot$ would make it hard to envision such
a star (and hence any lighter stars) having a quark matter core of any appreciable size.
Now that the mass of 
PSR J0751+1807
has been revised
downward from $(2.1 \pm 0.2) M_\odot$~\cite{Nice:2005fi}
to $(1.26^{+.14}_{-.12}) M_\odot$~\cite{NiceStairsPrivateCommunication},
the heaviest known
neutron star orbited by a white dwarf 
is PSR J0621+1002,
whose mass 
is $(1.69^{+0.11}_{-0.16})M_\odot$~\cite{Splaver:2002rj,NiceStairsPrivateCommunication}.
Also, one of the two pulsars Ter5I and Ter5J  (in a globular cluster)
must have a mass that is $>1.68 M_\odot$
at the 95\% confidence level~\cite{Ransom:2005ae}, and the 
mass of the X-ray pulsar Vela X-1 is 
above $1.6~M_\odot$~\cite{Barziv:2001ad}.

Given our lack of knowledge of the equations of
state for nuclear and quark matter,
measuring neutron star masses and radii alone 
do not allow us to reach our goals.

\subsection{Signatures of the compactness of neutron stars}
\label{subsec:compactness}

If we could detect gravity waves from neutron stars spiraling into
black holes in binary systems, the gravitational wave form during the
last few orbits, when the neutron star is being tidally disrupted, will encode
information about the density profile of the neutron star.
For example, upon assuming a conventional density profile,
the gravity wave form encodes information about the ratio
$M/R$~\cite{Faber:2002zn}, essentially via encoding the value
of the orbital frequency at which tidal deformation becomes significant.
This suggests a scenario in which
the presence of an interface separating a denser
quark core from a less-dense nuclear mantle could manifest itself
via the existence of two orbital frequency scales in the wave form, the first being
that at which the outer layers are deformed while the denser quark
core remains spherical and the second being the time at which even
the quark core is disrupted~\cite{Alford:2001zr}.  
This idea must be tested in
numerical relativity calculations, and it may turn out to
be better formulated in some other way.  For example,
perhaps the gravity wave form can be used to constrain the first few moments
of the density profile, and this information can then be used to contrast 
neutron stars with standard
density profiles characterized by a single length scale $R$ with 
those which are anomalously compact because they have a ``step'' in
their density profile.
Whatever the best formulation turns out to be, it seems clear that 
if LIGO sees events in which the tidal disruption of 
a neutron star  occurs within the LIGO band-width,
the gravity wave data will 
constrain the ``compactness'' of the neutron star, providing information
about the density profile that is complementary to that obtained from
a mass-radius relation.

If there is a ``step'' in the density profile at an interface, LIGO gravity waves
may provide evidence for its presence. But, should a density step be expected
if color superconducting quark matter is found in the core of a neutron star?
There are two qualitatively distinct possibilities for the density profile,
depending on the surface tension of the quark matter/nuclear matter interface $\sigma$.
If $\sigma$ is large enough, there will be  a stable, sharp, interface between two 
phases having different densities (but the same chemical potential).
If $\sigma$ is small enough, it becomes  favorable instead to form a macroscopic
volume filled with a net-neutral mixture of droplets of negatively charged quark matter and 
positively charged nuclear matter, see Sec.~\ref{sec_mixed},
which allows a continuous density profile.  
The distinction between these two scenarios has been analyzed quantitatively
for the case of a first order phase transition from nuclear matter to CFL 
quark matter~\cite{Alford:2001zr}. This is the simplest possible phase diagram of QCD,
with a single transition between the phases known to exist at nuclear 
density and at asymptotically high density.  We have seen earlier in this
review that this simple QCD phase diagram is obtained if $\Delta_{\rm CFL}$
is large enough, allowing CFL pairing to fend off stresses that seek to
split Fermi surfaces, all the way down in density until the nuclear matter
takes over from quark matter.
A sharp interface between the (electrically insulating)
CFL phase and (electrically conducting) nuclear matter
features charged boundary layers on either side of the interface,
which play an important role in determining the $\sigma$ above which
this step in the density profile is stable~\cite{Alford:2001zr}.
The critical $\sigma$ 
is about 40 MeV$/$fm$^3$,  lower than dimensional analysis would indicate
should be expected, meaning that the sharp interface with a density step is 
more likely than a mixture of charged components.  
The increase in the density at the interface can easily
be by a factor of two.    
The critical $\sigma$ above which
a sharp interface is favored has not been evaluated for the case of a
first order phase transition between nuclear matter and color superconducting
phases other than the CFL phase.

It is also possible that the long term analysis
of the binary 
double pulsar PSR J0737-3039A~\cite{Burgay:2003jj,Lyne:2004cj} may yield
a measurement of the moment of inertia of this 1.34 solar mass
neutron star~\cite{Kramer:2004gj,Morrison:2004df,Lattimer:2004nj,Kramer:2006nb}.
This could be another route to constraining the compactness of a neutron star,
and perhaps gaining evidence for or against a step in the density profile
of this star.

\subsection{Cooling}
\label{sec:cooling}

The avenues of investigation that we have described so far 
may constrain the possible existence of quark matter within
neutron star cores, but they are not sensitive to the differences among
different color superconducting phases of quark matter. We turn now to 
the first of three observational signatures that have
the potential to differentiate between 
CFL quark matter and other color superconducting phases. 

Within less than a minute of its birth in a supernova, a neutron star cools
below about 1 MeV and becomes transparent to neutrinos.  For the next
million years or so it cools mainly via neutrino emission from its interior.
Photon emission from the surface becomes
dominant only later than that. This means that information about
properties of the interior, in particular its
neutrino emissivity and heat capacity, can be
inferred from measurements of the temperature and age of neutron stars.
Because all forms of dense
matter are good heat conductors \cite{Shovkovy:2002kv},
neutron star interiors are isothermal and the
rate at which they cool is determined by the volume integrals
over the entire interior
of the local emissivity and the local specific heat.  This means
that the cooling tends to be dominated by 
the properties of whichever phase
has the highest neutrino emissivity and whichever phase has
the highest specific heat.

Different forms of dense matter fall into three categories,
ordered by decreasing neutrino emissivity.
The first  category includes any phase of matter that
can emit neutrinos via direct Urca processes,
yielding an emissivity $\epsilon_\nu \propto T^6$.  
Examples include unpaired quark matter, phases of quark matter
with some unpaired quarks including the crystalline phases and
the phases with single flavor pairing in Table III, baryonic
matter containing hyperons, nucleonic matter augmented by either
a pion or a kaon condensate, and even ordinary nuclear matter
at sufficiently high densities that the proton fraction exceeds about $0.1$.
For the specific
case of unpaired quark matter, the emissivity is
given by (\ref{emissivity})~\cite{Iwamoto:1980eb,Iwamoto:1982}, which can be written
as
\begin{eqnarray}
&\epsilon_\nu &\simeq (4  \times 10^{25} {\rm erg}~{\rm cm}^{-3}~{\rm s}^{-1})\times\nonumber\\
&\quad&\frac{\alpha_s}{0.5} \left( \frac{M_s^2/\mu}{100~{\rm MeV}}\right) 
\left(\frac{\mu}{500~{\rm MeV}}\right)^2
\left(\frac{T}{10^9 ~{\rm K}}\right)^6,
\label{unpairedemissivity}
\end{eqnarray}
where we have taken $\mu_e=M_s^2/(4\mu)$, appropriate for neutral unpaired 
quark matter. (Note that $\alpha_s\sim 0.5$ is comparable to the value $c\sim 0.3$
that we used in Sec.~\ref{subsec:massradius}, according to the lowest order relation
$c=2\alpha_s/\pi$. The $\alpha_s/0.5$ factor in (\ref{unpairedemissivity})
could be replaced by $c/0.3$.)  The emissivity of other phases of quark matter in
which only some quarks are unpaired, including the crystalline phases, is
reduced relative to (\ref{unpairedemissivity}), but only by factors of order
unity.  

Ordinary
nuclear matter at densities not too far above $n_0$,
where the proton fraction is less than 0.1, falls into a second category in which
there is no phase space for direct Urca processes and
neutrino emission occurs only via modified Urca processes like
$n+X\rightarrow p + X + e +\bar\nu$ with  $X$ some spectator nucleon,
giving the much lower emissivity
\beq
\epsilon_\nu^{nm}=(1.2\times 10^{20}{\rm erg}~{\rm cm}^{-3}~{\rm s}^{-1})
\left(\frac{n}{n_0}\right)^{2/3}\left(\frac{T}{10^9 ~{\rm K}}\right)^8\ .
\label{modifiedURCA}
\eeq
Neutron stars whose interiors emit neutrinos at this rate, perhaps modified
by effects of nucleon superfluidity,  cool following
a family of  standard cooling curves~(see \cite{Yakovlev:2000jp,Page:2004fy,Yakovlev:2004yr}
and references therein), taking $10^5$ to $10^6$ years
to cool below $10^8$ K.  

CFL quark matter constitutes a third category. As we have seen in 
Sec.~\ref{sec_trans}, it is 
unique among all phases of dense matter in having 
an emissivity $\propto T^{15}$ that is many orders of magnitude smaller than (\ref{modifiedURCA}).
Furthermore, whereas all other phases of dense matter have a specific heat $\propto T$,
in the CFL phase the specific heat is controlled by bosonic excitations making it
$\propto T^3$.  This means that if a neutron star has a CFL core, the total neutrino
emissivity and the total heat capacity of the star are both utterly dominated by
the contributions of the outer layers, whether these are made of nuclear matter
or of some phase that admits direct Urca reactions.  The CFL core holds little
heat, and  emits few neutrinos, but is a good conductor and so stays at the
same temperature as the rest of the star. The rest of the star controls how
the star cools.    

Finally, the single-flavor color superconducting phases
are interesting because they represent a potential transition from the first to
the third categories~\cite{Grigorian:2004jq,Aguilera:2005tg}:  their critical temperatures are so low that if some quarks
can only pair in spin-one channels, they will not pair until after the star
has cooled through an initial epoch of direct Urca emission; and, in
certain cases~\cite{Aguilera:2005tg,Schmitt:2005wg,Marhauser:2006hy}
all quarks can be gapped
below the
critical temperature for color-spin locked pairing, meaning
that these phases ultimately become like CFL quark matter, playing no role in the cooling
of the star which at late times will be controlled by the modified Urca processes
in the nuclear matter mantle.

We can now describe a possible future path to the discovery of CFL quark matter cores
within neutron stars.  Suppose that LIGO detects the gravitational waves
from the tidal disruption of a neutron star with some known mass
spiralling into a black hole and,
as we discussed in Sec.~\ref{subsec:compactness}, suppose that evidence is found that
the density profile of the neutron star has 
a denser core within a less dense mantle, consistent with the existence
of a step in the density profile.  Suppose furthermore that it
was understood by then that neutron stars with that mass cool following
one of the family of standard cooling curves, 
meaning that there can be no component
of their interior within which direct Urca processes are allowed at any time.  This 
combination of
observations would rule out the possibility that the dense core, inside
the density step, contained any of the color superconducting phases that
we have discussed except CFL.

The scenario above may be unlikely, because there are a growing number
of lines of evidence that although the cooling of many neutron stars  is
broadly consistent with the standard cooling curves, some fraction of neutron
stars cool much more quickly.  Examples of neutron stars that are too cold
for their age 
include those in the supernova remnants
3C58 and CTA1~\cite{Page:2004fy,Page:2006ud}.  A second, less direct, piece of evidence
is provided by an unsuccessful search for the X-ray emission from a
cooling neutron star in 15 other supernova 
remnants~\cite{Kaplan:2004tm,Kaplan:2006mb,KaplanPrivateCommunication}.  Although some
of these supernovae may have been Type IA supernovae which do not
produce neutron stars, and although some may have produced black holes,
it is likely that many of these supernovae remnants do contain neutron stars.
Their nonobservation results in an upper limit on their temperatures, and
in all cases this upper limit falls below the standard cooling curves.
A third line of evidence comes from neutron stars  that undergo transient
bouts of accretion~\cite{Brown:1998qv}. X-ray observations of one of these, SAX J1808.4-3658, during its
quiescent phase yield an upper limit on the thermal luminosity of the 
neutron star~\cite{Heinke:2006ie}.  The mean accretion rate averaged over many transient
accretion episodes is known, meaning that the average accretion heating
of the star is known.  The fact that the thermal luminosity is as low as it 
is means that the accretion heating of the star must be balanced by cooling
by neutrino emission at a rate that far exceeds (\ref{modifiedURCA}).
The emissivity for unpaired quark matter (\ref{emissivity}) is consistent with the data, as are the
direct Urca rates for sufficiently dense nuclear matter and for hyperon matter.
Pion condensation or kaon condensation yield emissivities that are proportional
to $T^6$ but with prefactors that are about two orders of magnitude
smaller than that in (\ref{emissivity}), and are ruled out as explanations
for the ability of SAX J1808.4-3658 
to keep cool~\cite{Heinke:2006ie}.  Similar conclusions can also
be inferred from the (even lower) limit on the quiescent luminosity
of the soft X-ray transient 
1H 1905+000~\cite{Jonker:2006td,Jonker:2007ef,DeeptoPrivateCommunication}, 
although in this instance
the time-averaged accretion rate is not as well known.

By now it certainly seems clear that some neutron stars cool much faster
than others.
It is then reasonable to speculate that
lighter neutron stars cool following the standard cooling curve and are
composed of nuclear matter throughout whereas, based on
the three lines of evidence above, heavier neutron stars
cool faster because they contain some form of dense matter that can
radiate neutrinos via the direct Urca process.   This could be quark
matter in one of the non-CFL color  superconducting phases,
but there are other, baryonic, possibilities.  If this speculation is correct,
then if neutron stars contain CFL cores they must be ``inner cores'',
within an outer core made of whatever is responsible for the rapid
neutrino emission.

\subsection{$r$-modes limiting pulsar spins}

A rapidly spinning neutron star
will quickly slow down if it is unstable with respect to  
bulk flows known as Rossby modes, or
$r$-modes, whose restoring force is due to the Coriolis effect and
which transfer the star's
angular momentum into gravitational 
radiation~\cite{Andersson:1997xt,Friedman:1997uh,Andersson:1998qs,Andersson:1998ze,Andersson:2001bz}.
For any given interior composition and temperature, above some
critical spin frequency there is an instability which leads to an exponentially
growing $r$-mode.  This means that as a neutron star is spun up by accretion,
its spin will be limited by a value very slightly above 
this critical frequency, at which the accretion torque is balanced by
gravitational radiation from the $r$-mode 
flows~\cite{Lindblom:1998wf,Owen:1998xg,Bildsten:1998ey,Andersson:1998ze,Andersson:1998qs}.
From a microphysical point of view, the $r$-mode instability is limited
by viscous damping: the greater the damping, the higher the critical
spin above which $r$-modes become unstable.  The critical frequency is
controlled by the shear viscosity in some regimes of temperature (typically lower)
and by the bulk viscosity in others (typically higher).
This means that the existence of pulsars with a given spin,
as well as any observational evidence for an upper limit on pulsar spins,
can yield constraints on the viscosities of neutron star interiors.

There is observational evidence for a physical limit on pulsar spins.
The fastest known pulsar is a  recently discovered radio pulsar
spinning at 716 Hz~\cite{Hessels:2006ze}. 
However, it is not
easy to draw inferences from the
distribution of spins of the many known 
radio pulsars as to whether 716 Hz
is close to some physical limit on the spin frequency
because there are 
significant observational biases that make it harder to
find faster radio pulsars.  
The most rapid pulsars are ``recycled'', meaning that they were spun-up
during an episode of accretion from a binary companion.  During such
accretion, a neutron star may be visible as an X-ray pulsar.  The
spin frequencies of the 13 known millisecond X-ray pulsars lie between
270 and 619 Hz.  
What makes this significant is first of all that
the episodes of accretion have long enough durations that they could
easily spin a neutron star up beyond 1000 Hz, and second of all
that there are no 
selection biases that preclude the discovery of X-ray pulsars with
frequencies as large as 2000 Hz~\cite{Chakrabarty:2003kt,Chakrabarty:2004tp}.  
Analysis of
the observed distribution of X-ray pulsar 
spin frequencies leads to two conclusions: first, the distribution 
is consistent with being uniform;\footnote{These data
thus rule out a proposal for how small quark matter cores could have
been detected~\cite{Glendenning:2000zz}.
If slowly-rotating neutron stars just barely reach quark-matter densities
in their center, then rapidly spinning oblate neutron stars, which
have slightly lower central density, will not contain quark matter.
This ``spinning out'' of a quark matter core could
be detected either by anomalies in braking indices of
radio pulsars that are slowing down~\cite{Glendenning:1997fy} or by 
anomalous population statistics of X-ray pulsars that are being
spun up by accretion~\cite{Glendenning:2000zz}.  The data on X-ray
pulsars show
no sign of such an effect~\cite{Chakrabarty:2003kt,Chakrabarty:2004tp}
indicating that,
if quark matter is present, spinning the star and making
it oblate does not get rid of it.
If neutron stars do have quark matter cores, therefore, the quark matter
must occupy a reasonable fraction of the star.  
}
and, second, 
there is some physical effect that sets a limit on
the allowed spin of a pulsar which (with 95\% confidence) 
is at 730 Hz or lower~\cite{Chakrabarty:2003kt,Chakrabarty:2004tp}.   
It is unlikely that a spin-limit in this vicinity can be
attributed to centrifugal break-up of the spinning neutron star: unless
neutron star radii are larger than anticipated, this ``mass shedding limit''
is significantly higher, above 1 kHz.  
On the other hand, if the observed limit on pulsar spin frequencies
is attributed to the onset of the $r$-mode instability,
the resulting
constraint on the viscosities of neutron star interiors 
is broadly consistent with 
the viscosities of nuclear matter, 
although this consistency is somewhat loose given 
the uncertainties in neutron star densities and in their 
temperatures while being spun up~\cite{Andersson:1998qs,Levin:2000vq}.

The physics of the $r$-mode instability definitively rules out the possibility
that accreting X-ray binary pulsars 
are strange stars that are composed of CFL quark matter
throughout~\cite{Madsen:1999ci}.  
From the results of   Sec.~\ref{sec_trans}, we can conclude that CFL quark matter
has negligible shear damping, and significantly smaller
bulk viscosity than nuclear matter.   (See
\cite{Haensel:2001mw,Haensel:2000vz,Sawyer:1989dp}, 
for calculations of bulk viscosity in nuclear matter, 
\cite{Haensel:2001em,Lindblom:2001hd} for baryonic matter
containing hyperons, and 
\cite{Chatterjee:2007qs} for baryonic matter
containing a kaon condensate.) 
A CFL strange star would therefore have a critical
frequency at which the $r$-mode instability sets in measured in Hz or fractions of Hz,
in gross disagreement with the data on spin frequencies of both X-ray and radio
pulsars.\footnote{Strange stars made of unpaired quark matter or of 2SC quark
matter can be consistent with the data~\cite{Madsen:1999ci}.}

It is a very interesting question, at present unresolved, whether the presence of
a CFL quark matter core within an ordinary neutron star introduces unstable $r$-modes
at low spin frequencies.  If there is a density step at the nuclear/CFL interface, there
may be oscillation modes localized near that interface.  The question is whether there
are $r$-modes that are sufficiently well localized on the CFL side of the interface
that they are undamped, or whether the tails of the mode wave functions that
extend into the nuclear matter side of the interface result in enough damping to 
prevent
the modes from becoming unstable.  Nobody has solved for the $r$-mode wave functions
for a rotating star whose density profile has a
step at an interface, with viscous dissipation
occurring on one side of the interface only.\footnote{Certain other oscillation modes
(``f-modes'' and ``g-modes'') of a nonrotating neutron star whose density profile includes a density
step have been computed~\cite{Sotani:2001bb}.}
If it were to turn out that a star with
a CFL core is even close to as unstable with respect to $r$-modes
as a star that is made entirely of CFL matter,
the existence of pulsars spinning with hundreds of Hz frequencies would immediately
rule out the possibility that these neutron stars have CFL cores.

\subsection{Supernova neutrinos}


The only time when a neutron star
emits enough neutrinos to be detectable on earth as a neutrino source
is during the first few seconds after the supernova explosion.
The time-of-arrival distribution
of supernova neutrinos could teach us about possible phase
transitions to CFL quark 
matter~\cite{Carter:2000xf,Reddy:2002xc,Reddy:2003ap,Kundu:2004mz,Jaikumar:2002vg}. 
All phases of quark matter and nuclear matter except CFL have short enough
mean free paths that the neutrinos detected from a supernova are emitted from a surface
of last scattering called the neutrinosphere, 
inside of which they were diffusing.  This surface of last
scattering moves inward to higher densities
during the first seconds after the supernova, as the 
protoneutron star cools.
Suppose that a volume in the core of the protoneutron star
has made a transition into the CFL phase, in which neutrinos scatter
only off Goldstone bosons which are less numerous (number density $\propto T^3$
rather than $\propto \mu^2 T$ for ungapped quark excitations).
As this core cools, the neutrino
mean free path within a CFL core becomes longer than in any
phase of matter in which there are unpaired quarks (or nucleons) off which
the neutrinos can scatter. The
last supernova neutrinos to arrive could 
carry information about conditions when 
the neutrinosphere reaches the CFL core.
Perhaps there may even be enhanced neutrino luminosity
at the end of an otherwise dropping time-of-arrival distribution,
as all those neutrinos that were previously trapped 
within the transparent core
fly out unimpeded~\cite{Carter:2000xf}. 
Determining whether this proposed signature can arise
requires implementing the transition to a CFL core, with its
long neutrino mean free paths, within a full-fledged simulation of neutrino
transport during a supernova.

\subsection{Rigid quark matter and pulsar glitches}
\label{sec:astro_glitches}

The existence of a rigid crystalline color superconducting core within
neutron stars may have a variety of observable consequences.   
For example,
if some agency (like magnetic
fields not aligned with the rotation axis) could maintain the
rigid core in a shape that has a nonzero quadrupole moment, gravity 
waves would be emitted.  The LIGO non-detection of such gravity waves
from nearby neutron stars \cite{Abbott:2007ce} already limits the possibility that they have rigid cores
that are deformed to the maximum extent allowed by the shear
modulus (\ref{shearmodulus2}), upon assuming 
a range of possible breaking strains,
and this constraint will tighten as LIGO continues 
to run~\cite{Haskell:2007sh,Lin:2007rz}.
(The 
analogous constraint
on strange stars that are rigid throughout was obtained in \cite{Owen:2005fn}.)
Perhaps the most exciting
implication of a rigid core, however, is the possibility that (some) pulsar
``glitches'' could originate deep within a neutron star, in its quark matter core.

A spinning neutron star observed
as a  pulsar gradually spins down as it loses rotational energy
to electromagnetic radiation.
But, every once in a while the angular velocity
at the crust of the star is observed to increase suddenly in a dramatic event called a 
glitch.
The standard 
explanation~\cite{Anderson:1975,Alpar:1977,Alpar:1984,AlparLangerSauls:1984,Epstein:1992,Link:1993,Jones:1997,Alpar:1996,Pines:1985,Link:1995wy}
requires the presence of
a superfluid in some region of the star which also
features a rigid array of spatial inhomogeneities which
can pin the vortices in the rotating superfluid.
In the standard explanation of pulsar glitches, these
conditions are met in the inner crust of a neutron star
which features a neutron superfluid coexisting with
a rigid array of positively
charged nuclei that may serve as vortex pinning sites.  
We shall see below that a rigid core 
made of crystalline color superconducting quark matter
also meets the basic requirements.

The viability of the standard scenario for the origin of
pulsar glitches in neutron star crusts
has recently been 
questioned~\cite{LinkPrivateCommunication}.  Explaining the
issue requires understanding how the basic requirements 
come into play in the generation of a glitch.
As a spinning pulsar slowly loses angular momentum over years,
since the angular momentum of any
superfluid component of
the star is proportional to the density of vortices
the vortices ``want'' to move apart.
However, if within the inner crust 
the vortices are pinned to a rigid structure, these vortices
do not move and
after a time 
this superfluid component of the star is spinning faster
than the rest of the star.  When the ``tension'' built up
in the array of pinned vortices
reaches a critical value, there
is a sudden ``avalanche'' in which vortices unpin, move outwards 
reducing the angular momentum of the superfluid, {\it and then re-pin}.  
As this superfluid suddenly loses angular momentum, the rest 
of the star, including in particular the surface whose angular
velocity is observed, speeds up --- a glitch.   
We see that this 
scenario requires superfluidity coexisting with a
rigid structure to which vortices can pin that
does not easily deform
when vortices pinned to it are under tension.
In very recent work, Link has questioned whether this scenario
is viable
because once neutron vortices
are moving through the inner crust, as must happen during a glitch,
they  are so resistant to bending
that they may never re-pin~\cite{LinkPrivateCommunication}.  
Link concludes that we do not 
have an understanding of any dynamics that could lead to
the re-pinning of moving vortices, and hence 
that we do not currently understand
glitches as a crustal phenomenon.

We have seen in Sec.~\ref{subsec:NJLCrystalline} that
if neutron star cores
are made of quark matter but $\Delta_{\rm CFL}$ is not large enough for
this quark matter to be in the CFL phase, then all of the
quark matter core --- and hence a significant fraction of the moment of inertia of the star ---
may be in one of the crystalline phases described in Sec.~\ref{subsec:NJLCrystalline}.
By virtue of being simultaneously superfluids
and rigid solids, the crystalline
phases of quark matter provide all the necessary conditions
to be the locus in which (some) pulsar glitches
originate.     
Their shear moduli
(\ref{shearmodulus}), namely
\begin{equation}
\nu =3.96\times 10^{33} {\rm erg}/{\rm cm}^3 
\left(\frac{\Delta}{10~{\rm MeV}}\right)^2 \left(\frac{\mu}{400~\rm{MeV}}\right)^2
\label{shearmodulus2}
\end{equation}
with $\Delta$ the gap parameter in the crystalline phase as
in Fig. \ref{fig:deltavsx},
make this form of quark matter 20 to 1000 times more rigid than the crust of a 
neutron star~\cite{Strohmayer:1991,Mannarelli:2007bs}, and hence more
than rigid enough for glitches to originate within them.
The crystalline phases are at the same time superfluid, and
it is reasonable to expect that the superfluid vortices
that will result when a neutron star with such a core rotates
will have lower free energy if they are centered
along the intersections of the nodal planes of the underlying
crystal structure, i.e. along lines along which the condensate
already vanishes even in the absence of a rotational vortex.
A crude estimate of the pinning force on vortices
within crystalline color superconducting quark matter indicates
that it is sufficient~\cite{Mannarelli:2007bs}.
So, the basic requirements for superfluid vortices pinning to
a rigid structure are all present.
The central questions that remain to be addressed are
the explicit construction of vortices in the crystalline phase
and the calculation of their pinning force, as well
as the calculation of the timescale over which sudden changes in the
angular momentum of the core are communicated
to the (observed) surface, presumably either via
the common electron fluid or via magnetic stresses.

Much theoretical work remains before the hypothesis
that pulsar glitches originate within a crystalline color superconducting
neutron star core is developed fully enough to allow it to
confront data on the magnitudes, relaxation
timescales, and repeat rates that characterize glitches.  Nevertheless,
this hypothesis offers one immediate advantage over the conventional
scenario that relied on vortex pinning in the neutron star crust.  
It is impossible for a neutron
star anywhere within which rotational vortices 
are pinned to precess on $\sim$ year time 
scales~\cite{Link:2006nc,Sedrakian:1998vi,Link:2007},
and yet there is now evidence that
several pulsars are precessing~\cite{Stairs:2000,Shabanova:2001ud,Chukwude}.
Since {\it all} neutron stars have crusts, the precession
of any pulsar is inconsistent with the pinning of vortices
within the crust, a requirement in the standard explanation of
glitches.  On the other hand, perhaps not all neutron stars have crystalline
quark matter cores --- 
for example, perhaps the lightest
neutron stars have nuclear matter cores.
Then, if vortices are never
pinned in the crust but are pinned within
a crystalline quark matter core,
those neutron stars that do have a crystalline quark matter core can glitch
but cannot precess while those that don't can precess but cannot glitch.

\section*{Acknowledgments}

We acknowledge helpful conversations with
N.\ Andersson, M.\ Braby, D.\ Chakrabarty, T.\ Hatsuda, S.\ Hughes, D.L.\ Kaplan, B.\ Link,
M.\ Mannarelli, C.\ Manuel, D.\ Nice, D.\ Page, A.\ Rebhan, S.\ Reddy, R.\ Sharma,
I.\ Stairs, Q.\ Wang, and F.\ Wilczek. This research was
supported in part by the Offices of Nuclear Physics and High
Energy Physics of the Office of
Science of the U.S.~Department of Energy under contracts
\#DE-FG02-91ER40628,  
\#DE-FG02-05ER41375 (OJI), 
\#DE-FG02-94ER40818, 
\#DE-FG02-03ER41260. 


\bibliographystyle{apsrmp}
\bibliography{rmp}

\end{document}